\documentclass[twocolumn]{aastex701}  
\usepackage{xcolor, bm, amsmath, enumitem, mfirstuc, float}
\usepackage[T1]{fontenc}
\usepackage{dcolumn}
\usepackage{pifont}
\newcommand{\cmark}{\ding{51}}
\newcommand{\xmark}{\ding{55}}

\definecolor{christi}{rgb}{0.0, 0.58, 0.71}

\newcommand{\trans}[3]{\capitalisewords{#1}\,\textsc{#2}\,#3\,\AA{}}
\newcommand{\spec}[2]{\capitalisewords{#1}\,\textsc{#2}}
\newcommand{\msun}{$M_\odot$}
\newcommand{\mdot}{$\dot{M}_\star$}
\newcommand{\vsini}{$v \,\mathrm{sin}(i)$}
\newcommand{\vinf}{$v_\infty$}
\newcommand{\vinfbe}{$v_\infty^\mathrm{BE}$}
\newcommand{\vinfsei}{$v_\infty^\mathrm{SEI}$}

\newcommand{\vrad}{$v_\mathrm{rad}$}
\newcommand{\vsys}{$v_\mathrm{sys}$}
\newcommand{\vrot}{$v_\mathrm{rot}$}

\newcommand{\teff}{$T_\mathrm{eff}$}

\newcommand{\logg}{$\log(g)$}

\newcommand{\lstar}{$L_\star$}
\newcommand{\mstar}{$M_\star$}

\newcommand{\zsun}{$Z_\odot$}

\newcommand{\ebv}{$E(B-V)$}

\newcommand{\kms}{km\,s$^{-1}$}

\defcitealias{bresolin06}{BPU} 
\defcitealias{bresolin07}{BUG} 
\defcitealias{evans07}{EBU} 
\defcitealias{evans19}{ECG}
\defcitealias{garcia09}{GHV}  
\defcitealias{gull22}{GWS} 
\defcitealias{lorenzo22}{LGN} 

\newcommand{\utah}{Department of Physics and Astronomy, University of Utah, 270 S 1400 E, Salt Lake City, UT 84112, USA} 
\newcommand{\princeton}{Department of Astrophysical Sciences, Princeton University, 4 Ivy Lane, Princeton, NJ 08544, USA} 
\newcommand{\rutgers}{Department of Physics and Astronomy, Rutgers University, 136 Frelinghuysen Road, Piscataway, NJ 08854, USA}
\newcommand{\utaustin}{Department of Astronomy, The University of Texas at Austin, 2515 Speedway, Stop C1400, Austin, TX 78712-1205, USA}
\newcommand{\cfc}{Cosmic Frontier Center, The University of Texas at Austin, 2515 Speedway, Stop C1400, Austin, TX 78712-1205, USA}
\newcommand{\cau}{Institut f{\"u}r Theoretische Physik und Astrophysik, Christian-Albrechts-Universit{\"a}t zu Kiel, Leibnizstr.\ 15, 24118 Kiel, Germany}
\newcommand{\heidelberg}{Zentrum f\"ur Astronomie der Universit\"at Heidelberg, Astronomisches Rechen-Institut, M\"onchhofstr.\ 12-14, 69120 Heidelberg, Germany}
\newcommand{\stsci}{Space Telescope Science Institute, 3700 San Martin Drive, Baltimore, MD 21218, USA}
\newcommand{\rpi}{Department of Physics, Applied Physics, and Astronomy, Rensselaer Polytechnic Institute, 110 8th St, Troy, NY 12180}
\newcommand{\notredame}{Department of Physics and Astronomy, University of Notre Dame, Notre Dame, IN 46556, USA}

\shorttitle{The Treasury of Extremely Metal-Poor O Stars}
\shortauthors{Telford et al.}

\begin{document}

\title{The Treasury of Extremely Metal-Poor O Stars}

\correspondingauthor{O.\ Grace Telford}
\author[0000-0003-4122-7749]{O.\ Grace Telford}
\affiliation{\utah}
\email[show]{grace.telford@utah.edu}

\author[0000-0003-1299-8878]{Christiana Erba}
\email{christi.erba@gmail.com}
\affiliation{\stsci}

\author[0000-0001-5538-2614]{Kristen B.\ W.\ McQuinn}
\affiliation{\stsci}
\affiliation{\rutgers}
\email{kmcquinn@stsci.edu}

\author[0000-0003-0145-8964]{Calum Hawcroft}
\affiliation{\stsci}
\email{chawcroft@stsci.edu}

\author[0000-0002-2090-9751]{Andreas A.\ C.\ Sander}
\affiliation{\cau}
\affiliation{\heidelberg}
\email{andreas.sander@uni-heidelberg.de}

\author[0000-0001-6326-7069]{Julia Roman-Duval}
\affiliation{\stsci}
\email[hide]{duval@stsci.edu}

\author[0000-0002-0302-2577]{John Chisholm}
\affiliation{\utaustin}
\affiliation{\cfc}
\email{chisholm@austin.texas.edu}

\author[0000-0002-4153-053X]{Danielle A.\ Berg}
\affiliation{\utaustin}
\affiliation{\cfc}
\email{daberg@austin.utexas.edu}

\author[0000-0001-5205-7808]{Varsha Ramachandran}
\affiliation{\heidelberg}
\email{vramachandran@uni-heidelberg.de}

\author[0000-0003-4158-5116]{Yong Zheng}
\affiliation{\rpi}
\email{zhengy14@rpi.edu}

\author[0000-0003-2685-4488]{Claus Leitherer}
\affiliation{\stsci}
\email{leitherer@stsci.edu}

\author[0000-0002-9816-9300]{Abby Mintz}
\affiliation{\princeton}
\email{abby.mintz@princeton.edu}

\author[0000-0001-6196-5162]{Evan N.\ Kirby}
\affiliation{\notredame}
\email{ekirby@nd.edu}


\begin{abstract}

The Treasury of Extremely Metal-Poor O Stars (TEMPOS) is a Hubble Space Telescope survey of hot and massive O-type stars in nearby, low-metallicity galaxies ($\lesssim$\,20\% of the solar metallicity, \zsun{}).
Understanding massive-star physics in this regime is essential to interpret observations of metal-poor galaxies, including both low-mass dwarf galaxies and chemically unevolved galaxies in the early Universe.
Yet, few far-ultraviolet (FUV) spectra of O stars of sufficient quality to characterize their fundamental properties and stellar winds exist below 20\%\,\zsun{}, and heterogeneous observation design and incomplete coverage of parameter space pose significant barriers to progress.
To remedy this, TEMPOS obtained new Cosmic Origins Spectrograph (COS) FUV spectra of 12 very metal-poor O stars, building upon archival data to assemble a spectroscopic atlas of 29 homogeneously observed stars that efficiently samples a wide range of spectral types and luminosity classes.
Here, we describe the motivation, sample selection, and observation design for TEMPOS and present the first data release of reduced and coadded COS spectra.
We then present initial results on the empirical properties of FUV O-star spectra below 20\%\,\zsun{}, including radial velocities, equivalent widths of photospheric lines, and terminal wind velocities (\vinf{}).
We show that \vinf{} correlates with host galaxy metallicity across $\sim$\,5--50\%\,\zsun{} and find tentative evidence of a steeper decline in wind strength below $\sim$\,10\%\,\zsun{}.
The combined dataset of FUV spectra and planned releases of photometry and optical spectra from the TEMPOS Treasury program will advance our understanding of both stellar astrophysics and the interstellar medium in the extremely metal-poor regime.

\end{abstract} 


\section{Introduction\label{sec:intro}}

Massive stars strongly influence the evolution of their host galaxies via their production of energetic feedback and metals. 
Despite being formed less frequently than low-mass stars and having much shorter lifetimes (millions versus billions of years), they dominate the overall luminosity of star-forming galaxies and deposit mechanical energy into their surroundings by driving stellar winds and ending their lives as supernova explosions. 
Together, these stellar feedback processes regulate the formation of new stars by heating and ejecting gas from the host galaxy via galactic outflows (e.g., \citealt{naab17}). 
Due to their shallow potential wells, low-mass dwarf galaxies are the most strongly affected by feedback \citep{collins22}.
These also tend to be the most metal-poor galaxies, as both stellar and gas-phase metallicity ($Z$) are tightly correlated with a galaxy's stellar mass \citep[e.g.,][]{tremonti04, berg12}.

Stars more massive than $\sim$20\,\msun{} appear as O-type stars during their main sequence lifetimes, when they have high enough effective temperatures (\teff{}) that their spectral energy distributions (SEDs) peak in the extreme ultraviolet (EUV).
Thus, massive O stars are the primary producers of radiation that can ionize hydrogen (at wavelengths $\lambda$\,$\leq$\,$912$\,\AA{}).
Some of that ionizing radiation escapes from galaxies into the intergalactic medium (IGM) when stellar feedback clears pathways through the dense interstellar medium (ISM; \citealt{erb15}).
At high redshift ($z$\,$\gtrsim$\,6), this process drove cosmic reionization, the Universe's last major phase change from neutral to ionized \citep{becker01}.
Observations of high-redshift galaxies from the James Webb Space Telescope (JWST) have provided evidence that low-mass and very metal-poor galaxies played an important role in reionization \citep[e.g.,][]{topping22, atek24}.
But because the IGM is opaque to ionizing photons at high redshift, we cannot directly measure the ionizing flux escaping from early galaxies.
Our ability to determine which galaxies produced the majority of photons that reionized the Universe therefore depends heavily on our understanding of low-metallicity massive stars' production of ionizing photons and other forms of feedback, which remains highly uncertain \citep{eldridge22}.

Accurate models of massive-star spectra and evolution at low metallicities are essential to infer the properties of both dwarf and high-redshift galaxies and to quantify the stellar feedback governing their evolution.
As the predictions of these models depend sensitively on the implementation of a wide range of physical processes that are not fully understood, observations of individual stars are required to test and calibrate the input physics.
\added{Efforts have been made to constrain dwarf galaxy evolution via the stellar mass--stellar metallicity relation in nearby galaxies (e.g., \citealt{kirby13, kudritzki21}; see also references in Table~\ref{tab:galaxies}), but such scaling relations are insufficient to distinguish among different treatments of the physics underlying massive-star feedback. In particular, uncertainties in} the radiation-driven winds of O and B-type stars impact their \added{predicted} evolution, rotation, nucleosynthesis, and ionizing photon production \citep[e.g.,][]{meynet02, vink22}.
Different approaches to modeling radiation-driven winds predict a wide range of mass-loss rates (\mdot{}). 
Moreover, the resulting mass-loss prescriptions, which are commonly derived from a regression analysis of a sample of models, yield significantly different dependencies on stellar luminosity (\lstar{}) and $Z$ \citep[e.g.,][]{vink01, bjorklund21, krticka25}.
Discrepancies among these theoretical \mdot{} predictions are most pronounced when they are extrapolated to the low-$Z$ regime.

Empirical \mdot{} measurements from wind-sensitive features in far-ultraviolet (FUV) spectra of individual stars are required to test and anchor theoretical mass-loss recipes.
To inform population-wide models of massive-star evolution, spectra, and feedback, FUV observations are needed across the full range of stellar properties, often parameterized by spectral type (SpT, which correlates with \teff{}) and luminosity class (LC, an indicator of evolutionary stage).
Historically, the largest samples of OB stars with FUV spectra available to calibrate stellar models have been in the Milky Way at solar metallicity (\zsun{}; e.g., \citealt{prinja90, walborn90}).
The nearest star-forming dwarf galaxies, the Large and Small Magellanic Clouds (LMC and SMC, respectively) also provide observational access to OB stars at 50\%\,\zsun{} (LMC) and 20\%\,\zsun{} (SMC; \citealt{russell92}).
Yet, FUV spectroscopy was long limited to a relatively small number of low-$Z$ OB stars in the Magellanic Clouds (e.g., \citealt{leitherer2010}) because their extragalactic distances make such observations expensive.

The recent advances of extragalactic observational astronomy into the low-mass and high-redshift galaxy regimes have created strong interest in the physics of massive stars at low $Z$.
The community's need for a comprehensive, empirical spectral atlas informed the implementation of the UV Legacy Library of
Young Stars as Essential Standards (ULLYSES; \citealt{roman-duval25}).
This Hubble Space Telescope (HST) Director's Discretionary program invested 500 orbits in each of two stellar mass regimes: low-mass T-Tauri stars in the Milky Way, and metal-poor, massive OB stars primarily in the Magellanic Clouds.
ULLYSES has provided the community with a FUV spectral library of $>$\,230 massive stars at 20--50\%\,\zsun{} that contains multiple examples of each SpT and LC combination across the full range accessible in the Magellanic Clouds, which was later supplemented with optical spectroscopy by the X-Shooting ULLYSES program \citep{vink23}.
ULLYSES also devoted 10\% of its orbits allocated for high-mass stars to observe low-resolution FUV spectra of six OB stars in dwarf galaxies more metal-poor and distant than the SMC. 
The much smaller number of lower-$Z$ stars that could be observed underscores the challenge of obtaining high-quality spectra of individual stars beyond the Magellanic Clouds. 
Still, the sub-20\%\,\zsun{} regime is highly important for understanding the early phases of galaxy evolution and the contribution of chemically unevolved galaxies to cosmic reionization.

Various HST general observer (GO) programs have observed FUV spectra of very metal-poor O stars in dwarf galaxies more distant than the SMC (e.g., \citealt{garcia14, telford21}), as well as in the the Magellanic Bridge (the gas between the SMC and LMC). However, O stars in the latter represent only a limited range of late SpTs and were found to have diverse abundances, with some reaching LMC-like metallicities \citep{ramachandran21, schosser25}. 
To date, \mdot{} measurements from detailed modeling of FUV spectra have only been reported for 11 O stars in very metal-poor dwarf galaxies \citep{bouret15, telford24, furey25}---too few to sample the full range of O-star spectral classifications, masses, and evolutionary stages below 20\%\,\zsun{}.
Moreover, existing FUV spectra are inhomogeneous in spectral resolution and wavelength coverage, preventing a fully consistent analysis of all the available observations of very low-$Z$ O stars.

Here, we present the Treasury of Extremely Metal-Poor O Stars (TEMPOS; GO-17491; PI: O.\ G.\ Telford), a HST Treasury program that strategically observed additional FUV spectra of O stars in nearby, very metal-poor dwarf galaxies to optimize the sampling of O-star parameter space below 20\%\,\zsun{}.
This survey was designed to efficiently build on archival HST/Cosmic Origins Spectrograph (COS) data and produce a uniformly observed dataset that will enable consistent analysis across a wide range of stellar and wind properties in the very metal-poor regime.
Ultimately, the goal of TEMPOS is to assemble a dataset that can anchor models of O stars' mass loss, evolution, and ionizing photon production at the low metallicities relevant for dwarf and reionization-era galaxies.

The remainder of the paper is organized as follows. 
Section~\ref{sec:science_goals} explains the TEMPOS science objectives, which inform the sample selection and survey design described in Section~\ref{sec:survey}. 
Section~\ref{sec:data} presents the new HST/COS observations obtained by TEMPOS and explains the vetting, reduction, and coaddition of both new and archival COS spectra. Section~\ref{sec:products} presents the resulting data products included in the first TEMPOS data release, which will become publicly available upon publication of this manuscript, and describes planned future releases of ancillary data.
Section~\ref{sec:properties} reports initial results from the TEMPOS HST/COS spectra, including stellar radial velocities, the strength of photospheric iron absorption in the FUV spectra, and the metallicity dependence of stellar wind features.
We summarize in Section~\ref{sec:summary}.
Throughout, we adopt the solar abundances recommended by \citet{lodders25} as our reference.

\section{TEMPOS Science Goals\label{sec:science_goals}}

The primary objective of the TEMPOS program is to provide an empirical benchmark for massive-star astrophysics in the very low-metallicity regime ($\lesssim$\,20\%\,\zsun{}). 
The observations and anticipated results from analyzing the O-star spectra will also enable broader investigations of astrophysics at low metallicity.
Here we provide an overview of the scientific opportunities that motivated the TEMPOS survey design. 

\subsection{Stellar and Wind Properties\label{sec:atm_modeling}} 

First and foremost, TEMPOS was designed to determine the stellar and wind properties of metal-poor O stars across the widest possible range of spectral classifications (i.e., SpTs and LCs).
FUV spectroscopy is essential to measure key properties of O stars, including their abundances and stellar wind parameters like \mdot{} and terminal wind velocity (\vinf{}).
However, additional observations are also needed to fully characterize these stars: optical spectra to cover diagnostics of the fundamental parameters \teff{} and surface gravity (\logg{}; e.g., \citealt{vink23}), and multi-wavelength photometry, ideally spanning the ultraviolet through infrared, to determine \lstar{}, stellar mass (\mstar{}), and line-of-sight dust extinction due to both the Milky Way and host galaxy (e.g., \citealt{gordon16}). 
Thus, TEMPOS is not only a FUV spectroscopic survey; it also includes new and archival optical spectroscopy and HST photometry, as described in Section~\ref{sec:ancillary_products} below.

Together, FUV spectra, optical spectra, and photometry of individual O stars enable quantitative spectroscopic analysis via stellar atmosphere modeling \citep[e.g.,][]{simon-diaz20, sander24}.
\added{With the full TEMPOS dataset, we will infer stellar and wind properties from a homogeneous joint analysis of the FUV and optical spectra} across a wide range of O-star spectral classifications (and therefore masses and evolutionary stages) below 20\%\,\zsun{}, aiming to synthesize a more coherent picture than existing individual efforts \citep{bouret15, telford24, furey25} that analyzed subsets of the TEMPOS sample with different methods.
Moreover, the atmosphere models fit to the observations can predict the O stars' emission in parts of the spectrum not covered by the data (e.g., the unobservable EUV, or infrared wavelengths accessible with JWST).

\subsection{Stellar Population Synthesis and Galaxy Evolution} 

The broader motivation for the TEMPOS program is the lack of empirical constraints at very low $Z$ on the models of stellar mass loss, evolution, spectra, and feedback that are inputs to all stellar population synthesis (SPS) models \citep{tinsley80, leitherer99, conroy13, hawcroft25}.
SPS modeling is an essential tool to infer the physical properties of galaxies across cosmic time from observations of their integrated light.
While the stellar model inputs to SPS codes have historically been calibrated at $\sim$\zsun{} using observations of massive stars in the Milky Way, the greater technical challenge and expense of observing stars in the Magellanic Clouds and beyond has left our SPS model ingredients much more uncertain in the metal-poor regime.

TEMPOS will provide much-needed stellar and wind parameters for a representative sample of very metal-poor O stars, which can then be compared to the predictions of mass-loss and stellar evolution models to constrain the uncertain aspects of the input stellar astrophysics in the low-$Z$ regime.
In particular, analysis of the TEMPOS dataset will constrain the relationship between the strength of radiation-driven stellar winds and fundamental properties of O stars below 20\%\,\zsun{}. 
The best-fit atmosphere models resulting from that analysis will also produce empirically motivated spectral templates for extremely metal-poor massive stars that, importantly, will extend into the ionizing EUV. 

Feedback from hot stars' winds and ionizing radiation sets the state of the ISM where supernovae explode, influencing how efficiently that energy couples to the ISM and drives metal-enriched gas out of their host galaxies \citep{smith21}.
Therefore, the implementation of these processes in SPS models at low $Z$ can have far-reaching implications for both the interpretation of metal-poor galaxy observations and feedback prescriptions in galaxy formation models \citep{naab17}.
Guided by results from TEMPOS, updated models of stellar winds and ionizing spectra informed by observations at very low $Z$ can be integrated into SPS models to enable robust analysis of low-mass galaxies across redshifts and better constrain the ionizing spectra of metal-poor galaxies during the epoch of reionization.

\subsection{Interstellar Medium Abundances and Kinematics\label{sec:ism}}

In addition to the massive-star science, the HST/COS spectra from TEMPOS will enable investigations of the gas in the host dwarf galaxies.
The stellar continua contain various ISM metal absorption lines, including those from key constituents of dust such as iron and volatile metallicity tracers such as sulfur.
In the moderate-resolution G130M and G160M COS gratings, they can be separated into components associated with the Milky Way, the host dwarf galaxy, and even outflows \citep{zheng20, zheng24, poudel25}. 
In fact, many of the archival COS spectra included in TEMPOS are drawn from HST programs with ISM-focused science goals (e.g., GO-15156, PI: Y.\ Zheng; GO-15880, PI: J. Roman-Duval).

The new observations obtained by TEMPOS expand the number of sightlines that can be used to probe ISM physics at very low $Z$. 
For example, detections of outflowing metals along multiple sightlines within individual dwarf galaxies constrain their metal-ejection efficiencies \citep{zheng20}, which can then be compared to the predictions of galaxy formation models to assess the validity of feedback implementations.
The gas-phase abundances can also be determined from ISM absorption lines and compared to the O stars' metal abundances (see Section~\ref{sec:atm_modeling}) to determine the depletion of ISM metals onto dust grains \citep{hamanowicz24}.
How these depletion measurements vary with environment (e.g., gas density, $Z$) constrains models of dust production, evolution, and emission in low-metallicity galaxies \citep{tchernyshyov15, roman-duval22a, roman-duval22b, hamanowicz24}. 


\section{Survey Design and Target Selection\label{sec:survey}}

\movetableright=-1.15in
\begin{table*}
\tabcolsep=0.075cm
\begin{center}
\caption{Properties of the Dwarf Galaxies Hosting TEMPOS Targets\label{tab:galaxies}} 
\begin{tabular}{llD{,}{\pm}{-1}D{,}{\pm}{-1}ccD{,}{\pm}{-1}D{,}{\pm}{-1}c}
Dwarf & Alternate &  \multicolumn{1}{c}{O/H} &  \multicolumn{1}{c}{Fe/H} & Stellar Mass & Distance & \multicolumn{1}{c}{\vsys{}} & \multicolumn{1}{c}{\vrot{}} & \ebv{}$_\mathrm{MW}$ \\
Galaxy & Name(s) &  \multicolumn{1}{c}{((O/H)$_\odot$)}  &  \multicolumn{1}{c}{((Fe/H)$_\odot$)} & ($M_\odot$) & (Mpc) & \multicolumn{1}{c}{(\kms{})} & \multicolumn{1}{c}{(\kms{})} & (mag)\\
\hline
Leo~P & AGC~208583 & 0.03,0.01 & \multicolumn{1}{c}{\nodata} & 2.9$\times 10^5$ & 1.62$\pm$0.15 & 260.8,2.5 & 15,5 & 0.026 \\
Leo~A & UGC~5364, DDO~69 & 0.04,0.01* & 0.04,0.01* & 6.0$\times 10^6$ & 0.74$\pm$0.05 & 22.3,2.9 & \multicolumn{1}{c}{\nodata} & 0.021\\ 
Sextans A & UGCA~205, DDO~75 & 0.08,0.01 & 0.09,0.01 & 4.4$\times 10^7$ & 1.44$\pm$0.05 & 324,2 & 20.0,2.0 & 0.045 \\ 
WLM & UGCA~444, DDO~221 & 0.12,0.03 & 0.13,0.02* & 4.3$\times 10^7$ & 0.98$\pm$0.04 & -130,1 & 30.0,3.0 & 0.038 \\ 
NGC~3109 & UGCA~194, DDO~236 & 0.10,0.02 & 0.21,0.06* & 7.6$\times 10^7$ & 1.34$\pm$0.05 & 403,2 & 72.4,0.5 & 0.067 \\ 
IC~1613 & UGC~668, DDO~8 & 0.14,0.03 & 0.21,0.04 & 1.0$\times 10^8$ & 0.76$\pm$0.02 & -233,1 & 5.9,1.0 & 0.025 \\ \hline
\end{tabular}
\end{center}
\vspace{-10pt}
\tablecomments{Abundances (or bulk [M/H], indicated by *) measured from the spectra of blue supergiants in the host galaxies are adopted from the following sources and converted to a fraction of the solar value, rescaled where possible to $12+\log$(O/H)$_\odot = 8.76$ and $12+\log$(Fe/H)$_\odot = 7.51$ \citep{lodders25}: \citet{urbaneja23} for Leo~A; \citet{kaufer04} (who report [$\alpha$/H], but not an oxygen abundance) for Sextans~A; \citet{bresolin06} and \citet{urbaneja08} for WLM; \citet{evans07} and \citet{hosek14} for NGC~3109; and \citet{bresolin07} and \citet{tautvaisiene07} (who analyzed M supergiants) for IC~1613.
Distances are drawn from the Extragalactic Distance Database (\citealt{jacobs09, anand21}; \url{https://edd.ifa.hawaii.edu/}), and stellar masses, heliocentric systemic velocities (\vsys{}), and maximum observed rotational velocities (\vrot{}) are adopted from the compilation of \citet{mcconnachie12}.
The exception is Leo~P, for which we adopt the gas-phase oxygen abundance from \citet{skillman13}; stellar mass from \citet{mcquinn24}; distance from \citet{mcquinn15}; and \vsys{} and \vrot{} from \citet{bernstein-cooper14}.  \ebv{}$_\mathrm{MW}$ is the Milky Way foreground extinction toward each galaxy from the \citet{schlafly11} dust map.}
\end{table*}

Here, we describe the sample selection and observation design of HST-GO-17491 (PI: O.\ G.\ Telford), a Cycle 31 Large Treasury program targeting metal-poor O-type stars.
These choices were made largely in May 2023, informed by the sixth (and penultimate) ULLYSES data release\footnote{\url{https://ullyses.stsci.edu/ullyses-dr6.html}} from March 2023.

\subsection{Nearby, Star-Forming Dwarf Galaxies with Sub-SMC Metallicities\label{sec:host_galaxies}}

The TEMPOS sample of metal-poor O stars is drawn from dwarf galaxies in the nearby universe that have the following properties: 
\begin{itemize}[noitemsep, topsep=0pt] 
\item contain confirmed O stars with spectral classifications (i.e., SpTs and LCs) determined from ground-based optical spectroscopy;
\item lie at distances $\lesssim$\,1.6\,Mpc (beyond which individual stars become prohibitively faint); 
\item have oxygen and iron abundances $\lesssim$\,20\%\,\zsun{}; and 
\item are minimally reddened by foreground dust in the Milky Way ($E(B-V)_\mathrm{MW} < 0.1$\,mag).
\end{itemize}

\begin{figure*}
\begin{centering}
\includegraphics[width=\linewidth]{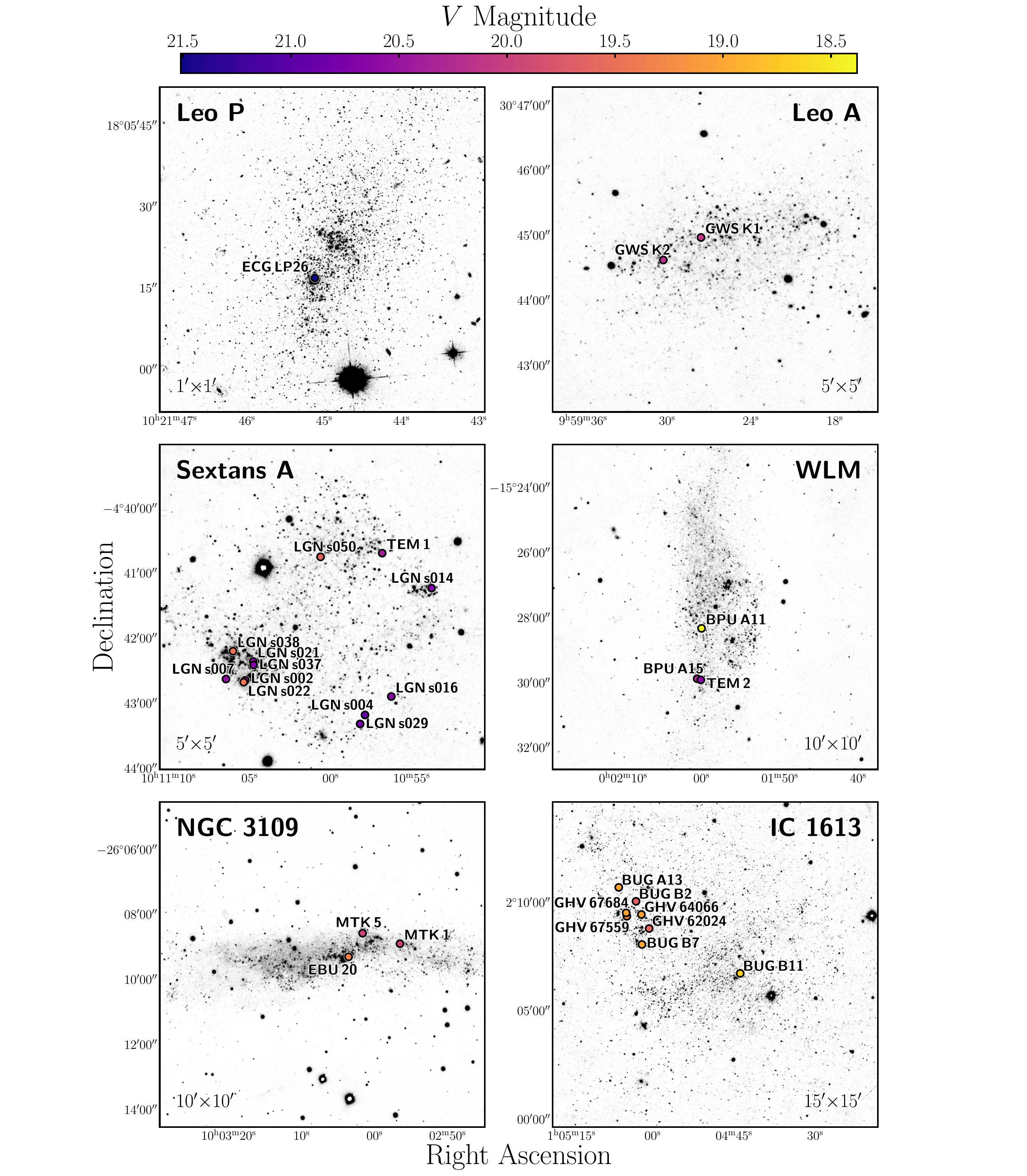}
\caption{\textbf{TEMPOS O-star targets in the context of their host galaxies.} Grayscale images show the six metal-poor dwarf galaxies (Table~\ref{tab:galaxies}) with the angular size of each annotated in the bottom-left or bottom-right corners. All except the top-left panel are Pan-STARRS1 $g$-band images, while the F475W image of Leo~P was observed with HST/ACS. Circles show the locations of the 29 O stars in TEMPOS (both New and Archival samples) and the color-coding indicates the $V$-band magnitude of each star (Table~\ref{tab:targets}). \label{fig:galaxies}}
\end{centering}
\end{figure*}

These requirements yield a sample of 6 gas-rich, star-forming dwarf galaxies whose properties are summarized in Table~\ref{tab:galaxies}.
Figure~\ref{fig:galaxies} shows images of the six metal-poor dwarf galaxies that host TEMPOS targets in grayscale.
Most panels show $g$-band images that were downloaded from the Pan-STARRS1 \citep{chambers16} Image Cutout Server\footnote{\url{https://ps1images.stsci.edu/cgi-bin/ps1cutouts}}.
Leo~P is not easily seen in the Pan-STARRS1 imaging, so the top-left panel of Figure~\ref{fig:galaxies} instead shows a F475W image of that galaxy from the Advanced Camera for Surveys (ACS) onboard HST \citep{mcquinn15}.
Colored circles with black outlines show the locations of the 29 metal-poor O stars in TEMPOS; the sample selection is described in Section~\ref{sec:selection} below.
The color-coding shows the apparent $V$ magnitude, and each star is labeled with its adopted TEMPOS star name (see Table~\ref{tab:targets} below). 

From Figure~\ref{fig:galaxies}, it is apparent that the O stars are not uniformly distributed across the six galaxies: Sextans~A and IC~1613 host the vast majority of the TEMPOS targets. 
The host galaxies' distances also vary widely, from $\sim$12--26 times the distance of the SMC (62\,kpc; \citealt{scowcroft16}).
As a result, the O stars span a range of more than 3~mag in the $V$ band, with the brightest targets in Wolf–Lundmark–Melotte (WLM) and IC~1613 and the faintest in the more distant galaxies Leo~P and Sextans~A. 
Of course, the apparent magnitudes also depend on the intrinsic \lstar{}; stars in NGC~3109 are brighter though its distance is similar to that of Sextans~A, and the two O stars in Leo~A appear faint despite it being one of the closest TEMPOS host galaxies.

Abundances of oxygen and iron have been inferred from optical spectra of massive supergiant stars in most of the TEMPOS host galaxies. 
Though all six host galaxies are metal-poor, they still span a factor of $\gtrsim4$ in both oxygen and iron abundances reported for their later-type, evolved massive stars (or gas-phase oxygen abundance for Leo~P only; Table~\ref{tab:galaxies}). 
We therefore divide them into two groups: the ``extremely metal-poor'' ($\lesssim$\,10\%\,\zsun{}) and the ``metal-poor'' ($\sim$10$-$20\%\,\zsun{}) host galaxies, taking both the oxygen and iron abundances into account.
Our use of the term extremely metal-poor follows the definition of ``extremely metal-deficient'' galaxies as those with metallicities below 10\%\,\zsun{} from \citet{kunth00}, while metal-poor \added{or very metal-poor} in this work refers to galaxies more metal-poor than the SMC.

Generally, the oxygen abundances of these young stars agree with their host galaxies' nebular oxygen abundances, as expected for the gas from which they formed.
The oxygen and iron abundances expressed as a fraction of the solar values, however, do not always agree; supergiants in IC~1613 and NGC~3109 in particular have subsolar oxygen-to-iron ratios. 
For O stars, COS spectra in IC~1613 confirm the expected low oxygen-to-iron ratios \citep{garcia14, bouret15}, but suggest a higher iron abundance (and lower oxygen-to-iron ratio) in WLM than the supergiant results \citep{bouret15, telford24}. 
Similarly detailed iron abundance measurements have not yet been done for O stars in the other TEMPOS galaxies, leaving the relative importance of individual metals for setting their wind strengths and ionizing spectra unclear. 
In future work, we will use atmosphere modeling to constrain the detailed stellar abundances for the TEMPOS sample (see Section~\ref{sec:atm_modeling}), but in this paper we simply group the stars into the metal-poor and extremely metal-poor categories based on the oxygen and iron abundances of their host galaxies' supergiant populations.

\subsection{O Star Target Selection\label{sec:selection}}

\movetableright=-0.4in
\begin{table*}
\caption{TEMPOS Targets with New and/or Archival HST/COS Spectra}
\label{tab:targets}
\tabcolsep=0.15cm
\begin{tabular}{rlcccD{.}{.}{-1}cc}
Host & Star & Spectral & R. A. & Decl. & \multicolumn{1}{c}{$V$} & New COS & Archival \\
Galaxy & Name & Classification & (J2000) & (J2000) & \multicolumn{1}{c}{(mag)} & Data & COS Data \\
\hline
\multicolumn{8}{c}{\textbf{Extremely Low Metallicity ($\bm{\lesssim10\%\,Z_\odot}$)}}\\ \hline

Leo P & ECG LP26$^a$ & O7-8 V & 10:21:45.12 & $+$18:05:16.93 & 21.51* &  & \cmark \\
Leo A & GWS K1 & O9 V & 09:59:27.52 & $+$30:44:57.75 & 20.10* & \cmark & \cmark \\
Leo A & GWS K2 & O9.7 V & 09:59:30.22 & $+$30:44:37.08 & 20.20* & \cmark & \cmark \\
Sextans A & LGN s002 & O4 Vz & 10:11:05.28 & $-$04:42:38.36 & 20.37 & \cmark & \cmark \\
Sextans A & LGN s004 & O5 III & 10:10:57.89 & $-$04:43:10.20 & 20.92 & \cmark & \cmark \\
Sextans A & LGN s007 & O6 V & 10:11:06.47 & $-$04:42:37.23 & 20.52 & \cmark & \cmark \\
Sextans A & LGN s014 & O7.5 III((f)) & 10:10:53.80 & $-$04:41:13.11 & 20.69 &  & \cmark \\
Sextans A & LGN s016 & O8 II & 10:10:56.27 & $-$04:42:53.14 & 20.64 & \cmark &  \\
Sextans A & LGN s021 & O8 V & 10:11:04.79 & $-$04:42:20.96 & 20.60 &  & \cmark \\
Sextans A & LGN s022 & O8 V & 10:11:05.38 & $-$04:42:40.15 & 19.46 &  & \cmark \\
Sextans A & LGN s029 & O8.5 III & 10:10:58.19 & $-$04:43:18.46 & 20.80 &  & \cmark \\
Sextans A & LGN s037 & O9 I & 10:11:04.77 & $-$04:42:24.24 & 20.68 &  & \cmark \\
Sextans A & LGN s038 & O9 I((f)) & 10:11:06.04 & $-$04:42:11.49 & 19.49 &  & \cmark \\
Sextans A & LGN s050 & O9.7 I & 10:11:00.65 & $-$04:40:44.40 & 19.61 &  & \cmark \\
Sextans A & TEM 1 & O9.7-B0 V & 10:10:56.85 & $-$04:40:40.90 & 20.40 &  & \cmark \\\hline \multicolumn{8}{c}{\textbf{Low Metallicity ($\bm{\sim10-20\%\,Z_\odot}$)}}\\ \hline
WLM & BPU A11 & O9.7 Ia & 00:01:59.95 & $-$15:28:19.58 & 18.38 &  & \cmark \\
WLM & BPU A15 & O7 V((f)) & 00:02:00.53 & $-$15:29:52.41 & 20.25 &  & \cmark \\
WLM & TEM 2 & O9.7-B0.2 III-V & 00:02:00.06 & $-$15:29:54.64 & 20.60 & \cmark &  \\
NGC 3109 & EBU 20 & O8 I & 10:03:03.30 & $-$26:09:21.34 & 19.33 & \cmark & \cmark \\
NGC 3109 & MTK 1 & O5 If & 10:02:56.24 & $-$26:08:58.23 & 19.99* & \cmark &  \\
NGC 3109 & MTK 5 & O8 II & 10:03:01.34 & $-$26:08:38.06 & 19.97* & \cmark &  \\
IC 1613 & BUG A13 & O3-4 V((f)) & 01:05:06.26 & $+$02:10:43.26 & 19.02 &  & \cmark \\
IC 1613 & BUG B11 & O9.5 I & 01:04:43.83 & $+$02:06:45.16 & 18.68 &  & \cmark \\
IC 1613 & BUG B2$^b$ & O7.5 III-V((f)) & 01:05:03.07 & $+$02:10:04.64 & 19.68 &  & \cmark \\
IC 1613 & BUG B7$^b$ & O9 II & 01:05:01.98 & $+$02:08:05.16 & 18.99 &  & \cmark \\
IC 1613 & GHV 62024$^c$ & O6.5 IIIf & 01:05:00.65 & $+$02:08:49.35 & 19.60 &  & \cmark \\
IC 1613 & GHV 64066$^d$ & O3 III((f)) & 01:05:02.08 & $+$02:09:28.17 & 19.03 & \cmark & \cmark \\
IC 1613 & GHV 67559$^b$ & O8.5 III((f)) & 01:05:04.77 & $+$02:09:23.28 & 19.24 &  & \cmark \\
IC 1613 & GHV 67684$^d$ & O8.5 I & 01:05:04.89 & $+$02:09:32.67 & 19.02 & \cmark & \cmark \\\hline
\end{tabular}
\tablecomments{We adopt star names following the convention of the ULLYSES Low-$Z$ sample. Stellar catalog papers: ECG -- \citet{evans19}; GWS -- \citet{gull22}; LGN -- \citet{lorenzo22}; TEM -- this paper; BPU -- \citet{bresolin06}; EBU -- \citet{evans07}; MTK -- \citet{mintz25}; BUG -- \citet{bresolin07}; GHV -- \citet{garcia09}. Most spectral classifications are taken from the stellar catalog papers, except for the following references (indicated after the star name): $^a$\,\citet{mcquinn15}; $^b$\,\citet{garcia14}; $^c$\,\citet{herrero12}; $^d$\,\citet{garcia13}.
We note that updated spectral classifcations will be reported in future TEMPOS papers using new ground-based optical spectroscopy (see Section~\ref{sec:ancillary_products}).  
$V$ magnitudes (or HST F475W magnitudes, indicated by * in the $V$ column) were drawn from \citet{massey07} for all stars in Sextans~A and WLM, and from the stellar catalog papers for all other stars except for LP26, which was reported by \citet{telford21}.}
\end{table*}

A major goal of TEMPOS is to provide uniform FUV spectroscopy for metal-poor O stars of as many spectral classifications as possible.
We discuss the reasoning behind our grating choices in Section~\ref{sec:gratings} below, but the key requirement is that every star in the sample be observed with the G130M grating and either G160M or G140L; so, at least two gratings are required for a star to be included in the TEMPOS sample.
\added{Here, we describe the process by which O stars in the six dwarf galaxies in Table~\ref{tab:galaxies} were selected, then discuss the impacts of those selection criteria on the resultant TEMPOS sample.}

\added{\subsubsection{Selection Criteria}}

First, we checked the available HST/COS spectra in the Mikulski Archive for Space Telescopes\footnote{\url{https://mast.stsci.edu}} (MAST), which were compiled into the ULLYSES Low-$Z$ sample\footnote{\url{https://ullyses.stsci.edu/ullyses-targets-lowz.html}}.
This resulted in a sample of 17 O stars with existing observations that met our grating requirements (see Section~\ref{sec:gratings} below) across the six nearby, metal-poor dwarf galaxies, which we include in TEMPOS and refer to as the ``Archival'' sample.
ULLYSES includes additional stars that we rejected from the TEMPOS Archival sample for various reasons:
\begin{itemize}[noitemsep, topsep=0pt] 
\item ULLYSES includes both O and B stars, whereas TEMPOS is focused on only the hottest O stars that dominate ionizing photon production;
\item a few O stars in the Magellanic Bridge sample only a limited range of late-O SpTs and are known to have diverse iron abundances \citep{ramachandran21, schosser25}; 
\item one He star in Leo~A is likely a product of binary mass transfer \citep{senchyna21}; and
\item many ULLYSES O stars have been observed with only one COS grating, which is insufficient to achieve the science goals of TEMPOS.
\end{itemize}

For stars in the last category above, we checked whether their spectral classifications were already represented in the TEMPOS Archival sample. 
If they added to the coverage of O-star parameter space in either the metal-poor or extremely metal-poor regime (considered separately), then we obtained new COS spectra of that star in a different grating to complement the archival data.
We also checked O stars in Sextans~A that were targeted by a Cycle 30 HST/COS program (GO-17111; PI: M.\ Garcia) that executed too late to be included in the ULLYSES sample, and from these, identified two stars with only G140L data that would expand the parameter space coverage of TEMPOS.
We did not observe any stars that are known binaries or that lie within the 2.5$''$ diameter of the COS primary science aperture of a UV-bright star in archival HST Wide-Field Camera 3 (WFC3) imaging.
For the latter check, we inspected NUV F275W imaging taken by HST programs GO-15275 (LUVIT; PI: K. Gilbert), GO-15880 (METAL-Z; PI: J. Roman-Duval), and GO-16104 (ULLYSES; PI: J. Roman-Duval).
This resulted in a total of eight stars that were previously observed with COS in only one grating for which we obtained new COS data in a second grating.

Finally, we drew additional targets for new COS observations from two sources: the recent catalog of OB stars in Sextans~A presented in \citet{lorenzo22}, and our own observations\footnote{The Keck observations used in this work are available in the Keck Observatory Archive: \url{https://koa.ipac.caltech.edu/}} of massive stars in TEMPOS galaxies with the Keck Cosmic Web Imager (KCWI; \citealt{morrissey18}) and DEep Imaging Multi-Object Spectrograph (DEIMOS; \citealt{faber03}) instruments on the Keck II Telescope (programs 2021B\_N194 and 2023A\_N048; PI: O.\ G.\ Telford; \citealt{telford24, mintz25}).
To ensure that stars would be good candidates for COS spectroscopy, we again checked that none had bright companions within 1.25$''$ in archival HST F275W imaging. 
Only stars for which COS spectra reaching a signal-to-noise ratio (SNR) of 10 per resolution element in two gratings could be obtained in $\leq$\,20 orbits were considered.
Within these requirements, we identified 4 additional O stars, not in the ULLYSES sample or targeted by any previous HST program, for new COS observations to maximize coverage of the metal-poor O-star parameter space.
We refer to the 12 stars with new COS spectra from TEMPOS as the ``New'' sample.

Table~\ref{tab:targets} presents the basic properties of the full (New and Archival) TEMPOS O-star sample.
We follow the naming scheme used in the ULLYSES Low-$Z$ sample, where the star identifier is preceded by the initials of the first three authors of the catalog paper from which that identifier is adopted. 
For reference, Table~\ref{tab:aliases} in Appendix~\ref{app:aliases} provides alternate names for the TEMPOS targets to facilitate matching with archival sources. 
Spectral classifications are generally adopted from the same catalog paper, with exceptions noted in the Table~\ref{tab:targets} comments.
Right ascension (R.A.) and declination (Decl.) are the coordinates reported for each star in  \textit{Gaia} DR3\footnote{\url{https://gea.esac.esa.int/archive/}} \citep{gaia16b, gaia23j}.
$V$ magnitudes for most stars are compiled from ground-based imaging (sources given in the caption of Table~\ref{tab:targets}).
For the following stars, the F475W magnitude measured from HST imaging with either ACS or the Wide Field Camera 3 (WFC3) is given instead: ECG~LP26 \citep{telford21}, GWS~K1, K2 \citep{gull22}, and MTK~1, 5 \citep{mintz25}.
All magnitudes are in the Vega system.
The final two columns indicate whether new COS observations from HST-GO-17491 and/or archival COS data are included in the TEMPOS data products presented in Section~\ref{sec:products} below.

Two new spectral classifications are reported in this paper, for stars TEM~1 in Sextans~A and TEM~2 in WLM (alternate names are given in Table~\ref{tab:aliases} in Appendix~\ref{app:aliases}). 
TEM~1 is part of the Archival sample, with COS G130M and G160M spectra from program HST-GO-16767 (PI: O.\ G.\ Telford). 
An optical spectrum of this star was observed with the DEIMOS spectrograph by Keck program 2022B\_N011 (PI: O.\ G.\ Telford) and will be presented in Hawcroft et al.\ (in preparation). 
TEM~2 is part of the New sample, and was identified in the same Keck/KCWI observation of WLM described in \citet{telford24} targeting the star BPU~A15 in the Archival sample.
That spectrum of TEM~2 will be presented alongside new DEIMOS spectra of OB stars in WLM (Erba et al.\ in preparation) as part of the followup optical spectroscopy component of the TEMPOS program described in Section~\ref{sec:ancillary_products} below.
Finally, we emphasize that the spectral classifications in Table~2 are drawn from the literature, and will be updated based on our new optical spectra in future TEMPOS papers.

\begin{figure*}
\begin{centering}
\includegraphics[width=\linewidth]{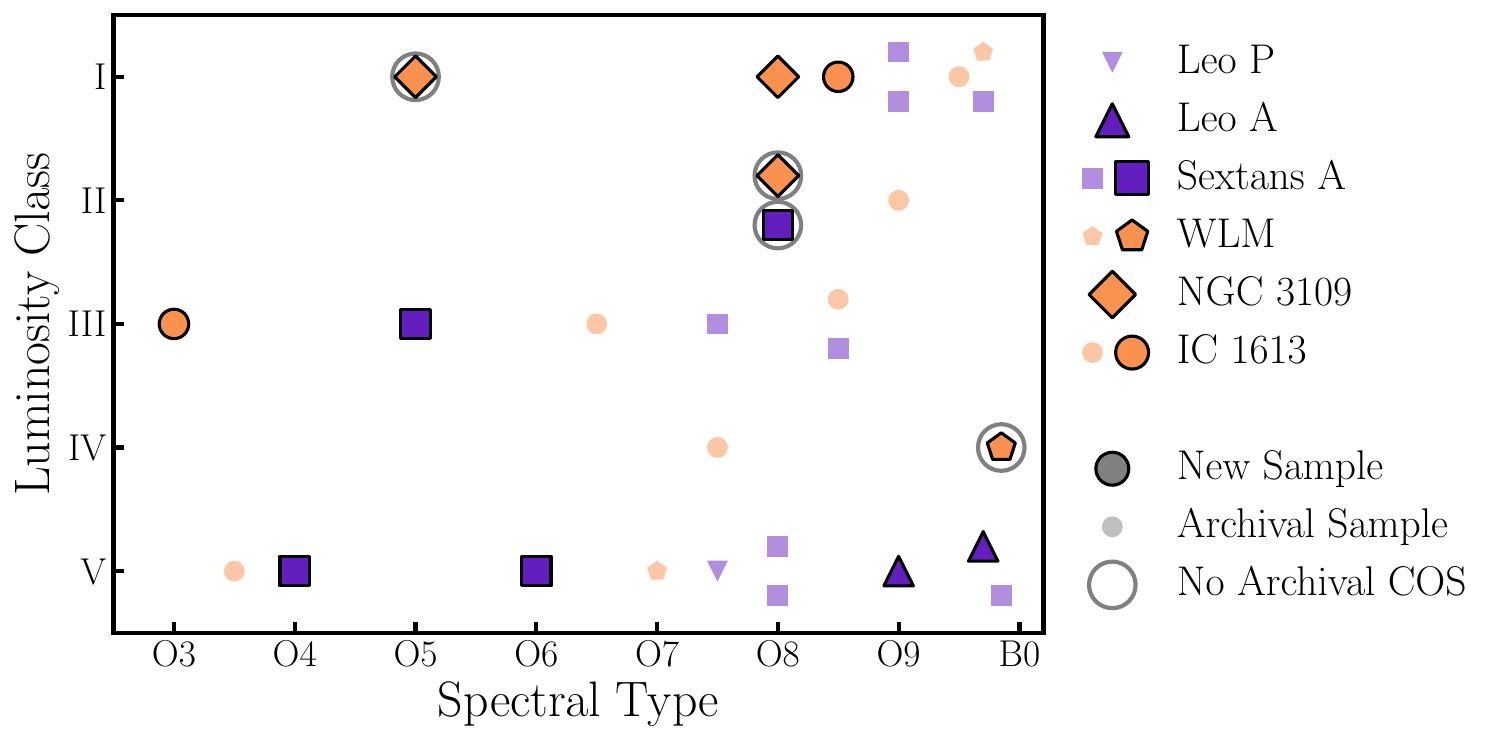}
\caption{\textbf{TEMPOS coverage of O-star parameter space at very low metallicity.} LC is plotted as a function of SpT for O stars in the New (large, dark points outlined in black) and Archival (small, light points) samples, as reported in Table~\ref{tab:targets}.
Marker shape encodes the host galaxy, while the color corresponds to whether that galaxy is in the extremely metal-poor (purple) or metal-poor (orange) category.
Points encircled in gray indicate stars that were never observed with COS before TEMPOS (HST-GO-17491).
Stars assigned a range of SpT or LC are plotted at the central value of that range, and for overlapping points, a small vertical offset is applied for clarity. New TEMPOS observations improve the sampling of this parameter space at both $\sim$10--20\%\,\zsun{} (orange) and $\lesssim$\,10\%\,\zsun{} (purple).\label{fig:paramspace}}
\end{centering}
\end{figure*}

Figure~\ref{fig:paramspace} shows the distribution of the TEMPOS sample in both SpT and LC.
Each point represents a star, where the marker shape corresponds to the host galaxy, and the color indicates whether the host galaxy belongs to the metal-poor (orange) or extremely metal-poor (purple) category.
Stars in the New sample are shown as larger, darker points with black outlines, while the Archival sample stars are smaller and lighter points.
From Figure~\ref{fig:paramspace}, it is clear that the Archival sample spans only a small range of this parameter space, primarily covering SpTs of O7 or later (but few LC III--V stars later than O8). 

Before TEMPOS, data meeting the wavelength coverage and resolution requirements described in Section~\ref{sec:gratings} did not sample the full O-star parameter space below 20\%\,\zsun{}.
The rarity of O stars in these nearby dwarf galaxies, particularly the hottest, early SpTs, results in sparser coverage than was achieved in the SMC and LMC with ULLYSES. 
Still, the COS observations of the New sample fill out this parameter space as much as possible at very low $Z$, given our current knowledge of O stars in these nearby dwarf galaxies.
In particular, stars earlier than O7 across the full range of LCs are now represented; several O8~I--II stars sample intermediate SpTs for supergiants; and three late-O~III--V stars probe the most common, lowest-mass O stars. 

\added{\subsubsection{Possible Selection Effects}

The aim in sample construction was to include one example of a metal-poor O star of every spectral classification possible (though multiple examples of some stellar types exist in the Archival sample).
Because the host galaxies have low star-formation rates and their massive stellar populations have not been exhaustively studied with ground-based optical spectroscopy, not all spectral classifications are represented in the  TEMPOS sample.
In these local metal-poor dwarf galaxies, early-O stars are extremely rare, which is reflected in the denser sampling of late-O types to the right side of Figure~\ref{fig:paramspace}. 
If anything, the early-O stars may be over-represented in TEMPOS compared to their abundance relative to late-O stars in their host galaxies, particularly because only a handful of the known late-O stars in these galaxies have been observed with HST/COS to date.
The coverage of Figure~\ref{fig:paramspace} is similar across the two metallicity categories, except that there are no early-O supergiants in the extremely metal-poor sample (purple) and no late-O dwarfs in the metal-poor sample (orange).
The Magellanic Clouds, in contrast, have higher star-formation rates and correspondingly more examples of early-O stars that allowed ULLYSES to obtain UV spectroscopy of multiple O stars of nearly every spectral classification \citep{roman-duval25}. 

The practical requirements of UV-bright stars and low dust extinction for COS observations could plausibly impart biases in the types of stars represented in TEMPOS.
However, examples of both early and late spectral types were excluded from the sample due to their faintness; there is not a monotonic relationship between spectral type and UV apparent magnitude, particularly due to the varying distances of the host galaxies. 
Also, despite the fact that we excluded known multiples from the sample, it is entirely possible that many TEMPOS targets have unresolved companions. One signpost of multiplicity is an inferred \lstar{} higher than expected for the star's spectral classification. This may be more common in brighter stars, but examples of unexpectedly high \lstar{} among the faintest stars in the TEMPOS Archival sample are already known (ECG~LP26 and LGN~s029; \citealt{telford24}). Thus, while the sample selection preferred UV-bright stars, that does not obviously imply a bias in the properties of the stars ultimately included in the TEMPOS sample.
}

\subsection{Choice of Gratings\label{sec:gratings}}

To measure the wind properties (\mdot{}, \vinf{}) and detailed metal abundances of O stars requires fitting stellar atmosphere models to the observed helium and metal lines in their FUV spectra.
The most important wind-sensitive features are the N\,\textsc{v}\,1240\,\AA{}, Si\,\textsc{iv}\,1400\,\AA{}, and C\,\textsc{iv}\,1550\,\AA{} doublets, and observing multiple wind diagnostics is essential to break degeneracies between \mdot{} and other model parameters.
Important photospheric metal lines that constrain the abundances and ionization state of stellar atmospheres include C\,\textsc{iii}\,1176\,\AA{}, O\,\textsc{iv}\,1341\,\AA{}, He\,\textsc{ii}\,1640\,\AA{}, N\,\textsc{iv}\,1719\,\AA{}, and forests of Fe\,\textsc{iv} and Fe\,\textsc{v} lines between 1300--1730\,\AA.
Thus, we require spectra covering the full $\sim$1150--1750\,\AA{} range.

Because photospheric metal lines are weak in metal-poor stellar spectra, moderate resolution ($R$\,$\gtrsim$\,4000) aids in detecting those essential diagnostics.
Moderate resolution also enables better constraints on the wind-line profiles than what is possible with low-resolution data, particularly in the separation of the contributions of the broad wind component and narrow ISM lines to the total observed profile.
Spectrally resolved line profiles also provide information on the projected rotation speed, \vsini{}.
In the FUV, the complex of closely spaced C\,\textsc{iii} transitions near 1176\,\AA{} is often used for this purpose \citep{heap06}, though it can be impacted by the stellar wind in late-O and B supergiants.

The ideal strategy would be to observe moderate-resolution spectra across the full 1150--1750\,\AA{} range, using both the G130M and G160M COS gratings ($R$$\sim$12,000$-$16,000)\footnote{\url{https://hst-docs.stsci.edu/cosihb/chapter-13-cos-reference-material/13-3-gratings}}.
However, we found that G160M observations would require prohibitively long integration times due to the lower throughput of that grating, particularly at the redder end. 
Therefore, we opted to use the G130M grating at a central wavelength (cenwave) of 1291~\AA{}, combined with the low-resolution G140L grating ($R$$\sim$1,500$-$4,000) at cenwave of 800~\AA{} to cover redder wavelengths ($\sim$1425--1750\,\AA{}).
Though its lower resolution will be less sensitive to weak lines and detailed line profiles, G140L can accurately measure the strength and extent of the broad C\,\textsc{iv}\,1550\,\AA{} wind feature (e.g., \citealt{furey25}), and also help to constrain the Fe abundance via the amount of continuum removed in the Fe\,\textsc{iv} forest redward of $\sim$1425\,\AA{}.
Fortunately, multiple transitions of each C, N, O, and Fe fall within the G130M grating, so detections of those weak absorption lines will enable constraints on the detailed metal abundances even without moderate-resolution data redward of the G130M wavelength coverage.
Many key ISM diagnostic lines are also covered by G130M, so the TEMPOS dataset will enable ISM science even though G160M spectra are not available for all stars (Section~\ref{sec:science_goals}). 

Ultimately, TEMPOS obtained G130M and/or G140L data for the 12 targets in the New sample, complementing archival data in one grating for 8 of those stars (see Table~\ref{tab:targets}). 
New G130M spectra were obtained for any stars that had not been previously observed with that grating, but new G140L spectra were obtained only if neither G140L nor G160M data existed in MAST for that star. 
We targeted a SNR of 10 per resolution element at the lowest-throughput extremes of the wavelength range covered by the G130M grating (specifically, at 1430\,\AA{}) to ensure secure detections of the key diagnostic lines C\,\textsc{iii}\,1176\,\AA{} and Si\,\textsc{iv}\,1400\,\AA{}.
This also results in sufficient SNR that all important ISM lines covered by G130M are detectable, including the only Fe lines at 1142--1144\,\AA{}.
For the G140L grating, we followed the ULLYSES observing strategy for low-$Z$ stars and required SNR=15 at 1130\,\AA{} \citep{roman-duval25}, which yields SNR\,$\geq$\,10 per resolution element at the C\,\textsc{iv}\,1550\,\AA{} wind line. 
Altogether, the TEMPOS design ensures consistent wavelength coverage, resolution, and SNR to enable uniform spectroscopic analysis across the full sample of very metal-poor O stars.


\section{HST/COS Observations and Data Reduction\label{sec:data}}

Here, we describe the acquisition, reduction, and coaddition of the FUV spectra in the New and Archival samples that are included in the TEMPOS data products (described in Section~\ref{sec:products} below). 

\subsection{New HST/COS Spectra Obtained by TEMPOS\label{sec:new_obs}}

Program HST-GO-17491 acquired COS spectroscopy of all 12 O stars in the TEMPOS New sample (see Table~\ref{tab:targets}) between 2023 December 26 and 2026 February 18.
The 110 awarded orbits of COS time were divided into 42 visits of $\leq3$ orbits, many of which had to be re-observed due to failed guide star reacquisition. 
We used the G130M grating with a central wavelength of 1291\,\AA{} (FP-POS=3,\,4) and the G140L grating with a central wavelength of 800\,\AA{} (FP-POS=all).
One exception to that strategy was the one-orbit G140L observation of GHV~64066, for which we used only FP-POS=1,\,4 to minimize the impact of overheads on SNR.
All observations were taken in TIME-TAG mode through the 2\farcs5 primary science aperture (PSA).
Because precise coordinates from \textit{Gaia} were available for all TEMPOS targets, we used a NUV imaging target acquisition (TA) strategy using the PSA with MIRRORA in most cases.
We opted to use MIRRORB due to UV-bright objects near two of our targets (GHV~64066 and GWS~K2).

\subsection{New and Archival Data Reduction and Quality Assessment\label{sec:reduction}}

All available COS observations of the New and Archival targets were downloaded from MAST. 
We collected all \texttt{x1d.fits} files, which contain the reduced spectrum from each science exposure  produced by the CalCOS pipeline (v3.6.0), for all FUV COS gratings (G130M, G160M, and/or G140L) from all HST GO programs that observed a given target. 
We also downloaded raw acquisition (\texttt{rawacq.fits}) files for all targets and visits from MAST.
The details of all COS spectroscopy we used for each TEMPOS target are provided in Table~\ref{tab:obsinfo} in Appendix~\ref{app:observations}.

We checked all files for data quality issues.
First, we checked whether TA was successful using the \texttt{ACQSTAT} header keyword of all \texttt{rawacq.fits} files.
In cases where imaging (rather than spectroscopic) TA was used, we plotted the acquisition image to verify that the targets were well-centered in the 2\farcs5-diameter COS aperture, and that an acquisition image had been obtained.
We then checked the \texttt{EXPFLAG}, \texttt{EXPTIME}, and \texttt{PLANTIME} header keywords in the \texttt{x1d.fits} files to assess whether data were successfully acquired and what fraction of the planned exposure time executed.
Any datasets for which the exposure time was $\leq80\%$ of that planned, or that contained no data, were rejected. 
We also calculated the median SNR of each spectrum across 1200--1700\,\AA{} and found that several \texttt{x1d.fits} files contained data so dominated by noise that the median SNR was less than 0; such files were rejected. 
Table~\ref{tab:obsinfo} lists all \texttt{x1d.fits} files that were not used in the TEMPOS data products.

To verify that the wavelength and flux calibration were consistent across all observations of a given target, we plotted and visually compared all spectra from individual \texttt{x1d.fits} files.
We also inspected spectra that were observed with the same grating by different GO programs (which may have been years apart) to confirm that there are no qualitative differences across those observations.

\subsection{Coadding and Binning the Spectra\label{sec:coadds}}

\begin{figure*}
\begin{centering}
\includegraphics[width=0.8\linewidth]{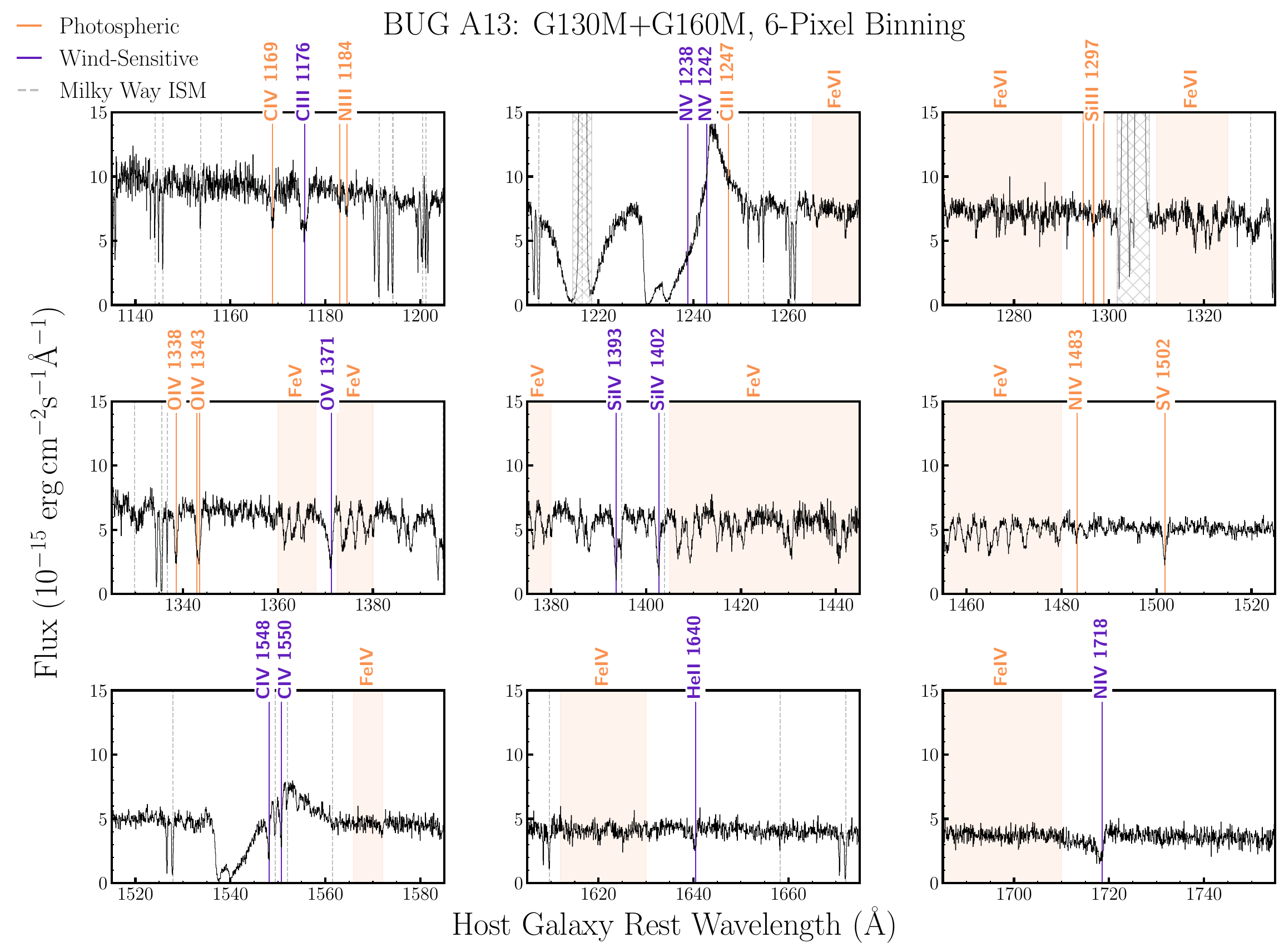}
\includegraphics[width=0.8\linewidth]{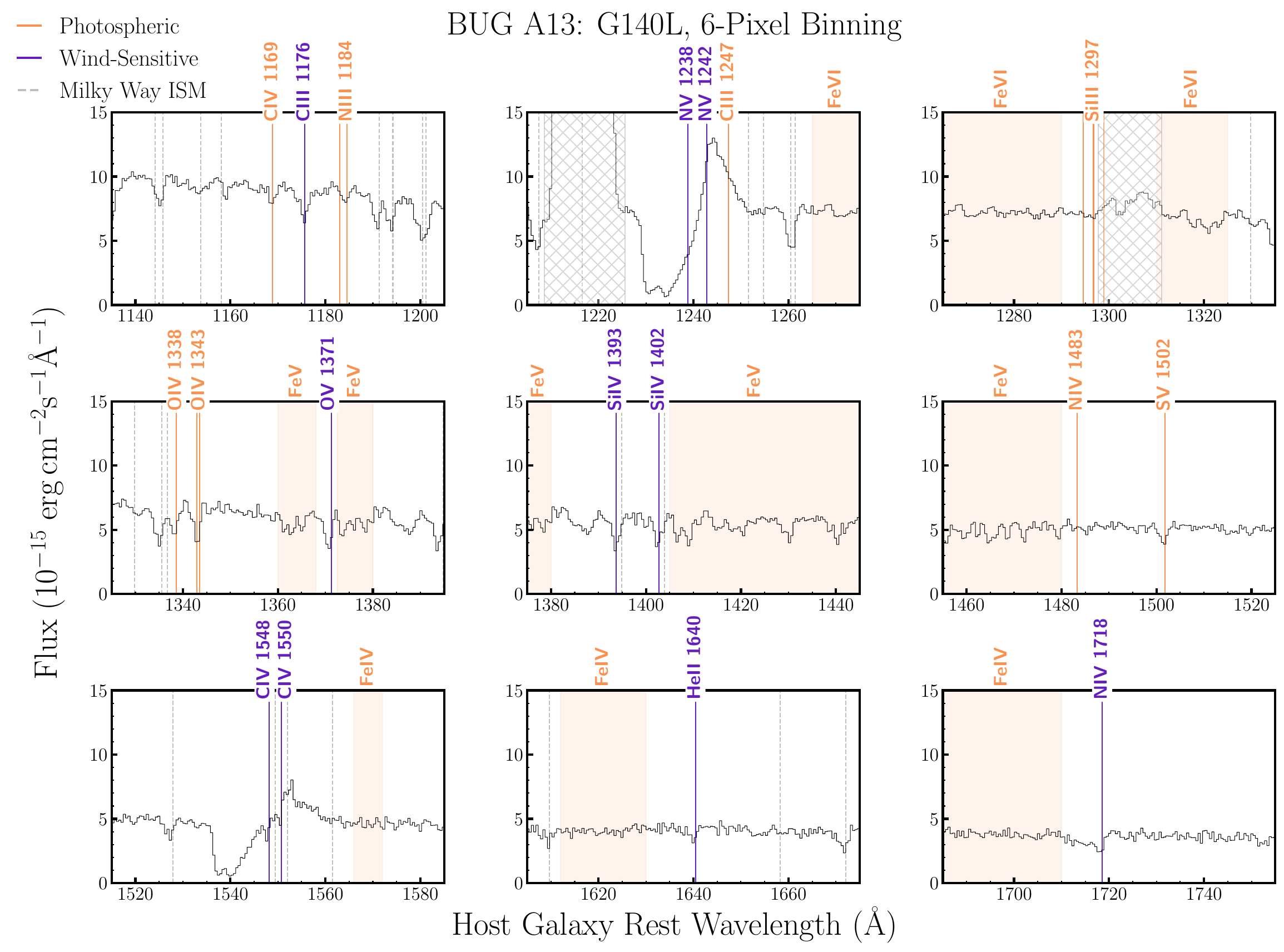}
\caption{\textbf{Overview of diagnostic features in the TEMPOS COS spectra.} Observed flux is plotted as a function of wavelength in the rest frame of the host galaxy (black) for 9 wavelength ranges in the coadded COS spectra of the example star BUG~A13. The top 3 rows show the moderate-resolution G130M and G16M gratings, and the bottom 3 rows show the low-resolution G140L grating. Solid vertical lines highlight important FUV photospheric (orange) and wind (purple) diagnostic features, though some wind-sensitive features can appear as purely photospheric (e.g., \trans{c}{iii}{1176} here). Orange shading highlights regions of the spectra that contain many closely spaced iron transitions. Gray dashed vertical lines show the location of Milky Way ISM absorption features, and gray hatched regions cover contamination by geocoronal emission (Ly$\alpha$, \spec{o}{i}). \label{fig:keylines}}
\end{centering}
\end{figure*}

We combined all individual reduced \texttt{x1d.fits} exposures that passed our quality cuts for a given star and grating to produce a coadded COS spectrum, or coadd.
We used the software\footnote{\url{https://github.com/spacetelescope/ullyses}} developed by the ULLYSES program \citep{roman-duval25}, with minor modifications to enable combining COS spectra that were collected by different HST programs and handle different target naming conventions and fits header keywords in TEMPOS versus ULLYSES (see Section~\ref{sec:products} below).
This package includes the coaddition algorithm now used by Hubble Advanced Spectral Products\footnote{\url{https://archive.stsci.edu/missions-and-data/hst/hasp}} (HASP), as well as code to produce the high-level science products (HLSPs) that were released by ULLYSES.
In brief, the individual COS spectra taken with a given grating are first resampled onto a common wavelength grid using the largest dispersion and a nearest-neighbor interpolation approach, which avoids correlated errors. 
\added{In practice, because we only coadd spectra observed with the same COS grating, this procedure preserves the native COS sampling.}
The resampled spectra are then averaged, weighted by the throughput multiplied by the exposure time at each wavelength sample.
Further details are provided on the ULLYSES website\footnote{\url{https://ullyses.stsci.edu/ullyses-data-description.html\#CoaddSpectra}} and in Roman-Duval et al. (in preparation). 

All COS data for the New and Archival samples were downloaded from MAST following a major update of reference files for geometric distortion and walk corrections in 2025 July \citep{indriolo25}. 
These updates, together with a new high-voltage-dependent sensitivity correction, have improved the wavelength and flux calibration of all COS FUV spectra\footnote{\url{https://www.stsci.edu/contents/news/cos-stans/july-2025-stan}}. 
Therefore, the TEMPOS coadds for the Archival sample may differ from the ULLYSES HLSP spectra for the same stars.
It is recommended to use the most up-to-date data from MAST, so the TEMPOS coadds represent an improvement over the ULLYSES data products that used the older reference files.

Because the TEMPOS targets are quite faint, even the coadded spectra have modest SNR (typically $<$\,10 per pixel across the full wavelength range).
To make the stellar FUV line profiles easier to see and measurements of line strengths more robust, we binned the coadded spectra to boost the SNR.
Coadds are binned within individual gratings only---ensuring uniform wavelength sampling across the input spectrum---by six pixels, equal to the size of a theoretical COS resolution element across all gratings\footnote{\url{https://hst-docs.stsci.edu/cosihb/chapter-3-description-and-performance-of-the-cos-optics/3-2-size-of-a-resolution-element}}.
These binned and coadded spectra are used in all figures and measurements throughout this paper and are available in the TEMPOS data products (described in Section~\ref{sec:products} below).
To provide a sense of the data quality, Table~\ref{tab:snrs} in Appendix~\ref{app:observations} reports the continuum SNR per resolution element at several wavelengths spanning the full range covered by the COS spectra.

Figure~\ref{fig:keylines} presents a summary of the key diagnostic lines covered by the TEMPOS COS observations, using the star BUG~A13 in IC~1613 as an example. 
This star has been observed in all three gratings, enabling a visual comparison of the moderate- and low-resolution spectra. 
The top half of the figure shows nine wavelength ranges in the moderate-resolution gratings (G130M and G160M), and the bottom half shows the same wavelength ranges for the G140L spectrum of the same star. 
In each panel, the observed flux as a function of wavelength (corrected to the host galaxy heliocentric systemic velocity, \vsys{}) is shown in black, and gray hatched regions indicate prominent airglow emission lines that are often seen across the TEMPOS sample.
Orange and purple vertical lines indicate the locations of photospheric absorption lines and of lines that can be stellar wind diagnostics, respectively.
Not all potentially wind-sensitive lines will show wind signatures for a given star, as the utility of various wind diagnostics depends on the stellar SpT and LC.
Orange shaded regions in some panels highlight regions of the COS spectra with large numbers of iron transitions, known as ``iron forests'' (\spec{Fe}{vi}, \spec{Fe}{v}, and \spec{Fe}{iv} from shorter to longer wavelengths).
Finally, dashed gray vertical lines show the expected locations of strong absorption features due to the Milky Way ISM (at zero velocity).

In Figure~\ref{fig:keylines}, well-developed P-Cygni wind profiles are visible for the \trans{n}{v}{1238,\,1242} and \trans{c}{iv}{1548,\,1550} doublets, and blueshifted wind absorption is also present for the \trans{o}{v}{1371}, \trans{he}{ii}{1640}, and \trans{n}{iv}{1718} lines.
BUG~A13 drives a strong stellar wind with a high \mdot{}, as found by previous quantitative analyses of the COS spectra \citep{garcia14, bouret15, furey25}.
Moreover, various photospheric metal lines are clearly detected in this relatively high-SNR spectrum, including prominent \spec{Fe}{vi} and \spec{Fe}{v} lines.

Comparing the top and bottom halves of Figure~\ref{fig:keylines}, the advantages of moderate-resolution spectra are obvious.
Though longer exposure time is required to achieve the same SNR at higher spectral resolution, the $R$$\sim$12,000$-$16,000 G130M and G160M spectra enable more precise analysis of line profiles; detection of weaker absorption lines that may not be apparent in the lower-resolution G140L spectra, particularly structure in the Fe forests; and clear separation of narrow ISM absorption from broader stellar wind features (see, for example, the \trans{s}{iv}{1393,\,1402} profiles).
The ability to detect and separate ISM absorption makes the moderate-resolution TEMPOS coadds valuable for a wide range of scientific applications beyond the analysis of the massive stars themselves (Section~\ref{sec:ism}). 
Still, the same stellar features are identifiable in both the moderate- and low-resolution spectra of BUG~A13, particularly the shape and strength of the \trans{c}{iv}{1548,\,1550} wind doublet and \spec{fe}{iv} forest redward of the G130M wavelength coverage.
Thus, we expect to constrain key stellar and wind properties even for the subset of TEMPOS targets for which only G130M and G140L spectra are available. 


\section{TEMPOS Data Products\label{sec:products}}

Here, we describe the public data products from TEMPOS, which will be released as high-level science products (HLSPs) in MAST. 
We present the HLSPs in the first data release, available upon publication of this paper via \dataset[10.17909/fcda-fn73]{\doi{10.17909/fcda-fn73}} and on the TEMPOS HLSP website\footnote{\url{https://archive.stsci.edu/hlsp/tempos/}}.
We then provide a summary of planned ancillary data products from TEMPOS that will be made public in future data releases.

\subsection{First Data Release: Coadded COS Spectra\label{sec:cos_products}}

\movetableright=-0.2in
\begin{table}
\caption{Contents of TEMPOS HLSP \texttt{*cspec.fits} Files}
\label{tab:filestructure}
\tabcolsep=0.15cm
\begin{tabular}{ll}
\hline
\multicolumn{2}{l}{\textbf{Extension 0 (`PRIMARY') Header Keywords}}\\ \hline
Keyword & Description \\ \hline
TELESCOP & Telescope used to acquire data \\                   
INSTRUME & Instrument used to acquire data \\               
DETECTOR & Detector used to acquire data \\
DISPERSR & Identifier of disperser \\
CENWAVE & Central wavelength setting \\
APERTURE & Identifier of entrance aperture \\
OBSMODE & Instrument operating mode \\
HLSPTARG & TEMPOS target name \\
RADESYS & World coordinate reference frame\\
TARG\_RA  & Target right ascension (degrees) \\
TARG\_DEC & Target declination (degrees) \\ 
CENTRWV & Central wavelength of the coadd \\
MINWAVE & Minimum wavelength in coadd \\
MAXWAVE & Maximum wavelength in coadd \\ \hline
\multicolumn{2}{l}{\textbf{Ext. 1 (`SCIENCE') and 3 (`BINNED') Columns}}\\ \hline
Name & Units \\ \hline
WAVELENGTH & \AA{} \\
FLUX & erg\,s$^{-1}$\,cm$^{-2}$\,\AA{}$^{-1}$ \\
ERROR & erg\,s$^{-1}$\,cm$^{-2}$\,\AA{}$^{-1}$ \\
SNR & \\
EFF\_EXPTIME$^\dag$ & s \\ \hline
\multicolumn{2}{l}{\textbf{Ext. 2 (`PROVENANCE') Columns}}\\ \hline
Name & Description \\ \hline
FILENAME & Names of *x1d.fits files \\
PROPOSID & Proposals that acquired data \\
TELESCOPE & Telescopes used to acquire data \\                   
INSTRUMENT & Instruments used to acquire data \\               
DETECTOR & Detectors used to acquire data \\
DISPERSER & Identifiers of dispersers \\
CENWAVE & Central wavelength settings \\
APERTURE & Identifiers of entrance apertures \\
SPECRES & Resolving power at cenwaves \\
CAL\_VER & Versions of CalCOS pipeline \\
MJD\_BEG & Date-times of observation start \\
MJD\_MID & Date-times of observation midpoint \\
MJD\_END & Date-times of observation end \\
XPOSURE & Exposure times of observation \\
MINWAVE & Minimum wavelengths in spectra \\
MAXWAVE & Maximum wavelengths in spectra \\ \hline
\end{tabular}
\tablecomments{$^\dag$EFF\_EXPTIME is only in the SCIENCE extension; this column does not exist in the BINNED extension.}
\end{table}

We release the TEMPOS COS spectra as multi-extension fits files, which are designed to integrate seamlessly with the ULLYSES HLSPs so that the astronomical community can consistently analyze observations across the full range of metallicity from $\sim$5--50\%\,\zsun{}.
We only release single-grating coadded spectra, analogous to the ULLYSES \texttt{*cspec.fits} files, as these are the recommended products for any science applications. 
The ULLYSES \texttt{*aspec.fits} and \texttt{*preview-spec.fits} files abut spectra with different wavelength sampling and resolution, which are intended only for ``quick look'' data visualization.

The HLSP files are named as: \texttt{hlsp\_tempos\_hst\_cos\_} \texttt{<galaxy-star-name>\_<grating>\_v1\_cspec.fits}, where \texttt{<galaxy-star-name>} gives the galaxy and star names listed in Table~\ref{tab:targets} and \texttt{<grating>} is the COS grating.
Table~\ref{tab:filestructure} lists the metadata in the primary header (fits extension 0) for the coadds, as well as metadata for the individual \texttt{*x1d.fits} files that contributed to each coadd, which are provided as arrays in the columns of fits extension 2.
The columns of fits extensions 1 (`SCIENCE') and 3 (`BINNED') contain the following arrays for the coadds output by the ULLYSES software at native sampling and binned by six pixels (equal to one COS spectral resolution element; see Section~\ref{sec:coadds}), respectively: wavelength, flux, error, and SNR per pixel. For the coadds at native sampling in extension 1 only, the effective exposure time (i.e., the total exposure time for good pixels contributing to the coadd) is also reported as a function of wavelength.  
The first three fits extensions are organized identically to the ULLYSES \texttt{*cspec.fits} HLSPs to facilitate uniform analysis of the TEMPOS and ULLYSES datasets, and we additionally provide the spectra binned to one COS resolution element in a fourth extension for convenience. 

\subsection{Planned Releases of Ancillary Data: Ground-Based Optical Spectra and HST Photometry\label{sec:ancillary_products}}

To enable quantitative analysis of the TEMPOS COS spectra (see Section~\ref{sec:atm_modeling}), we obtained complementary ground-based optical spectroscopy from the W.\ M.\ Keck Observatory (programs 2024B\_N014 and 2024B\_N015; PI: O.\ G.\ Telford). 
We used the multi-object spectrograph DEIMOS to observe four custom slitmasks targeting the TEMPOS O stars in four of the six host galaxies for which moderate-resolution ($R$$\sim$4000) optical spectra from Keck were not already available.
Similar-quality spectra were previously observed for all TEMPOS targets in the remaining two galaxies, Leo~P \citep{telford23, telford24} and NGC~3109 \citep{mintz25}, using KCWI and DEIMOS, respectively. 

We also included likely O and B stars in the metal-poor host galaxies selected from NUV HST photometry on our DEIMOS slit masks, following a similar selection method to that described in \citet{mintz25}. 
This observation design gave us the opportunity to characterize as many OB stars in these metal-poor dwarf galaxies as possible, and potentially identify new O stars with SpTs and LCs not yet represented in the TEMPOS sample. 
Catalogs of the observed OB stars, as well as assigned spectral classifications and fundamental parameters measured from the optical spectra, will be presented in a series of forthcoming papers.
Our DEIMOS observation design and reduction methodology will be presented in Erba et al.\ (in prep.), and the coadded optical spectra will be made public in a future data release. 

Finally, TEMPOS will also produce catalogs of stellar photometry measured from archival HST imaging in the NUV and optical (and also NIR where possible), closely following the methods of \citet{telford21} and \citet{mintz25}.
The HST photometry of all targets of our COS and DEIMOS observations will be presented in future papers and made publicly available via the TEMPOS HLSP website hosted by MAST. 


\section{Initial Results: the FUV Properties of Extremely Metal-Poor O Stars\label{sec:properties}}

Here, we present an empirical characterization of the TEMPOS COS spectra.
We measure the stellar radial velocities (\vrad{}) and equivalent widths (EWs) of photospheric absorption lines, then estimate the stellar wind terminal velocities (\vinf) from the \trans{c}{iv}{1548} transition. 

\subsection{Stellar Radial Velocities\label{sec:rvs}}

\movetableright=0.2in
\begin{table*}
\caption{Radial Velocity Measurements from Coadded COS Spectra}
\label{tab:rvs}
\tabcolsep=0.1cm
\begin{tabular}{rlcccccc}
Galaxy & Star Name & $v_\mathrm{rad}^\mathrm{G130M}$ & $v_\mathrm{rad}^\mathrm{G130M+G160M}$ & $v_\mathrm{rad}^\mathrm{G140L}$ & Adopted $v_\mathrm{rad}$ & $v_\mathrm{rad}-v_\mathrm{sys}$ \\
 & & (\kms{}) & (\kms{}) & (\kms{}) & (\kms{}) & (\kms{}) \\
\hline
\multicolumn{7}{c}{\textbf{Extremely Low Metallicity ($\bm{\lesssim10\%\,Z_\odot}$)}}\\ \hline
Leo P & ECG LP26 & \nodata & 273 & \nodata & 273 & 12 \\
Leo A & GWS K1 & \nodata & 20 & \nodata & 20 & -3 \\
Leo A & GWS K2 & \nodata & 28 & \nodata & 28 & 6 \\
Sextans A & LGN s002 & 331 & \nodata & 151 & 331 & 7 \\
Sextans A & LGN s004 & 305 & \nodata & 115 & 305 & -19 \\
Sextans A & LGN s007 & 398 & \nodata & 96 & 398 & 74 \\
Sextans A & LGN s014 & 281 & \nodata & 258 & 281 & -43 \\
Sextans A & LGN s016 & 320 & \nodata & 80 & 320 & -4 \\
Sextans A & LGN s021 & \nodata & 328 & \nodata & 328 & 4 \\
Sextans A & LGN s022 & 323 & \nodata & 289 & 323 & 0 \\
Sextans A & LGN s029 & \nodata & 289 & \nodata & 289 & -35 \\
Sextans A & LGN s037 & \nodata & 320 & \nodata & 320 & -4 \\
Sextans A & LGN s038 & 273 & \nodata & 281 & 273 & -51 \\
Sextans A & LGN s050 & 301 & \nodata & 276 & 301 & -23 \\
Sextans A & TEM 1 & \nodata & 331 & \nodata & 331 & 7 \\\hline \multicolumn{7}{c}{\textbf{Low Metallicity ($\bm{\sim10-20\%\,Z_\odot}$)}}\\ \hline
WLM & BPU A11 & \nodata & -128 & \nodata & -128 & 2 \\
WLM & BPU A15 & \nodata & -140 & \nodata & -140 & -10 \\
WLM & TEM 2 & -108 & \nodata & -263 & -108 & 22 \\
NGC 3109 & EBU 20 & 380 & \nodata & 321 & 380 & -23 \\
NGC 3109 & MTK 1 & 411 & \nodata & 279 & 411 & 8 \\
NGC 3109 & MTK 5 & 415 & \nodata & 430 & 415 & 12 \\
IC 1613 & BUG A13 & \nodata & -252 & -298 & -252 & -19 \\
IC 1613 & BUG B11 & \nodata & -240 & \nodata & -240 & -7 \\
IC 1613 & BUG B2 & -213 & \nodata & -254 & -213 & 20 \\
IC 1613 & BUG B7 & \nodata & -231 & -275 & -231 & 2 \\
IC 1613 & GHV 62024 & -237 & \nodata & -267 & -237 & -4 \\
IC 1613 & GHV 64066 & -230 & \nodata & -447 & -230 & 3 \\
IC 1613 & GHV 67559 & -222 & \nodata & -275 & -222 & 11 \\
IC 1613 & GHV 67684 & -277 & \nodata & -546 & -277 & -44 \\\hline
\end{tabular}

\tablecomments{Stellar radial velocities (\vrad{}) measured in different COS gratings. Uncertainties on the velocities are dominated by the wavelength calibration accuracy of 3\,\kms{} and 150\,\kms{} for the moderate- and low-resolution gratings, respectively. We adopt measurements from the moderate-resolution gratings (G130M only or G130M and G160M together, if available) as the star's \vrad{}, but use the measured $v_\mathrm{rad}^\mathrm{G140L}$ to analyze the G140L spectra in Sections~\ref{sec:photlines} and \ref{sec:windlines}. The last column reports the difference between the stellar \vrad{} and host galaxy \vsys{} (Table~\ref{tab:galaxies}).}
\end{table*}

First, we measure \vrad{} from the observed wavelengths of various photospheric absorption lines.
This is required to correct the observed spectra to the rest frame to ensure consistent EW measurements over fixed rest-frame wavelength ranges across the sample (see Section~\ref{sec:photlines}) and to measure the extent of the blueshifted wind absorption in the rest frame (see Section~\ref{sec:windlines}).
Moreover, it is interesting to quantify the occurrence rate of offsets larger than expected from the galaxies' rotation curves between \vrad{} and the host galaxy systemic velocity (\vsys{}; Table~\ref{tab:galaxies}), as such offsets may be due to a previous dynamical interaction or unseen companion. 

The velocity calibration accuracy differs for different COS gratings\footnote{\url{https://hst-docs.stsci.edu/cosihb/chapter-5-spectroscopy-with-cos/5-1-the-capabilities-of-cos}}. 
For the moderate-resolution gratings (G130M, G160M), velocities are accurate within 3\,\kms{}; but for the low-resolution G140L grating, velocities are only good to within 150\,\kms{}. 
All stars in TEMPOS have been observed with G130M grating, and a subset also have G160M data.
Thus, we measure \vrad{} for the moderate-resolution (G130M or G130M+G160M together) and G140L gratings separately, reported in Table~\ref{tab:rvs}, and adopt the measurement from the moderate-resolution gratings as the observed \vrad{} for each star.
While we do report \vrad{} measured from the G140L spectra, we emphasize that these are not reliable estimates of the stellar radial velocities; they are only needed for the EW and wind line extent measurements from the G140L spectra (in Sections~\ref{sec:photlines} and \ref{sec:windlines} below).

To measure \vrad{}, we model several commonly detected photospheric absorption lines as Gaussian functions. 
First, the observed coadded, binned spectrum is shifted to the host galaxy's rest frame (\vsys{}; see Table~\ref{tab:galaxies}) to ensure that each photospheric line falls between two continuum regions, or fixed wavelength ranges on either side of each line over which we determine the local continuum level (reported in Table~\ref{tab:windows} in Appendix~\ref{app:ews}). 
We then fit a linear model to the spectrum within the red and blue continuum regions on either side of the line (with known ISM or stellar lines masked out), and the observed line profile is normalized by dividing out the best-fit continuum level. 

We then fit Gaussian models to all normalized line profiles covered by the grating(s) under consideration simultaneously with a single \vrad{}. 
We prefer this approach over measuring \vrad{} for each line separately because it ensures that poorly constrained \vrad{} measurements for weak or undetected lines do not bias our results. 
We use lines that span a wide wavelength range and sample different ionization states to ensure that at least some lines are well-detected for a given star, regardless of its SpT: \trans{c}{iii}{1176}, \trans{n}{iii}{1184}, \trans{c}{iii}{1247}, \trans{o}{iv}{1342,\,1343}, \trans{s}{v}{1502}, and \trans{he}{ii}{1640}.
These lines were visually examined in each spectrum, and for a few stars, either \trans{c}{iii}{1176} or \trans{he}{ii}{1640} was excluded from the fitting due to obvious contamination by wind signatures. 

The resulting best-fit \vrad{} in each grating combination are reported in Table~\ref{tab:rvs}. 
The level of agreement between \vrad{} obtained from moderate-resolution and G140L gratings varies across the sample: though they are similar in many cases, we find differences up to $\sim$200\,\kms{} for a few stars.
Interestingly, the largest velocity offsets between the moderate- and low-resolution gratings occur for stars observed at a cenwave of 800\,\AA{} with the G140L grating, and are preferentially in the sense that G140L is blueshifted with respect to the moderate-resolution gratings.
\added{There is no obvious wavelength dependence of these offsets, as the same velocity reasonably matches all observed absorption line profiles across 1176--1640\,\AA{} used in our \vrad{} measurements.}
All such observations were executed in November 2021 or later, while all of the earlier G140L spectra were observed at a cenwave of 1105\,\AA{}. 
We highlight that the G140L wavelength calibration may be poorer than the nominal 150\,\kms{} accuracy for cenwave 800\,\AA{} and/or since late 2021, \added{and this observing mode should therefore not be relied upon for velocity measurements}.
A similar wavelength offset was confirmed to be present in other datasets observed at the same cenwave by the HST Help Desk\footnote{\url{https://stsci.service-now.com/hst}} in 2025 October, \added{and we recommend that interested G140L users contact the Help Desk for further information and advice.}
Still, there are substantial velocity differences from the moderate-resolution gratings for both G140L observing setups in the TEMPOS sample, as expected for the less precise wavelength calibration in that grating. 
For clarity, Table~\ref{tab:rvs} reports the adopted stellar \vrad{} from the moderate-resolution grating(s), and the last column reports the difference between the adopted stellar \vrad{} and the host galaxy \vsys{}.

Overall, the velocity offsets are quite small, with a median of $-0.7$\,\kms{} and median absolute value of 9.5\,\kms{}. 
For comparison, 30\,\kms{} is a common velocity threshold used to define ``runaway'' stars observationally (e.g., \citealt{blaauw61, hoogerwerf01}), corresponding to a radial velocity of 17\,\kms{} for an isotropic velocity distribution ($\left<v_\mathrm{rad} \right>= \left<v\right> / \sqrt{3}$). 
We find that 11 of the 29 stars, or 38\% of the TEMPOS sample, have velocity offsets larger than 17\,\kms{}, evenly distributed across the two metallicity regimes.
But these dwarf galaxies rotate (see maximum observed \vrot{} reported in Table~\ref{tab:galaxies}), so we must consider whether these large velocity offsets can be explained by each O star's position along its host galaxy's rotation curve. 
Of the 11 measured velocity offsets larger than 17\,\kms{}, seven are in the opposite sense of the host galaxy's rotation curve at the location of the star (BUG A13; GHV 67684; LGN s004, s007, s014, s029, s050; \citealt{skillman88, lake89});
one is $\sim$30\,\kms{} larger than the host galaxy's \vrot{} (LGN s038);
and the remaining three (BUG B2, EBU 20, and TEM 2) could be lower than 17\,\kms{} after accounting for the rotation curve \citep{jobin90, jackson04}.
However, TEM~2 and BPU~A15 in WLM are just a few arcseconds apart and should therefore have similar contributions of the host galaxy's rotation to their \vrad{}, but their measured \vrad{} are different by 32\,\kms{}.

Still, eight stars with large velocity offsets is a high occurrence rate (28\%) of candidate runaways among the relatively isolated, metal-poor O stars in TEMPOS.
Leading scenarios to explain such anomalous velocities include the supernova of a binary companion or dynamical cluster interactions \citep{blaauw61, poveda67}.
Moreover, the majority of binary supernovae are predicted to result in velocity offsets smaller than the canonical 30\,\kms{} threshold \citep{renzo19}, implying that an even larger fraction of the TEMPOS sample could have undergone such interactions.
It is also possible that some stars' \vrad{} may change over time due to one or more companions\added{; our planned detailed modeling of the spectra could potentially reveal anomalous line profiles or ratios indicative of hidden companions}.
Future spectroscopic monitoring of these stars would be valuable to determine whether they are more likely post-interaction single stars or are currently in multiple-star systems.

\subsection{FUV Photospheric Lines\label{sec:photlines}}

\begin{figure*}[!ht]
\begin{centering}
\includegraphics[width=\linewidth]{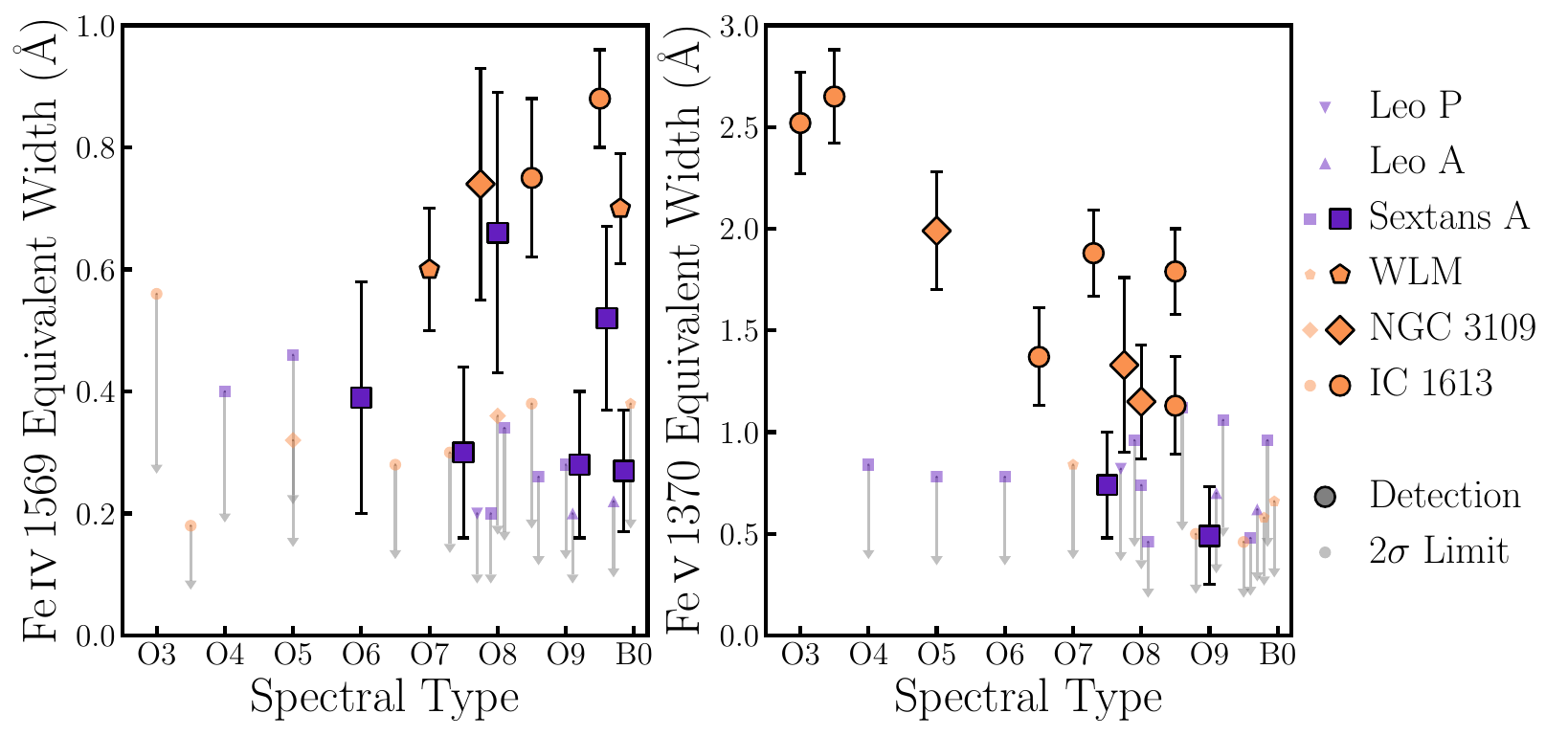}
\caption{\textbf{Strength of iron forest absorption as a function of spectral type and metallicity.} Equivalent widths of the Fe\,\textsc{iv}\,1569 (left panel) and Fe\,\textsc{v}\,1370 (right panel) ``forest'' regions are plotted as a function of SpT for the TEMPOS sample. Marker shape indicates the star's host galaxy, and color specifies whether that galaxy is in the metal-poor (orange) or extremely metal-poor (purple) sample. Larger, solid points with black error bars show EWs detected at the $>$\,2$\sigma$ level, while smaller, lighter points with gray arrows show 2$\sigma$ upper limits. Overlapping points are offset slightly in the horizontal direction for clarity. Overall, Fe\,\textsc{iv}\,1569 EWs increase and Fe\,\textsc{v}\,1370 EWs decrease toward later SpT stars with typically lower \teff{}, and stars in the metal-poor sample have stronger Fe forest absorption than stars of the same SpT in the extremely metal-poor sample. \label{fig:fe_ews_vs_spt}}
\end{centering}
\end{figure*}

Photospheric absorption features in FUV spectra encode information about key stellar properties, including their \teff{} and surface abundances.
While quantitative inference of those parameters requires detailed atmosphere modeling (which we defer to future papers), the relative strengths of photospheric features are qualitatively informative. 
The large sample assembled by TEMPOS enables us to explore trends in photospheric line strengths across a wide range of O-star SpTs and host galaxy metallicity.

To that end, we measure the equivalent widths (EWs) of ten photospheric absorption features due to various ions.
These include all of the lines used in our radial velocity measurements in Section~\ref{sec:rvs} above, as well as the \trans{c}{iv}{1169} and \trans{n}{iv}{1718} lines and two ``iron forest'' regions that probe the amount of continuum removed by \spec{Fe}{iv} and  \spec{Fe}{v}. 
First, we shift the observed spectrum to the stellar rest frame, using \vrad{} measured in that grating (Table~\ref{tab:rvs}). 
Then we fit a local continuum level, following the same procedure outlined in Section~\ref{sec:rvs} and using the continuum regions reported in Table~\ref{tab:windows} in Appendix~\ref{app:ews}. 
The EW is then measured as $\int (1 - f_\lambda/f_\lambda^\mathrm{cont}) \, d\lambda$, where $f_\lambda$ is the observed flux and $f_\lambda^\mathrm{cont}$ is the best-fit continuum level, and the integral is computed over a wavelength range covering each photospheric feature of interest, also reported in Table~\ref{tab:windows}, with any contaminating stellar or ISM lines masked out.

We calculate uncertainties on the EW measurements by resampling the observed spectra 1000 times, where the flux at each wavelength is drawn from a Gaussian centered on the measurement and standard deviation equal to the flux uncertainty. 
We then fit the continuum level and measure the EW for each of 1000 resampled spectra, and take the standard deviation of those EWs as the measurement uncertainty. 
We consider a feature to be detected if the measured EW is at least 2 times the uncertainty, and report 2$\sigma$ upper limits for lines that do not meet that threshold. 
\added{This metric does not capture uncertainty in the EWs due to the choice of wavelengths over which the continuum level is fit. For features that have relatively few stellar and ISM lines nearby, we confirm that the detailed choice of continuum region negligibly changes the measured EWs. In cases where there are many lines nearby (e.g., in the iron forests), this is challenging to evaluate because only narrow wavelength ranges appear line-free in all TEMPOS COS spectra. However, because we visually inspect the continuum regions in every spectrum to ensure that the normalization is not biased by obvious absorption or emission features, we do not expect the choice of continuum region to impact the results discussed below.}

Table~\ref{tab:ews} in Appendix~\ref{app:ews} reports the EW measurements, uncertainties, and upper limits for the ten photospheric features in all available gratings for each star.
Ellipses indicate that a given line is not covered by that grating.
We visually inspect all of these typically photospheric features to check for stellar wind signatures, which can bias the measured EWs.
We do not report EWs for features that were wind-affected for a given star in Table~\ref{tab:ews} (these cases are marked with \xmark).

Overall, there is reasonable agreement across the EW measurements for the same feature in overlapping gratings, though there are many cases where a feature is not detected in G140L despite being well-measured in a moderate-resolution grating (and a few examples of the inverse). 
The probability of detecting a feature depends not only on its intrinsic strength, but also on the SNR and resolution of the spectrum.
Visual inspection of the TEMPOS spectra reveals that weak absorption features (common for the metal lines in these very low-$Z$ stars) often appear even weaker or are no longer obvious in the G140L spectra compared to their appearance in a moderate-resolution grating.
Thus, we adopt EW measurements from G130M or G160M when available, and only use the G140L measurement when a feature is detected in that grating alone. 

Figure~\ref{fig:fe_ews_vs_spt} shows how the strength of absorption due to the \spec{Fe}{iv} and \spec{fe}{v} forests depends on both host galaxy metallicity and stellar SpT (a proxy for \teff{}).
The wavelength ranges over which these EWs are measured are shaded orange in Figure~\ref{fig:keylines}; Fe\,\textsc{v}\,1370 includes the two shaded regions on either side of \trans{o}{v}{1371}.
The Fe\,\textsc{iv}\,1569 (left panel) and Fe\,\textsc{v}\,1370 (right panel) EWs are plotted against SpT, where each point represents one star and the symbol indicates its host galaxy. 
Stars in the extremely metal-poor sample are colored purple, and those in the metal-poor sample are colored orange (as in Figure~\ref{fig:paramspace}). 
Dark points with error bars are detections, while the lighter, smaller points with gray arrows are 2$\sigma$ upper limits.

The increase in \spec{Fe}{iv} EW and decrease in \spec{Fe}{v} EW for later-type O stars observed in Figure~\ref{fig:fe_ews_vs_spt} is consistent with their lower \teff{}, which sets the ionization balance in the stellar atmospheres.
Beyond the overall trends with SpT, there is also a clear split in the sample such that at a fixed SpT, stars in the higher-$Z$ host galaxies have stronger iron forest absorption in both \spec{Fe}{iv} and \spec{Fe}{v}. 
This is expected if the iron abundances of the TEMPOS targets closely track those previously reported for more evolved blue and red supergiants in the same host galaxies (Table~\ref{tab:galaxies}), but iron abundances have so far been reported for only a handful of the O stars in the metal-poor galaxy sample \citep{bouret15, telford24}.
This observed split in iron forest absorption strength across the TEMPOS targets suggests that the extremely metal-poor galaxy sample (Leo~A, Leo~P, and Sextans~A) contains O stars that have lower Fe abundances than those in the metal-poor sample, or in the Magellanic Clouds. 
\added{These preliminary results will be followed up with iron abundance measurements from detailed atmosphere modeling in future work} (Section~\ref{sec:atm_modeling}), but this empirical metric \added{is consistent with the expectation} that the TEMPOS dataset probes iron abundances substantially below the 20\%\,\zsun{} level accessible in the SMC.

\subsection{FUV Wind Features\label{sec:windlines}}

\begin{figure}
\begin{centering}
  \includegraphics[width=\linewidth]{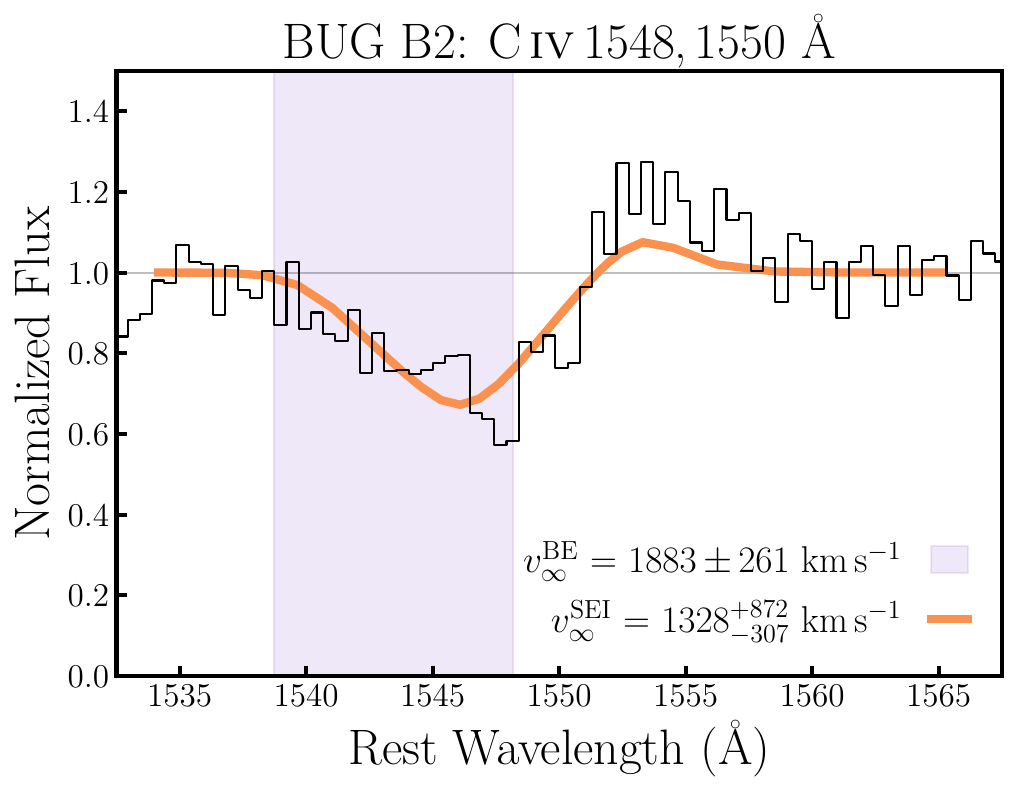}
    \includegraphics[width=\linewidth]{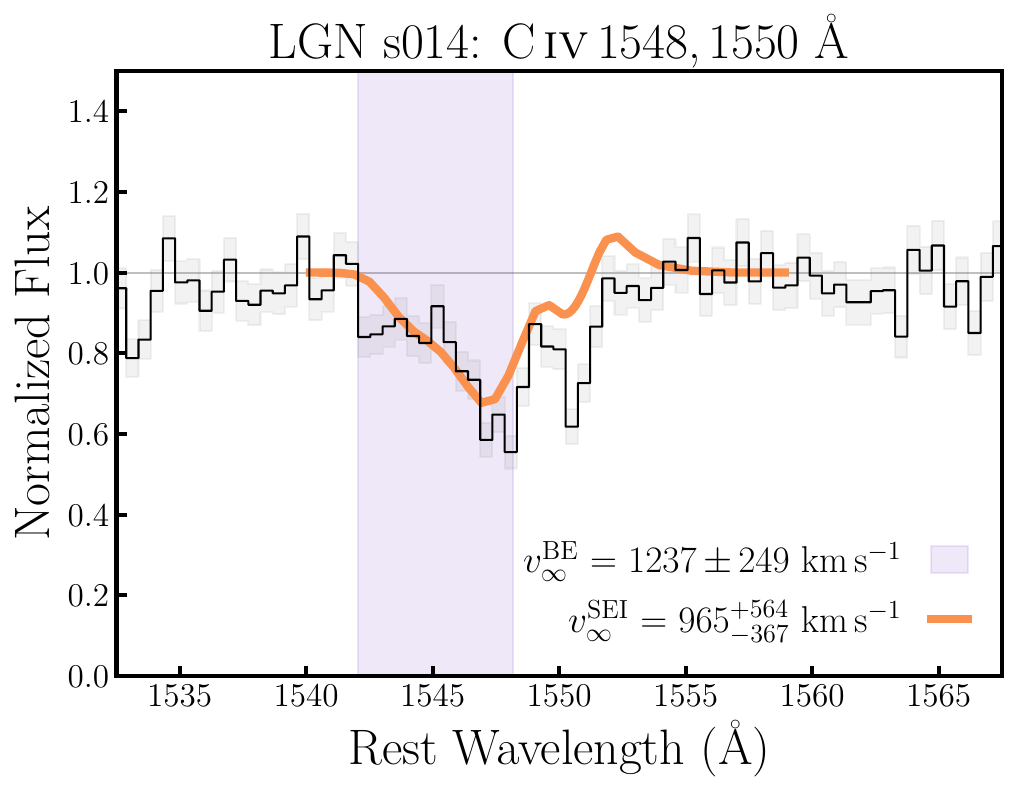}
\caption{\textbf{Example C\,\textsc{iv}\,1548\,\AA{} profiles.} The wind-sensitive \trans{c}{iv}{1548,\,1550} doublet in COS G140L spectra of two O7.5~III-V((f)) TEMPOS targets: BUG~B2 in IC~1613 (top) and LGN~s014 in Sextans~A (bottom). The observed, continuum-normalized fluxes binned to one resolution element are plotted as a function of wavelength in black, with uncertainties shown as gray shading. Spectra are shifted to the rest frame using the \vrad{} measured in the G140L grating (Table~\ref{tab:rvs}). Purple shading covers wavelengths between the rest wavelength of \trans{c}{iv}{1548} and the measured blueward velocity extent of the wind absorption feature (\vinfbe{}; Section~\ref{sec:vinf_be}), and the orange line shows the SEI model corresponding to the best-fit \vinfsei{} (Section~\ref{sec:vinf_sei}). Despite similar SpTs and LCs, the star in the more metal-rich galaxy IC~1613 shows a more developed \spec{c}{iv} P-Cygni profile: it has clear redshifted emission, and extends several hundred \kms{} bluer than the wind absorption seen in the Sextans~A star. 
\label{fig:civprofiles}}
\end{centering}
\end{figure}

A major goal of TEMPOS is to determine how stellar wind properties, particularly \mdot{} and \vinf{}, change in the very low-$Z$ regime.
Measuring those quantities requires atmosphere modeling of FUV and optical spectra, which we defer to future work, but \vinf{} can be estimated straightforwardly from the blueward velocity extent of wind absorption in FUV spectra.
Qualitatively, TEMPOS targets span a wide range in wind line strengths, from well-developed P-Cygni profiles extending to thousands of \kms{} to purely photospheric lines with no evidence of wind.
Nearly all of the stars ($\sim$10 total) with strong wind features are found in the more metal-rich subset of the host galaxies (i.e., IC~1613, NGC~3109, and WLM), and weaker wind lines are seen in the extremely low-$Z$ stars at fixed SpT.

To quantify that statement, we estimate \vinf{} from the \trans{c}{iv}{1548} transition in the TEMPOS spectra to assess empirically the dependence of \vinf{} on SpT, LC, and $Z$ in the metal-poor regime.
Of all the available wind-sensitive lines in the FUV spectra, \trans{c}{iv}{1548} allows for the most homogeneous measurements across the full sample: \trans{si}{iv}{1393} is not present in the hotter O stars, while \trans{n}{v}{1238} is weak in the cooler stars.
Below, we use two different techniques to measure \vinf{} from the \trans{c}{iv}{1548} profiles in the coadded TEMPOS spectra, then compare to published \vinf{} for O stars in the SMC and LMC measured from ULLYSES ultraviolet spectra.

\subsubsection{Empirical Terminal Wind Velocities\label{sec:vinf_be}}

\begin{figure*}
\begin{centering}
\includegraphics[width=\linewidth]{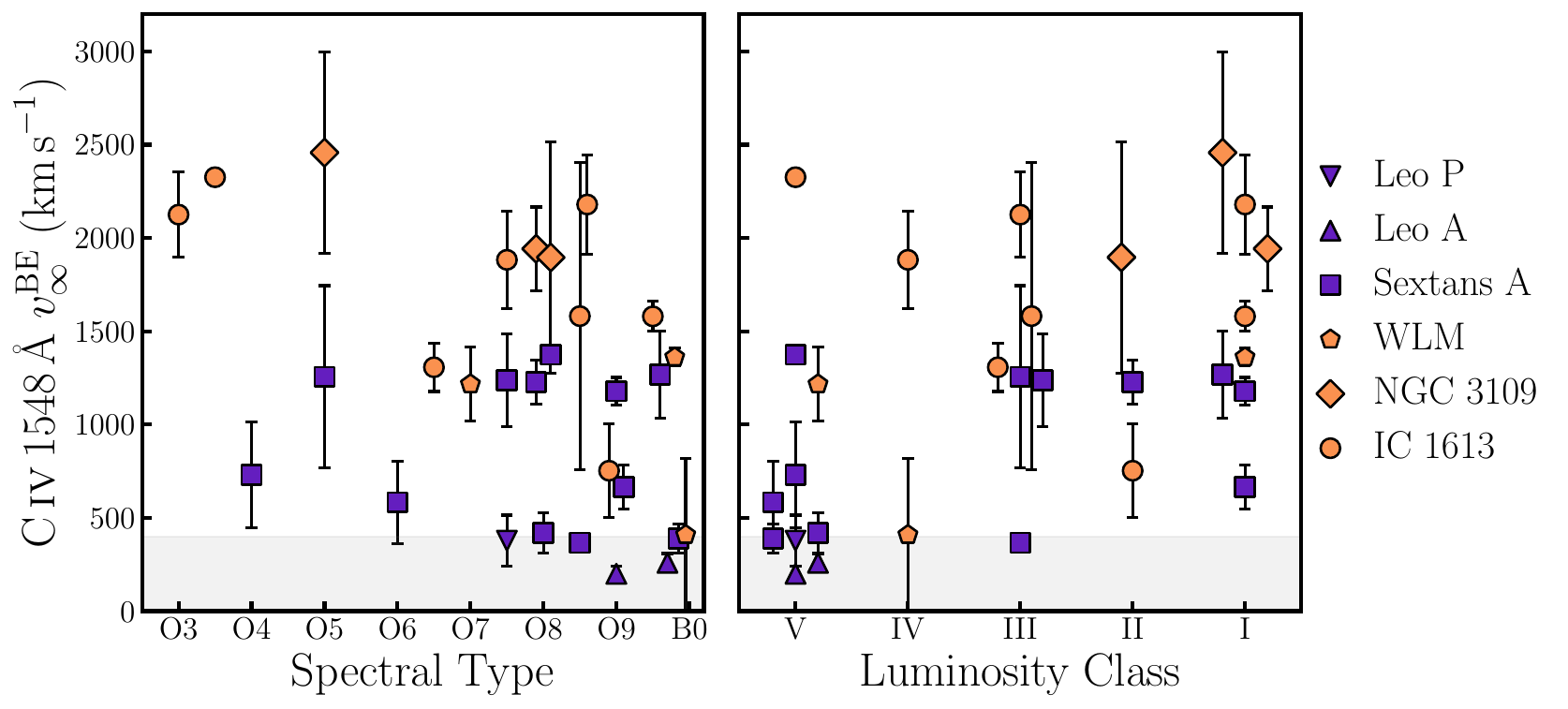}
\caption{\textbf{Terminal wind velocity as a function of spectral classification and metallicity.} Blueward velocity extent of the \trans{c}{iv}{1548} wind line absorption (Section~\ref{sec:vinf_be}, Table~\ref{tab:vinf}) for all O stars in TEMPOS as a function of SpT (left) and LC (right). Marker shape corresponds to the host galaxy and color indicates whether that galaxy is in the extremely metal-poor (purple) or metal-poor (orange) category, as in Figure~\ref{fig:paramspace}. For clarity, overlapping points have been offset slightly in the horizontal direction. The gray shaded region at \vinfbe{}\,$\leq$\,400\,\kms{} indicates where no stellar wind signatures are visually apparent in the \spec{c}{iv} profiles. Weak trends with SpT (a proxy for \teff{}) and LC (a proxy for evolutionary stage) are apparent, and the O stars in extremely metal-poor galaxies have systematically lower \vinf{} compared to the higher-metallicity TEMPOS targets. \label{fig:vinf_vs_spt_lc}}
\end{centering}
\end{figure*}

\movetableright=0.2in
\begin{table*}
\caption{Terminal Wind Velocity Estimates from \trans{c}{iv}{1548} Wind Profiles}
\label{tab:vinf}
\tabcolsep=0.2cm
\begin{tabular}{rlD{,}{\pm}{4.4}D{,}{}{4.4}D{,}{\pm}{4.4}D{,}{}{4.4}}
Galaxy & Star Name & \multicolumn{1}{c}{G160M $v^\mathrm{BE}_\infty$} & \multicolumn{1}{c}{G160M $v^\mathrm{SEI}_\infty$} & \multicolumn{1}{c}{G140L $v^\mathrm{BE}_\infty$} &\multicolumn{1}{c}{G140L $v^\mathrm{SEI}_\infty$} \\
 & & \multicolumn{1}{c}{(\kms{})} & \multicolumn{1}{c}{(\kms{})} & \multicolumn{1}{c}{(\kms{})} & \multicolumn{1}{c}{(\kms{})} \\
\hline
\multicolumn{6}{c}{\textbf{Extremely Low Metallicity ($\bm{\sim5\%\,Z_\odot}$)}}\\ \hline
Leo P & ECG LP26 & 378,137 & \multicolumn{1}{c}{\xmark{}} & \multicolumn{1}{c}{\nodata} & \multicolumn{1}{c}{\nodata} \\
Leo A & GWS K1 & 200,39 & \multicolumn{1}{c}{\xmark{}} & \multicolumn{1}{c}{\nodata} & \multicolumn{1}{c}{\nodata} \\
Leo A & GWS K2 & 259,50 & \multicolumn{1}{c}{\xmark{}} & \multicolumn{1}{c}{\nodata} & \multicolumn{1}{c}{\nodata} \\
Sextans A & LGN s002 & \multicolumn{1}{c}{\nodata} & \multicolumn{1}{c}{\nodata} & 730,285 & 727,^{+586}_{-60} \\
Sextans A & LGN s004 & \multicolumn{1}{c}{\nodata} & \multicolumn{1}{c}{\nodata} & 1255,489 & 700,^{+799}_{-251} \\
Sextans A & LGN s007 & \multicolumn{1}{c}{\nodata} & \multicolumn{1}{c}{\nodata} & 583,222 & 711,^{+87}_{-160} \\
Sextans A & LGN s014 & \multicolumn{1}{c}{\nodata} & \multicolumn{1}{c}{\nodata} & 1237,249 & 965,^{+564}_{-367} \\
Sextans A & LGN s016 & \multicolumn{1}{c}{\nodata} & \multicolumn{1}{c}{\nodata} & 1228,120 & 1258,^{+362}_{-128} \\
Sextans A & LGN s021 & 419,106 & 672,^{+326}_{-322} & \multicolumn{1}{c}{\nodata} & \multicolumn{1}{c}{\nodata} \\
Sextans A & LGN s022 & \multicolumn{1}{c}{\nodata} & \multicolumn{1}{c}{\nodata} & 1375,21 & 1039,^{+340}_{-287} \\
Sextans A & LGN s029 & 366,35 & 624,^{+228}_{-173} & \multicolumn{1}{c}{\nodata} & \multicolumn{1}{c}{\nodata} \\
Sextans A & LGN s037 & 666,118 & 866,^{+334}_{-270} & \multicolumn{1}{c}{\nodata} & \multicolumn{1}{c}{\nodata} \\
Sextans A & LGN s038 & \multicolumn{1}{c}{\nodata} & \multicolumn{1}{c}{\nodata} & 1179,73 & 747,^{+239}_{-84} \\
Sextans A & LGN s050 & \multicolumn{1}{c}{\nodata} & \multicolumn{1}{c}{\nodata} & 1268,232 & 404,^{+831}_{-88} \\
Sextans A & TEM 1 & 388,78 & 756,^{+191}_{-370} & \multicolumn{1}{c}{\nodata} & \multicolumn{1}{c}{\nodata} \\\hline \multicolumn{6}{c}{\textbf{Low Metallicity ($\bm{\sim10\%\,Z_\odot}$)}}\\ \hline
WLM & BPU A11 & 1361,48 & 1260,^{+31}_{-0} & \multicolumn{1}{c}{\nodata} & \multicolumn{1}{c}{\nodata} \\
WLM & BPU A15 & 1216,199 & 1259,^{+28}_{-47} & \multicolumn{1}{c}{\nodata} & \multicolumn{1}{c}{\nodata} \\
WLM & TEM 2 & \multicolumn{1}{c}{\nodata} & \multicolumn{1}{c}{\nodata} & 409,411 & \multicolumn{1}{c}{\xmark{}} \\
NGC 3109 & EBU 20 & \multicolumn{1}{c}{\nodata} & \multicolumn{1}{c}{\nodata} & 1942,224 & 1618,^{+231}_{-197} \\
NGC 3109 & MTK 1 & \multicolumn{1}{c}{\nodata} & \multicolumn{1}{c}{\nodata} & 2457,541 & 1977,^{+22}_{-42} \\
NGC 3109 & MTK 5 & \multicolumn{1}{c}{\nodata} & \multicolumn{1}{c}{\nodata} & 1896,619 & 1774,^{+1144}_{-423} \\
IC 1613 & BUG A13 & 2326,15 & 2203,^{+69}_{-38} & 2591,296 & 2229,^{+81}_{-115} \\
IC 1613 & BUG B11 & 1580,79 & 1067,^{+93}_{-21} & \multicolumn{1}{c}{\nodata} & \multicolumn{1}{c}{\nodata} \\
IC 1613 & BUG B2 & \multicolumn{1}{c}{\nodata} & \multicolumn{1}{c}{\nodata} & 1883,261 & 1328,^{+872}_{-307} \\
IC 1613 & BUG B7 & 753,249 & 789,^{+695}_{-132} & 646,318 & 738,^{+158}_{-0} \\
IC 1613 & GHV 62024 & \multicolumn{1}{c}{\nodata} & \multicolumn{1}{c}{\nodata} & 1307,130 & 949,^{+0}_{-134} \\
IC 1613 & GHV 64066 & \multicolumn{1}{c}{\nodata} & \multicolumn{1}{c}{\nodata} & 2125,229 & 1855,^{+67}_{-118} \\
IC 1613 & GHV 67559 & \multicolumn{1}{c}{\nodata} & \multicolumn{1}{c}{\nodata} & 1581,825 & 1303,^{+126}_{-340} \\
IC 1613 & GHV 67684 & \multicolumn{1}{c}{\nodata} & \multicolumn{1}{c}{\nodata} & 2179,266 & 1688,^{+152}_{-152} \\\hline
\end{tabular}

\tablecomments{Terminal wind velocities (\vinf{}) estimated from observed \trans{c}{iv}{1548} wind profiles using two methods: the observed blueward extent of the wind absorption trough (BE; Section~\ref{sec:vinf_be}), and Sobolev with exact integration (SEI; Section~\ref{sec:vinf_sei}). Measurements are reported separately for the G160M and G140L gratings with different velocity resolutions. \xmark{} symbols indicate that a reliable SEI fit could not be obtained due to a lack clear wind signature in the \trans{c}{iv}{1548} profile.}
\end{table*}

First, we apply a model-independent approach to estimate \vinf{} using the observed blueward extent (BE) of the \trans{c}{iv}{1548} wind absorption.
Though the extent of the saturated part of a P-Cygni profile is the best metric of \vinf{} \citep{prinja90}, this is not well defined for the weak, generally unsaturated wind profiles seen in the TEMPOS sample, so we adopt the velocity of the blue edge of the wind absorption as an empirical metric. 
To normalize the coadded COS spectra (either G160M or G140L, binned to one resolution element), we shift to the rest frame using the measured \vrad{} in that grating (see Section~\ref{sec:rvs}, Table~\ref{tab:rvs}), then fit a linear model to continuum regions on either side of \trans{c}{iv}{1548}. 
For most stars, we adopt rest-frame wavelengths of 1515-1520\,\AA{} and 1564-1566\,\AA{} for the blue and red continuum windows, respectively. 
These were chosen to avoid known absorption lines in the spectra, but we adjust these windows in some cases where those windows are contaminated by obvious absorption or overlap with a detector segment edge.
The continuum windows are also far enough from the \spec{c}{iv} doublet to avoid overlap with even the strongest P-Cygni profiles in the TEMPOS data.

To determine \vinfbe{} from a \trans{c}{iv}{1548} profile, we identify the wavelength where the blueshifted wind absorption profile reaches the continuum level, then convert to a velocity defining the rest wavelength of 1548.19\,\AA{} as zero velocity.
The normalized fluxes at two consecutive wavelength samples are required to be consistent with the continuum level, within the uncertainties, to ensure that fluctuations due to noise do not drive the BE measurements. 
For the G140L spectra of GHV~67684 and BUG~B7 only, that procedure results in an obvious overestimate of \vinfbe{}, so we drop the requirement of a second consecutive pixel consistent with the continuum level.
The velocity corresponding to the first wavelength sample consistent with the continuum level is then defined as \vinfbe{} of the \trans{c}{iv}{1548} profile. 
We resample the COS spectrum 1000 times, where each flux sample is drawn from a Gaussian centered on the measurement and with a standard deviation equal to the uncertainty, and repeat the continuum normalization and \vinfbe{} measurements. 
The standard deviation of the resulting \vinfbe{} distribution is then taken as the measurement uncertainty.
We note that pixels have a larger velocity width in G140L versus G160M data, so the \vinfbe{} uncertainties are correspondingly larger. 

Figure~\ref{fig:civprofiles} illustrates this method for two stars with similar SpTs, but in different host galaxies.
The top panel shows a COS G140L spectrum of the O7.5~III-V((f)) star BUG~B2 in IC~1613, while the bottom shows the O7.5~III((f)) star LGN-s014 in Sextans~A. 
The purple shaded region in each panel shows the wavelength range spanned by the \trans{c}{iv}{1548} absorption: the red limit is the rest wavelength of 1548.19\,\AA{} and the blue limit is the measured \vinfbe{}. 
Even though these stars are both O7.5 giants, they have markedly different \trans{c}{iv}{1548} profiles such that the star in the more metal-rich galaxy IC~1613 shows a more developed P-Cygni profile, with \vinfbe{} larger by over 600\,\kms{} and redshifted emission that is not seen in its Sextans~A counterpart.

The resulting terminal wind velocities estimated from the blueward extent method (\vinfbe{}) are presented in Table~\ref{tab:vinf}.
We report measurements and uncertainties for the two gratings covering the \trans{c}{iv}{1548} wind profile, G160M and G140L, separately because uncertainties are systematically larger for the measurements from G140L spectra compared to G160M, and because two stars (BUG~A13, B7) have observations in both gratings.

Figure~\ref{fig:vinf_vs_spt_lc} shows how \vinfbe{} measured from the \trans{c}{iv}{1548} wind absorption feature depends on various O-star properties for the TEMPOS sample. 
\vinfbe{} measurements are plotted as a function of SpT (tightly correlated with stellar effective temperature, \teff{}) in the left panel, and as a function of LC (a measure of the star's evolutionary status) in the right panel. 
Each marker represents a star, where the marker symbol corresponds to the host galaxy and the color indicates whether that galaxy falls in the extremely metal-poor (purple) or metal-poor (orange) category.
We adopt \vinfbe{} from the G160M grating for the two stars with measurements in both G160M and G140L, as the uncertainties are lower in the higher-resolution grating. 

Large observational datasets in the Milky Way, LMC, and SMC have established that \vinf{} is correlated with stellar \teff{} and luminosity (e.g., \citealt{prinja90, hawcroft24}).
Similar trends are apparent for the TEMPOS targets in Figure~\ref{fig:vinf_vs_spt_lc}, such that earlier-type and supergiant O stars preferentially drive faster winds, especially among stars in the metal-poor sample ($\sim$10--20\%\,\zsun{}; orange points).
Yet, those trends are weak and have large scatter, even within different metallicity categories.
This may be due to the qualitative nature of spectral classifications, which imperfectly trace the fundamental stellar properties most relevant to launching winds and can be biased by hidden stellar companions. 

On the other hand, there is a clear separation in the metallicity of the host galaxy, where stars in the metal-poor sample preferentially have larger \vinfbe{} than stars in the extremely metal-poor ($\lesssim$\,10\%\,\zsun{}) sample at fixed SpT or LC.
This is consistent with the expectation that line-driven winds should weaken at lower $Z$ (e.g., \citealt{vink21, bjorklund21, krticka25}).
Though the TEMPOS galaxies are only roughly split into metal-poor and extremely metal-poor categories based on the oxygen and iron abundances of more evolved supergiants (see Section~\ref{sec:host_galaxies}, Table~\ref{tab:galaxies}), the wind strength trends seen in Figure~\ref{fig:vinf_vs_spt_lc} suggest that the TEMPOS dataset spans a wide enough range in stellar parameters and abundances to enable interesting tests of $Z$-dependent wind physics.

\subsubsection{Terminal Wind Velocities from the Sobolev with Exact Integration Technique\label{sec:vinf_sei}}

\begin{figure}
\begin{centering}
  \includegraphics[width=\linewidth]{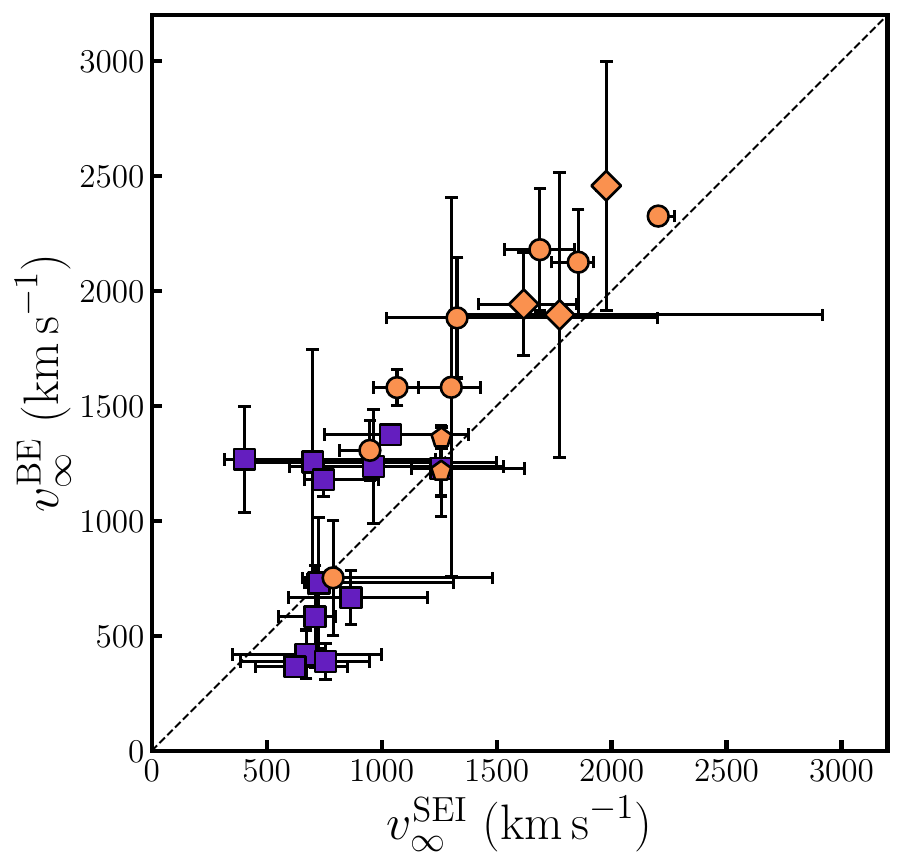}
\caption{\textbf{Comparing terminal wind velocities estimated from the blueward extent and SEI methods.} Terminal wind velocity estimated from the blueward extent of the \trans{c}{iv}{1548} wind profile (\vinfbe{}) is plotted as a function of the SEI-based terminal wind velocity (\vinfsei{}) for the 25 TEMPOS O stars for which \vinfsei{} could be measured, with symbol markers as in Figure~\ref{fig:vinf_vs_spt_lc}. The black dashed line shows where points would lie if the measurements from the two methods were equal. Overall, \vinf{} from the empirical blueward extent method are linearly correlated with those from SEI fitting down to \vinfsei{}\,$\sim$\,800\,\kms{}, but \vinfbe{} are typically larger by a median of 15\% (270\,\kms). \label{fig:vinf_sei_vs_be}}
\end{centering}
\end{figure}

\begin{figure*}[!htp]
\begin{centering}
\includegraphics[width=\linewidth]{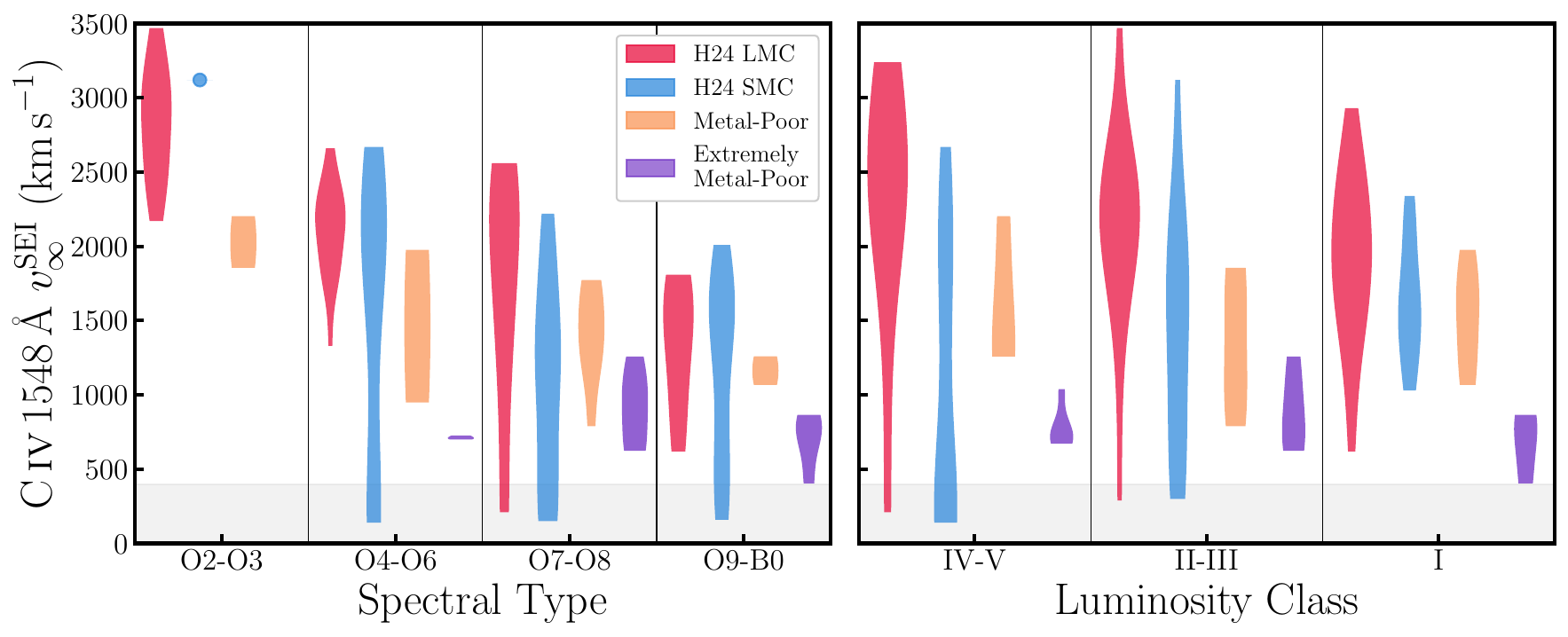}
\caption{\textbf{Terminal wind velocity distributions for O stars in TEMPOS and ULLYSES.} Violin plots illustrating the distribution of \vinfsei{} in bins of stellar SpT (left) and LC (right) for stars in four samples of varying metallicity: LMC (red, 50\%\,\zsun{}), SMC (blue, 20\%\,\zsun{}), TEMPOS metal-poor galaxies (orange, $\sim$10--20\%\,\zsun{}), and TEMPOS extremely metal-poor galaxies (purple, $\lesssim$\,10\%\,\zsun{}). \vinfsei{}, SpTs, and LCs are drawn from \citet{hawcroft24} for the LMC and SMC and reported in this work for the TEMPOS sample (Tables~\ref{tab:targets} and \ref{tab:vinf}). The blue point in the left panel shows \vinfsei{} for a single SMC star in the O2--O3 bin, and the gray shaded region at \vinfsei{}\,$\leq$\,400\,\kms{} indicates where stellar wind signatures in the \spec{c}{iv} profiles become very weak. The typical \vinfsei{} decreases with decreasing $Z$ within each SpT and LC bin, and the extremely metal-poor TEMPOS sample appears offset to lower \vinfsei{} than expected from the otherwise approximately smooth trend.
\label{fig:vinf_vs_spt_lc_violin}}
\end{centering}
\end{figure*}

Next, we compare the model-independent \vinfbe{} to \vinf{} measured from the same \trans{c}{iv}{1548} wind profiles using the Sobolev with exact integration (SEI) method (\vinfsei{}; \citealt{lamers87, haser95, sundqvist14}). 
We follow the same method as \cite{hawcroft24}, where we aim to primarily fit the bluest extent of the absorption profile and the gradient of the blue edge as it returns to the continuum level, which constrain the free parameters \vinfsei{} and the microturbulent velocity ($v_\mathrm{turb}$), respectively. 
In practice, this means that we fix the wind acceleration parameter $\beta$ to unity in the SEI modeling and vary two additional free parameters ($\kappa_{0}$ and $\alpha$, where $\kappa(v) = \kappa_{0}(v / v_{\infty}^\mathrm{SEI})^\alpha$) that describe the line opacity and set the absorption strength. 
Since we are not attempting to reproduce the full \trans{c}{iv}{1548} profile, the best-fit values of $\kappa_{0}$ and $\alpha$ should not be interpreted as physically meaningful quantities.
The orange lines in Figure~\ref{fig:civprofiles} shows the best-fit SEI models for the stars BUG~B2 (top) and LGN~s014 (bottom).
In both cases, the blue edge and gradient of the observed \spec{c}{iv} profile are well fit by the model (though the shape of the LGN~s014 profile is challenging to reproduce with any model), but the rest of the line profile (e.g., the redshifted emission) is not matched because those portions of the wind feature are not considered in the SEI modeling.

Our approach differs from that of \cite{hawcroft24} in two ways. 
First, due to the variation in SNR and spectral resolution across the TEMPOS sample, we find that including a larger portion of the line profile in the fit is necessary to constrain the opacity constants for a subset of the stars.
Second, to improve the uncertainty constraints we apply a genetic-algorithm based fitting approach (pyGA; e.g., \citealt{Abdul-masih19, hawcroft21}) instead of SciPy optimization routines. 
This allows us to better explore the degeneracy between parameters to place more robust physical uncertainties on \vinfsei{} due to the influence of turbulence than the statistical errors provided in previous works.
The resulting \vinfsei{} measurements are reported in Table~\ref{tab:vinf}.
No SEI fit could be obtained for four stars that have \trans{c}{iv}{1548} profiles consistent with being purely photospheric (and that have correspondingly low \vinfbe{}). 

Figure~\ref{fig:vinf_sei_vs_be} shows the relationship between \vinfbe{} and \vinfsei{} across the 25 TEMPOS stars for which \vinfsei{} could be measured. 
Marker shapes and colors indicate the host galaxy name and $Z$ as in Figure~\ref{fig:vinf_vs_spt_lc}, and the line of equality is shown as the black dashed line for reference.
Overall, there is a well-defined linear trend between the two \vinf{} measurements above \vinfsei{}\,$\sim$\,800\,\kms{}, but the scatter increases substantially at the lowest velocities.
There is also a clear offset such that the empirical \vinfbe{} measurements overpredict the SEI-based \vinfsei{}, by a median of 15\% (270\,\kms{}), consistent with the systematic offset between the velocities of the blue edge and saturated absorption found by \citet{prinja90}. 
This is likely due to the inclusion of microturbulence in the SEI modeling, which accounts for the contribution of the motions of shocked material to the wind profile and lowers \vinfsei{} \citep[e.g.,][]{hawcroft24}.
For the subset of TEMPOS stars that have wind parameters from detailed atmosphere modeling of FUV spectra in the literature, we confirm that \vinfsei{} is generally in better agreement with those previously reported \vinf{} than \vinfbe{} \citep{bouret15, telford24, furey25}. 
We conclude that, though the model-independent \vinfbe{} likely overestimate the true \vinf{}, the good correspondence with \vinfsei{} enables \vinfbe{} to be used as a reliable metric of metal-poor O-stars' relative wind strengths.

\subsubsection{The Metallicity Dependence of Wind Terminal Velocities Across ULLYSES and TEMPOS\label{sec:vinf_comparison}}

Finally, we compare to published \vinf{} measurements for O stars in the SMC (20\%\,\zsun{}) and LMC (50\%\,\zsun{}) from the ULLYSES program.
We use \vinfsei{} in this comparison, so the four TEMPOS O stars with the weakest \trans{c}{iv}{1548} wind profiles are excluded, but we confirm that our conclusions are not impacted by the choice of \vinf{} measurement technique.
We adopt \vinfsei{} measured via the same methodology employed in this paper and spectral classifications drawn from the literature for 43 and 47 O stars in the SMC and LMC, respectively, from \citet{hawcroft24}. 
Combining those measurements from ULLYSES FUV spectra with the \vinfsei{} reported here, we can compare trends in wind strength across four different groups of O stars spanning a wide range of $Z$ in the LMC, SMC, TEMPOS metal-poor galaxies, and TEMPOS extremely metal-poor galaxies.

Figure~\ref{fig:vinf_vs_spt_lc_violin} shows the distribution of \vinfsei{} within each of those four $Z$ categories, indicated by different colors, split into bins of SpT (left) and LC (right). 
Intermediate SpTs are rounded down; e.g., an O6.5 star is included in the O4--O6 SpT bin.
Wider parts of each colored shape indicate a higher density of \vinfsei{} measurements, and the vertical extent shows the range of \vinfsei{} within each sample. 
The number of stars contributing to each distribution varies, with a minimum of two.
There is just one SMC O star in the O2--O3 SpT bin, shown as a blue point in the left panel.
Again, gray shading below \vinfsei{}\,$\leq$\,400\,\kms{} indicates where stellar wind signatures become very weak or are absent from the \spec{c}{iv} profiles; no TEMPOS targets are shown in this regime because we could not obtain reliable \vinfsei{} measurements for stars with very low \vinfbe{} (see Section~\ref{sec:vinf_sei}).

The left panel of Figure~\ref{fig:vinf_vs_spt_lc_violin} shows a clear trend of \vinfsei{} decreasing toward later SpTs across all $Z$ samples, and decreasing to lower $Z$ within each SpT bin. 
There are stars with negligible wind signatures in the LMC and SMC for which \vinfsei{} could still be constrained, causing those distributions to extend to very low values in multiple SpT and LC bins; the maximum \vinfsei{} and widest part of the distribution are therefore more useful metrics than the minimum \vinfsei{}. 
Interestingly, the right panel shows no clear trend between \vinfsei{} and LC, but within each of the three LC bins, there is again a clear correlation between the typical \vinfsei{} and $Z$ (with the notable exception of many weak-wind stars in the SMC, especially in the LC IV-V bin).

Across the LMC, SMC, and TEMPOS metal-poor samples, both typical and maximum \vinfsei{} generally decrease smoothly with decreasing $Z$ within a given SpT or LC bin.
This is expected from radiation-driven wind theory \citep[e.g.,][]{castor75}, as the higher opacity of more metal-rich gas provides a larger radiative force that can accelerate faster winds.
The TEMPOS extremely metal-poor sample, however, sits apart from the rest toward much lower \vinfsei{} in most bins of SpT and LC. 
These O stars in galaxies $\lesssim$\,10\%\,\zsun{} \added {appear to} break with the otherwise smooth trend of decreasing \vinfsei{} toward lower $Z$, with their wind strengths falling off more dramatically.
We will further investigate this interesting \added {preliminary} result in future work using detailed abundances, \vinf{}, and \mdot{} from \added {atmosphere modeling of} the full TEMPOS dataset.


\section{Summary and Conclusions\label{sec:summary}}

We have presented the Treasury of Extremely Metal-Poor O Stars (TEMPOS), a HST/COS Large Treasury program designed to assemble a uniform and comprehensive spectroscopic atlas of massive O-type stars below 20\%\,\zsun{}. 
TEMPOS provides the best possible coverage of O-star spectral classifications in this regime, given our current knowledge of O stars in nearby ($\lesssim$\,1.6\,Mpc), very metal-poor dwarf galaxies that are feasible to observe with HST/COS (see Section~\ref{sec:survey}).

The first data release from TEMPOS  provides the community with coadded, science-ready FUV spectra for the entire sample of 29 O stars (see Sections~\ref{sec:data}--\ref{sec:products}).
The data products are designed to be fully consistent with the HLSPs produced by the ULLYSES HST Director's Discretionary program at 20--50\%\,\zsun{} \citep{roman-duval25} and will be made available in MAST via \dataset[10.17909/ fcda-fn73]{\doi{10.17909/ fcda-fn73}} upon publication of this paper. 
Future data releases will include $R$$\sim$4000 optical spectra observed with either the DEIMOS or KCWI spectrographs on Keck and HST photometry for all TEMPOS targets. 
Together, these data products will enable investigations of both massive-star and ISM astrophysics in the very metal-poor regime (see Section~\ref{sec:science_goals}).

Finally, we have presented initial results on measurements of various stellar photospheric and wind lines in the TEMPOS HST/COS spectra. 
Our conclusions are:
\begin{enumerate}
\item We report stellar radial velocities (\vrad{}) for the moderate-resolution (G130M, G160M) and low-resolution (G140L) COS gratings separately and find a previously unreported systematic blueshift (up to $\sim$200\,\kms{}) in the wavelength calibration for G140L observations at cenwave 800\,\AA{} (Section~\ref{sec:rvs}; Table~\ref{tab:rvs}).
At least 8 of the 29 O stars in TEMPOS (28\%) have \vrad{} at least 17\,\kms{} different from the velocity expected at their locations along the host galaxy's rotation curve, suggesting that post-interaction and/or multiple-star systems are common at very low $Z$.
\item We measure equivalent widths of various metal and helium photospheric lines (or groups of lines; Table~\ref{tab:ews}) and show that both \spec{fe}{iv} and \spec{fe}{v} absorption are stronger for stars in the metal-poor sample ($\sim$10--20\%\,\zsun{}) compared to the extremely metal-poor sample ($\lesssim$\,10\%\,\zsun{}), after accounting for the dependence on spectral type (Section~\ref{sec:photlines}, Figure~\ref{fig:fe_ews_vs_spt}). 
Though few iron abundance measurements for O stars in the host dwarf galaxies exist in the literature, this result suggests that the TEMPOS sample spans a wide enough range in iron abundance to enable interesting new tests of $Z$-dependent stellar astrophysics.
\item We measure the stellar wind terminal velocities via the empirical blueward extent of the \trans{c}{iv}{1548} wind absorption (\vinfbe{}) and SEI modeling of the same line (\vinfsei{}; Sections~\ref{sec:vinf_be}--\ref{sec:vinf_sei}; Figure~\ref{fig:civprofiles}; Table~\ref{tab:vinf}). 
We show that the two velocity measurements are linearly correlated, though with increased scatter at low velocities, and that \vinfbe{} is systematically larger than \vinfsei{} by 15\%, consistent with previous findings (Figure~\ref{fig:vinf_sei_vs_be}). 
\vinfbe{} weakly correlates with both stellar spectral type (SpT) and luminosity class (LC) across the TEMPOS sample, and O stars in the metal-poor sample ($\sim$10--20\%\,\zsun{}) have systematically higher \vinfbe{} than stars in the extremely metal-poor sample ($\lesssim$10\%\,\zsun{}) at fixed spectral classification (Figure~\ref{fig:vinf_vs_spt_lc}). 
\item Finally, we compare \vinfsei{} for the TEMPOS sample to published \vinfsei{} for the ULLYSES sample in the SMC and LMC (20--50\%\,\zsun{}; \citealt{hawcroft24}). 
Typical \vinfsei{} decrease with decreasing $Z$ across the four samples within bins of SpT and LC, and O stars in the TEMPOS extremely metal-poor sample have markedly lower \vinfbe{} (Section~\ref{sec:vinf_comparison}; Figure~\ref{fig:vinf_vs_spt_lc_violin}).
This is broadly consistent with the expected $Z$ dependence of radiation-driven winds, and suggests that wind driving may be very different in the extremely low-$Z$ regime ($\lesssim$10\%\,\zsun{}).
\end{enumerate}

In future work, we will consistently analyze all of the HST/COS and ancillary TEMPOS observations together to determine the fundamental parameters, abundances, and winds of very metal-poor O stars.
The public data products from TEMPOS will provide new insight into massive-star astrophysics at very low $Z$; guide the development of future stellar evolution and spectral models appropriate for low-$Z$ galaxies across cosmic time; and enable new investigations of massive stars and the ISM in the nearby, very metal-poor host galaxies.


\begin{acknowledgements}
Based on observations with the NASA/ESA Hubble Space Telescope obtained at the Space Telescope Science Institute (STScI), which is operated by the Association of Universities for Research in Astronomy, Incorporated, under NASA contract NAS5-26555. 
Support for this work was provided by NASA through grant numbers GO-16767, GO-16920, and GO-17491 from STScI. 
OGT acknowledges support from a Carnegie-Princeton Fellowship through Princeton University and the Carnegie Observatories.
AACS is supported by the Deutsche Forschungsgemeinschaft (DFG, German Research Foundation) in the form of an Emmy Noether Research Group – Project-ID 445674056 (SA4064/1-1, PI Sander).
This work was performed in part at Aspen Center for Physics, which is supported by National Science Foundation grant PHY-2210452.

This research used NASA's Astrophysics Data System, adstex\footnote{\url{https://github.com/yymao/adstex}}, and the arXiv preprint server. 
This work has made use of data from the European Space Agency (ESA) mission
{\it Gaia}\footnote{\url{https://www.cosmos.esa.int/gaia}}, processed by the {\it Gaia}
Data Processing and Analysis Consortium (DPAC)\footnote{\url{https://www.cosmos.esa.int/web/gaia/dpac/consortium}}. Funding for the DPAC
has been provided by national institutions, in particular the institutions
participating in the {\it Gaia} Multilateral Agreement.
\end{acknowledgements}

\facilities{HST (COS, ACS, WFC3), Gaia, PS1}

\software{Astropy \citep{astropy, astropy2, astropy3}, iPython \citep{ipython}, Matplotlib \citep{matplotlib}, NumPy \citep{numpy, numpy2}, SAOImageDS9 \citep{joye03}, SciPy \citep{scipy2}, ullyses \citep{roman-duval20, roman-duval25}, pandas \citep{pandas_2024} }


\appendix
\section{Dictionary of Alternate Names for the TEMPOS Targets\label{app:aliases}}
\setcounter{table}{0}
\renewcommand{\thetable}{A\arabic{table}}
\renewcommand{\theHtable}{A\arabic{table}}

To facilitate accessing the HST/COS data in MAST and matching TEMPOS to other programs, we provide alternate names for all targets in Table~\ref{tab:aliases}.
For each HST GO program that observed a given target, we provide the program ID and the name assigned to that star in the HST Phase~II file, which is used in MAST. 
We also provide the name for each target adopted by the HST Director's Discretionary ULLYSES program (excluding six stars that are not part of the ULLYSES Low-$Z$ sample) and an identifier that can be resolved by Simbad, where available. 

\movetableright=0.1in
\begin{table*}
\caption{Alternate Names of TEMPOS Targets}
\label{tab:aliases}
\tabcolsep=0.1cm
{\scriptsize
\begin{tabular}{rlllll}
Galaxy & Star Name & PID & MAST Name & ULLYSES Name & Simbad-Resolvable Name \\
\hline
\multicolumn{6}{c}{\textbf{Extremely Low Metallicity ($\bm{\lesssim10\%\,Z_\odot}$)}}\\ \hline

Leo P & ECG LP26 & 15967 & LEO-P & Leo-P ECG LP 26 & [ECG2019] LP 26 \\
Leo A & GWS K1 & 15921 & J095927.53+304457.7 & Leo A GWS K1 & [GWS2022] K1 \\
 &  & 17491 & LEOA-GWS-K1 &  &  \\
Leo A & GWS K2 & 15921 & J095930.22+304437.0 & Leo A GWS K2 & [GWS2022] K2 \\
 &  & 17491 & LEOA-GWS-K2 &  &  \\
Sextans A & LGN s002 & 17111 & TARGET-1 & \nodata & LGGS J000200.52-152951.8 \\
 &  & 17491 & SEXA-LGN-S002 &  &  \\
Sextans A & LGN s004 & 16930 & SEXTANS-A-GHN-S4 & Sextans A LGN s004 & [VPW98] 1805 \\
 &  & 17491 & SEXA-LGN-S004 &  &  \\
Sextans A & LGN s007 & 17111 & TARGET-2 & \nodata & LGGS J101106.48-044237.1 \\
 &  & 17491 & SEXA-LGN-S007 &  &  \\
Sextans A & LGN s014 & 14245 & STAR-J101053.81-044113.0 & Sextans A LGN s014 & LGGS J101053.81-044113.0 \\
 &  & 15880 & SEXTANS-A-OB326 &  &  \\
Sextans A & LGN s016 & 17491 & SEXA-LGN-S016 & \nodata & LGGS J101056.28-044253.0 \\
Sextans A & LGN s021 & 16920 & SA3 & Sextans A LGN s021 & LGGS J101104.79-044220.9 \\
Sextans A & LGN s022 & 14245 & STAR-J101105.38-044240.1 & Sextans A LGN s022 & [VPW98] 451 \\
 &  & 15880 & SEXTANS-A-OB521 &  &  \\
Sextans A & LGN s029 & 15967 & SEXTANS-A & Sextans A LGN s029 & LGGS J101058.19-044318.4 \\
Sextans A & LGN s037 & 16767 & SA1 & Sextans A LGN s037 & LGGS J101104.78-044224.1 \\
Sextans A & LGN s038 & 14245 & STAR-J101106.05-044211.4 & Sextans A LGN s038 & LGGS J101106.05-044211.4 \\
 &  & 15880 & SEXTANS-A-OB523 &  &  \\
Sextans A & LGN s050 & 14245 & STAR-J101100.66-044044.3 & Sextans A LGN s050 & LGGS J101100.66-044044.3 \\
 &  & 15880 & SEXTANS-A-OB321 &  &  \\
Sextans A & TEM 1 & 16767 & SA2 & Sextans A LGGS & LGGS J101056.86-044040.8 \\
 &  &  &  & J101056.86-044040.8 &  \\\hline \multicolumn{6}{c}{\textbf{Low Metallicity ($\bm{\sim10-20\%\,Z_\odot}$)}}\\ \hline
WLM & BPU A11 & 12867 & WLM-A11 & WLM BPU A 11 & [SC85b] 30 \\
WLM & BPU A15 & 15967 & WLM & WLM BPU A 15 & LGGS J000200.52-152951.8 \\
WLM & TEM 2 & 17491 & WLM-1 & \nodata & LGGS J000200.05-152954.1 \\
NGC 3109 & EBU 20 & 16511 & NGC-3109-EBU-20 & NGC 3109 EBU 20 & [EBU2007] 20 \\
 &  & 17491 & NGC3109-EBU-20 &  &  \\
NGC 3109 & MTK 1 & 17491 & NGC3109-EBU-48 & \nodata & [EBU2007] 48 \\
NGC 3109 & MTK 5 & 17491 & NGC3109-1 & \nodata & \nodata \\
IC 1613 & BUG A13 & 12587 & IC1613-010506-021043 & IC1613 BUG A13 & [BUG2007] A 13 \\
 &  & 12867 & IC1613-A13 &  &  \\
 &  & 15880 & IC1613-A13 &  &  \\
IC 1613 & BUG B11 & 12867 & IC1613-B11 & IC1613 BUG B11 & [BUG2007] B 11 \\
 &  & 15880 & IC1613-B11 &  &  \\
IC 1613 & BUG B2 & 12587 & IC1613-010503-021004 & IC1613 BUG B2 & [BUG2007] B 2 \\
 &  & 15880 & IC1613-B2 &  &  \\
IC 1613 & BUG B7 & 12587 & IC1613-010502-020805 & IC1613 BUG B7 & [BUG2007] B 7 \\
 &  & 15156 & IC1613-010502-020805 &  &  \\
IC 1613 & GHV 62024 & 12587 & IC1613-010501-020849 & IC1613 GHV 62024 & [GHV2009] Star 62024 \\
 &  & 15880 & IC1613-62024 &  &  \\
IC 1613 & GHV 64066 & 15880 & IC1613-64066 & IC1613 GHV 64066 & [GHV2009] Star 64066 \\
 &  & 17491 & IC1613-GHV-64066 &  &  \\
IC 1613 & GHV 67559 & 12587 & IC1613-010505-020923 & IC1613 GHV 67559 & [GHV2009] Star 67559 \\
 &  & 15880 & IC1613-67559 &  &  \\
IC 1613 & GHV 67684 & 15880 & IC1613-67684 & IC1613 GHV 67684 & [GHV2009] Star 67684 \\
 &  & 17491 & IC1613-GHV-67684 &  &  \\\hline
\end{tabular}}

\end{table*}

\section{Details of HST/COS Observations Contributing to TEMPOS Data Products\label{app:observations}}
\setcounter{table}{0}
\renewcommand{\thetable}{B\arabic{table}}
\renewcommand{\theHtable}{B\arabic{table}}

Table~\ref{tab:obsinfo} presents basic information about the COS observations contributing to the TEMPOS coadds in the first data release for each star in the sample, both new (from HST-GO-17491) and archival (from other GO programs). For each unique combination of star, proposal ID, and grating, we report the central wavelength(s) used; the resulting wavelength coverage of the data; and the total exposure time. We also list any \texttt{*x1d.fits} files that were identified as problematic (i.e., empty files or containing corrupted data) during our data quality checks, described in Section~\ref{sec:reduction}.
These ``rejected'' files were excluded from the TEMPOS coadds and should not be used in any future scientific analyses. 

We report the continuum SNR per resolution element at several representative wavelengths in Table~\ref{tab:snrs}.
Each spectrum was binned by 6 pixels, equal to one COS resolution element, following the procedure described in Section~\ref{sec:coadds}.
The continuum SNR was then measured from the binned spectrum within 7 wavelength regions, each 5--10\,\AA{} wide and spaced approximately 100\,\AA{} apart. These regions were chosen to be near, but avoid, strong photosphere and wind lines to provide a measure of the data quality close to key diagnostic features in the spectra.

\movetableright=-0.5in
\begin{table*}
\caption{TEMPOS DR1: New and Archival HST/COS Observations}
\label{tab:obsinfo}
\tabcolsep=0.15cm
{\scriptsize
\begin{tabular}{rllllccl}
Galaxy & Star Name & PID & Grating & Cenwave(s) (\AA{}) & Wavelengths (\AA{}) & ExpTime (s) & Rejected *x1d.fits \\
\hline
\multicolumn{8}{c}{\textbf{Extremely Low Metallicity ($\bm{\lesssim10\%\,Z_\odot}$)}}\\ \hline

Leo P & ECG LP26 & 15967 & G130M & 1291 & 1130--1430 & 21503 &   \\
  &   & 15967 & G160M & 1600 & 1405--1778 & 48398 &   \\
Leo A & GWS K1 & 17491 & G130M & 1291 & 1130--1430 & 15189 &   \\
  &   & 15921 & G160M & 1533 & 1338--1710 & 20857 & le5b03n7q \\
Leo A & GWS K2 & 17491 & G130M & 1291 & 1130--1430 & 17501 &   \\
  &   & 15921 & G160M & 1533 & 1339--1710 & 20857 &   \\
Sextans A & LGN s002 & 17491 & G130M & 1291 & 1130--1430 & 15896 & lf8y07w8q, lf8y08mcq, lf8y09ixq, \\
 & & & &   & & & lf8y09izq \\
  &   & 17111 & G140L & 800 & 784--1953 & 14365 & lezy01e8q \\
Sextans A & LGN s004 & 17491 & G130M & 1291 & 1130--1430 & 25155 & lf8y10m1q, lf8y10m3q, lf8y10m5q, \\
 & & & &   & & & lf8y11lpq, lf8y11m1q, lf8y13cjq \\
  &   & 16930 & G140L & 800 & 784--1953 & 11446 &   \\
Sextans A & LGN s007 & 17491 & G130M & 1291 & 1130--1430 & 20803 & lf8y14kcq, lf8y14kfq, lf8y14kjq, \\
 & & & &   & & & lf8y15p9q, lf8y64tbq, lf8y66pfq \\
  &   & 17111 & G140L & 800 & 784--1953 & 14794 & lezy05ztq, lezy06ekq \\
Sextans A & LGN s014 & 15880 & G130M & 1291, 1309 & 1132--1454 & 32082 &   \\
  &   & 14245 & G140L & 1105 & 1111--2283 & 24045 &   \\
Sextans A & LGN s016 & 17491 & G130M & 1291 & 1129--1430 & 34932 & lf8y17zhq, lf8y19q4q, lf8y19qsq, \\
 & & & &   & & & lf8y20eeq, lf8y20erq, lf8y20etq \\
  &   & 17491 & G140L & 800 & 784--1953 & 13993 & lf8y22gaq, lf8y22grq, lf8y22guq, \\
 & & & &   & & & lf8y23koq, lf8y23kqq, lf8y23ksq \\
Sextans A & LGN s021 & 16920 & G130M & 1291 & 1130--1430 & 9530 &   \\
  &   & 16920 & G160M & 1600 & 1406--1778 & 24770 &   \\
Sextans A & LGN s022 & 15880 & G130M & 1291, 1309 & 1132--1454 & 19077 & le4s12raq \\
  &   & 14245 & G140L & 1105 & 1112--2283 & 8065 &   \\
Sextans A & LGN s029 & 15967 & G130M & 1291 & 1130--1429 & 10726 &   \\
  &   & 15967 & G160M & 1600 & 1405--1778 & 26909 &   \\
Sextans A & LGN s037 & 16767 & G130M & 1291 & 1130--1430 & 12847 &   \\
  &   & 16767 & G160M & 1600 & 1406--1778 & 33230 &   \\
Sextans A & LGN s038 & 15880 & G130M & 1291, 1309 & 1132--1454 & 19422 &   \\
  &   & 14245 & G140L & 1105 & 1113--2283 & 10475 &   \\
Sextans A & LGN s050 & 15880 & G130M & 1291, 1309 & 1131--1454 & 27845 & le4s18h4q, le4s18h8q \\
  &   & 14245 & G140L & 1105 & 1114--2283 & 9811 &   \\
Sextans A & TEM 1 & 16767 & G130M & 1291 & 1130--1430 & 10181 &   \\
  &   & 16767 & G160M & 1600 & 1406--1779 & 25549 &   \\\hline \multicolumn{8}{c}{\textbf{Low Metallicity ($\bm{\sim10-20\%\,Z_\odot}$)}}\\ \hline
WLM & BPU A11 & 12867 & G130M & 1291, 1327 & 1135--1468 & 4625 &   \\
  &   & 12867 & G160M & 1577, 1589, 1611, 1623 & 1387--1796 & 9993 &   \\
WLM & BPU A15 & 15967 & G130M & 1291 & 1130--1430 & 7881 &   \\
  &   & 15967 & G160M & 1600 & 1405--1779 & 15699 &   \\
WLM & TEM 2 & 17491 & G130M & 1291 & 1131--1430 & 19216 &   \\
  &   & 17491 & G140L & 800 & 785--1954 & 13799 &   \\
NGC 3109 & EBU 20 & 17491 & G130M & 1291 & 1130--1430 & 13934 &   \\
  &   & 16511 & G140L & 800 & 786--1954 & 9631 &   \\
NGC 3109 & MTK 1 & 17491 & G130M & 1291 & 1130--1430 & 21116 & lf8y28uwq, lf8y28v0q \\
  &   & 17491 & G140L & 800 & 787--1953 & 13263 & lf8y31g6q, lf8y31ggq, lf8y32bxq \\
NGC 3109 & MTK 5 & 17491 & G130M & 1291 & 1131--1430 & 13783 & lf8y33xhq \\
  &   & 17491 & G140L & 800 & 786--1953 & 7121 &   \\
IC 1613 & BUG A13 & 12867 & G130M & 1291, 1327 & 1135--1468 & 4640 &   \\
  &   & 15880 & G130M & 1291 & 1130--1430 & 1680 &   \\
  &   & 12587 & G140L & 1105 & 1113--2282 & 6684 &   \\
  &   & 12867 & G160M & 1577, 1589, 1611, 1623 & 1387--1796 & 10006 &   \\
IC 1613 & BUG B11 & 12867 & G130M & 1291, 1327 & 1134--1468 & 4581 &   \\
  &   & 15880 & G130M & 1291 & 1131--1430 & 4211 &   \\
  &   & 12867 & G160M & 1577, 1589, 1611, 1623 & 1386--1796 & 9928 &   \\
IC 1613 & BUG B2 & 15880 & G130M & 1291, 1309 & 1132--1454 & 19302 &   \\
  &   & 12587 & G140L & 1105 & 1113--2282 & 8567 &   \\
IC 1613 & BUG B7 & 15156 & G130M & 1291 & 1130--1429 & 7902 &   \\
  &   & 12587 & G140L & 1105 & 1114--2283 & 4284 &   \\
  &   & 15156 & G160M & 1577 & 1382--1754 & 10749 &   \\
IC 1613 & GHV 62024 & 15880 & G130M & 1291, 1309 & 1132--1454 & 19402 &   \\
  &   & 12587 & G140L & 1105 & 1114--2285 & 11485 &   \\
IC 1613 & GHV 64066 & 15880 & G130M & 1291, 1309 & 1130--1453 & 7976 & le4s04dbq, le4s04dqq, le4s04dsq \\
  &   & 17491 & G140L & 800 & 788--1953 & 1952 &   \\
IC 1613 & GHV 67559 & 15880 & G130M & 1291, 1309 & 1130--1454 & 9433 & le4s11eaq, le4s11ecq, le4s11eeq \\
  &   & 12587 & G140L & 1105 & 1114--2283 & 6684 &   \\
IC 1613 & GHV 67684 & 15880 & G130M & 1291, 1309 & 1130--1454 & 9101 &   \\
  &   & 17491 & G140L & 800 & 785--1953 & 3864 &   \\\hline
\end{tabular}}

\end{table*}

\movetableright=-0.25in
\begin{table*}
\caption{Signal-to-Noise Ratio Per Resolution Element for HST/COS Spectra in TEMPOS DR1}
\label{tab:snrs}
\tabcolsep=0.1cm
{\scriptsize
\begin{tabular}{rllccccccc}
Galaxy & Star Name & Grating & SNR$_\text{1164\,\AA{}}$ & SNR$_\text{1268\,\AA{}}$ & SNR$_\text{1355\,\AA{}}$ & SNR$_\text{1475\,\AA{}}$ & SNR$_\text{1558\,\AA{}}$ & SNR$_\text{1650\,\AA{}}$ & SNR$_\text{1740\,\AA{}}$ \\
\hline
\multicolumn{10}{c}{\textbf{Extremely Low Metallicity ($\bm{\lesssim10\%\,Z_\odot}$)}}\\ \hline

Leo P & ECG LP26 & G130M & 11.8 & 11.3 & 11.0 & \nodata & \nodata & \nodata & \nodata \\
 &  & G160M & \nodata & \nodata & \nodata & 14.6 & 10.8 & 7.2 & 5.8 \\
Leo A & GWS K1 & G130M & 14.7 & 14.5 & 13.1 & \nodata & \nodata & \nodata & \nodata \\
 &  & G160M & \nodata & \nodata & 18.9 & 14.1 & 9.7 & 8.2 & \nodata \\
Leo A & GWS K2 & G130M & 16.1 & 15.6 & 13.8 & \nodata & \nodata & \nodata & \nodata \\
 &  & G160M & \nodata & \nodata & 18.7 & 13.7 & 9.3 & 7.8 & \nodata \\
Sextans A & LGN s002 & G130M & 11.1 & 10.5 & 10.2 & \nodata & \nodata & \nodata & \nodata \\
 &  & G140L & 22.8 & 25.3 & 23.4 & 17.0 & 14.0 & 10.6 & 8.4 \\
Sextans A & LGN s004 & G130M & 12.4 & 11.6 & 11.0 & \nodata & \nodata & \nodata & \nodata \\
 &  & G140L & 18.5 & 20.6 & 17.5 & 13.0 & 10.8 & 7.9 & 6.4 \\
Sextans A & LGN s007 & G130M & 11.6 & 10.6 & 10.9 & \nodata & \nodata & \nodata & \nodata \\
 &  & G140L & 21.6 & 23.6 & 21.3 & 15.6 & 12.6 & 9.6 & 7.3 \\
Sextans A & LGN s014 & G130M & 15.7 & 16.7 & 14.9 & \nodata & \nodata & \nodata & \nodata \\
 &  & G140L & 28.9 & 34.7 & 29.5 & 22.0 & 17.7 & 13.3 & 10.6 \\
Sextans A & LGN s016 & G130M & 13.4 & 12.4 & 12.4 & \nodata & \nodata & \nodata & \nodata \\
 &  & G140L & 17.8 & 19.3 & 16.9 & 12.5 & 10.0 & 7.1 & 5.4 \\
Sextans A & LGN s021 & G130M & 9.7 & 9.4 & 9.0 & \nodata & \nodata & \nodata & \nodata \\
 &  & G160M & \nodata & \nodata & \nodata & 13.4 & 10.2 & 6.7 & 5.7 \\
Sextans A & LGN s022 & G130M & 18.9 & 20.0 & 18.3 & \nodata & \nodata & \nodata & \nodata \\
 &  & G140L & 25.7 & 31.7 & 27.2 & 20.7 & 16.1 & 12.4 & 10.0 \\
Sextans A & LGN s029 & G130M & 8.2 & 8.0 & 7.9 & \nodata & \nodata & \nodata & \nodata \\
 &  & G160M & \nodata & \nodata & \nodata & 11.4 & 8.6 & 5.8 & 4.6 \\
Sextans A & LGN s037 & G130M & 8.5 & 8.8 & 8.3 & \nodata & \nodata & \nodata & \nodata \\
 &  & G160M & \nodata & \nodata & \nodata & 12.2 & 9.1 & 6.1 & 5.2 \\
Sextans A & LGN s038 & G130M & 17.5 & 18.7 & 17.0 & \nodata & \nodata & \nodata & \nodata \\
 &  & G140L & 25.9 & 32.9 & 28.5 & 21.6 & 17.1 & 13.0 & 10.7 \\
Sextans A & LGN s050 & G130M & 18.4 & 20.0 & 18.3 & \nodata & \nodata & \nodata & \nodata \\
 &  & G140L & 23.1 & 28.8 & 24.9 & 19.0 & 14.7 & 11.1 & 9.4 \\
Sextans A & TEM 1 & G130M & 9.5 & 9.2 & 8.8 & \nodata & \nodata & \nodata & \nodata \\
 &  & G160M & \nodata & \nodata & \nodata & 12.9 & 9.7 & 6.6 & 5.7 \\\hline \multicolumn{10}{c}{\textbf{Low Metallicity ($\bm{\sim10-20\%\,Z_\odot}$)}}\\ \hline
WLM & BPU A11 & G130M & 8.7 & 16.3 & 13.4 & \nodata & \nodata & \nodata & \nodata \\
 &  & G160M & \nodata & \nodata & \nodata & 17.9 & 13.6 & 10.4 & 8.7 \\
WLM & BPU A15 & G130M & 11.3 & 11.9 & 10.3 & \nodata & \nodata & \nodata & \nodata \\
 &  & G160M & \nodata & \nodata & \nodata & 13.0 & 9.9 & 6.6 & 5.6 \\
WLM & TEM 2 & G130M & 13.9 & 14.2 & 12.6 & \nodata & \nodata & \nodata & \nodata \\
 &  & G140L & 24.7 & 27.1 & 23.5 & 16.4 & 13.2 & 9.9 & 7.6 \\
NGC 3109 & EBU 20 & G130M & 9.6 & 9.3 & 9.7 & \nodata & \nodata & \nodata & \nodata \\
 &  & G140L & 17.7 & 21.0 & 18.8 & 14.4 & 13.1 & 9.6 & 8.1 \\
NGC 3109 & MTK 1 & G130M & 16.1 & 15.0 & 14.3 & \nodata & \nodata & \nodata & \nodata \\
 &  & G140L & 27.3 & 26.6 & 25.3 & 19.0 & 16.7 & 11.4 & 8.5 \\
NGC 3109 & MTK 5 & G130M & 16.3 & 14.8 & 14.3 & \nodata & \nodata & \nodata & \nodata \\
 &  & G140L & 24.6 & 26.1 & 23.4 & 17.1 & 14.3 & 10.1 & 8.0 \\
IC 1613 & BUG A13 & G130M & 14.0 & 20.3 & 16.5 & \nodata & \nodata & \nodata & \nodata \\
 &  & G140L & 41.5 & 49.0 & 40.9 & 29.8 & 26.3 & 17.2 & 14.8 \\
 &  & G160M & \nodata & \nodata & \nodata & 18.4 & 16.9 & 10.9 & 8.9 \\
IC 1613 & BUG B11 & G130M & 15.7 & 22.2 & 18.3 & \nodata & \nodata & \nodata & \nodata \\
 &  & G160M & \nodata & \nodata & \nodata & 18.3 & 13.3 & 9.6 & 8.3 \\
IC 1613 & BUG B2 & G130M & 20.4 & 22.4 & 18.6 & \nodata & \nodata & \nodata & \nodata \\
 &  & G140L & 29.1 & 36.0 & 30.6 & 22.3 & 17.8 & 12.9 & 11.1 \\
IC 1613 & BUG B7 & G130M & 19.0 & 19.5 & 17.8 & \nodata & \nodata & \nodata & \nodata \\
 &  & G140L & 30.1 & 37.3 & 31.8 & 22.8 & 17.4 & 12.9 & 10.7 \\
 &  & G160M & \nodata & \nodata & \nodata & 18.4 & 11.0 & 10.2 & 7.8 \\
IC 1613 & GHV 62024 & G130M & 19.5 & 21.4 & 17.8 & \nodata & \nodata & \nodata & \nodata \\
 &  & G140L & 32.4 & 40.0 & 34.2 & 24.6 & 19.3 & 13.9 & 12.3 \\
IC 1613 & GHV 64066 & G130M & 18.0 & 19.0 & 15.8 & \nodata & \nodata & \nodata & \nodata \\
 &  & G140L & 18.1 & 20.7 & 17.1 & 12.6 & 11.4 & 7.5 & 6.0 \\
IC 1613 & GHV 67559 & G130M & 17.4 & 19.0 & 15.9 & \nodata & \nodata & \nodata & \nodata \\
 &  & G140L & 32.5 & 39.6 & 33.8 & 24.3 & 18.7 & 13.3 & 11.5 \\
IC 1613 & GHV 67684 & G130M & 19.8 & 21.0 & 17.6 & \nodata & \nodata & \nodata & \nodata \\
 &  & G140L & 26.6 & 28.4 & 23.7 & 17.2 & 15.4 & 10.0 & 8.1 \\\hline
\end{tabular}}

\end{table*}

\section{Equivalent Widths and Continuum Regions for Line Measurements\label{app:ews}}
\setcounter{table}{0}
\renewcommand{\thetable}{C\arabic{table}}
\renewcommand{\theHtable}{C\arabic{table}}

Table~\ref{tab:windows} provides the rest-frame wavelength ranges used to normalize the continuum and measure equivalent widths (EWs).
A linear model was fit to the spectrum within two continuum wavelength ranges, one on each the blue and red side of each photospheric feature, to determine the local continuum level.
This continuum model was divided out to normalize the spectra before measuring radial velocities and EWs (Sections~\ref{sec:rvs} and \ref{sec:photlines}).
EWs were then calculated by integrating over the ``EW Integration'' regions given in Table~\ref{tab:windows}. 
The same wavelength regions were used across all gratings and stars in the TEMPOS COS dataset. 

Table~\ref{tab:ews} reports the EWs of photospheric lines described in Section~\ref{sec:photlines}. 
We provide measurements in all gratings that cover a given feature, including the red and blue continuum ranges on either side (Table~\ref{tab:windows}).
Features can cover more than one transition of the same ion: the C\,\textsc{iii}\,1176 EW covers six closely spaced transitions; N\,\textsc{iii}\,1184 covers the doublet at 1183,\,1184\,\AA{}; and O\,\textsc{iv}\,1342 covers transitions at 1342,\,1343\,\AA{} (but not the 1338\,\AA{} transition, which is often blended with a nearby ISM line in G140L data).
We also report EWs for two ``iron forest'' regions, Fe\,\textsc{v}\,1370 and Fe\,\textsc{iv}\,1569, which cover larger wavelength ranges and are meant to probe the impact of absorption by many closely spaced transitions of these ions.
We exclude the contribution of the wind-sensitive \trans{O}{v}{1371} line from the Fe\,\textsc{v}\,1370 EW measurements by masking out 1368.0--1372.5\,\AA{}. 
For the G140L spectra of GHV~64066 and GHV~67684 only, the mask must be extended to 1373.5\,\AA{} to fully exclude the \trans{O}{v}{1371} line, likely due to a combination of increased broadening and poorer wavelength calibration in that grating. 

In general, we were able to find wavelength ranges for the continuum normalization regions and the EW integrals that both avoided contamination from nearby ISM or stellar lines and fully covered the feature of interest across all COS gratings. 
This is non-trivial because the line-spread function of G140L results in much broader features and can cause blending with nearby lines.
In particular, the N\,\textsc{iii}\,1184 EWs in the G140L grating may be biased low because we would need to integrate over a wider wavelength range to fully cover the doublet, but choose not to because this would include absorption from nearby lines that are obvious in the G130M spectra. 
Finally, we note that some of the typically photospheric lines are clearly affected by stellar winds for a subset of stars.
We visually inspected all lines for wind signatures and do not report EWs for those few cases; they are instead marked with an \xmark{} symbol in Table~\ref{tab:ews}.

\movetableright=0.6in
\begin{table*}
\caption{Wavelength Ranges Used for Continuum Normalization and EW Measurements}
\label{tab:windows}
\tabcolsep=0.5cm
\begin{tabular}{lccc}
Feature & EW Integration & Blue Continuum & Red Continuum \\
 & (\AA{}) & (\AA{}) & (\AA{}) \\
\hline 
C\,\textsc{iv}\,1169 & $1167.5-1170.5$ & $1165.8-1167.5$ & $1170.5-1171.8$ \\
C\,\textsc{iii}\,1176 & $1173.0-1178.0$ & $1171.0-1173.0$ & $1178.0-1181.5$ \\
N\,\textsc{iii}\,1184 & $1182.5-1185.2$ & $1179.8-1181.2$ & $1186.5-1187.2$ \\
C\,\textsc{iii}\,1247 & $1246.4-1248.0$ & $1245.0-1246.4$ & $1248.0-1249.0$ \\
O\,\textsc{iv}\,1342 & $1341.5-1345.5$ & $1331.0-1333.0$ & $1345.5-1350.0$ \\
Fe\,\textsc{v}\,1370 & $1360.0-1381.0$ & $1346.0-1350.0$ & $1381.0-1383.5$ \\
S\,\textsc{v}\,1502 & $1500.5-1503.0$ & $1497.0-1500.0$ & $1505.0-1508.0$ \\
Fe\,\textsc{iv}\,1569 & $1566.0-1572.0$ & $1564.0-1566.0$ & $1572.0-1574.0$ \\
He\,\textsc{ii}\,1640 & $1638.6-1643.0$ & $1633.0-1636.5$ & $1644.5-1646.5$ \\
N\,\textsc{iv}\,1718 & $1716.6-1720.3$ & $1709.0-1714.0$ & $1725.0-1729.0$ \\\hline
\end{tabular}

\tablecomments{Wavelength ranges used to measure equivalent widths of various photospheric absorption lines in the COS spectra and to fit the local continuum around each feature (see Sections~\ref{sec:rvs} and \ref{sec:photlines}, Appendix~\ref{app:ews}, and Table~\ref{tab:ews}).}
\end{table*}

\movetableright=-1.2in
\begin{table*}
\caption{FUV Photospheric Line Equivalent Widths}
\label{tab:ews}
{\scriptsize
\tabcolsep=0.1cm
\begin{tabular}{rllD{,}{\pm}{2.2}D{,}{\pm}{2.2}D{,}{\pm}{2.2}D{,}{\pm}{2.2}D{,}{\pm}{2.2}D{,}{\pm}{2.2}D{,}{\pm}{2.2}D{,}{\pm}{2.2}D{,}{\pm}{2.2}D{,}{\pm}{2.2}}
Host & Star & Grating & \multicolumn{1}{c}{C\,\textsc{iv}\,1169} & \multicolumn{1}{c}{C\,\textsc{iii}\,1176} & \multicolumn{1}{c}{N\,\textsc{iii}\,1184} & \multicolumn{1}{c}{C\,\textsc{iii}\,1247} & \multicolumn{1}{c}{O\,\textsc{iv}\,1342} & \multicolumn{1}{c}{Fe\,\textsc{v}\,1370} & \multicolumn{1}{c}{S\,\textsc{v}\,1502} & \multicolumn{1}{c}{Fe\,\textsc{iv}\,1569} & \multicolumn{1}{c}{He\,\textsc{ii}\,1640} & \multicolumn{1}{c}{N\,\textsc{iv}\,1718}  \\
 Galaxy & Name & & \multicolumn{1}{c}{(\AA{})} & \multicolumn{1}{c}{(\AA{})} & \multicolumn{1}{c}{(\AA{})} & \multicolumn{1}{c}{(\AA{})} & \multicolumn{1}{c}{(\AA{})} & \multicolumn{1}{c}{(\AA{})} & \multicolumn{1}{c}{(\AA{})} & \multicolumn{1}{c}{(\AA{})} & \multicolumn{1}{c}{(\AA{})} & \multicolumn{1}{c}{(\AA{})}  \\\hline
\multicolumn{13}{c}{\textbf{Extremely Low Metallicity ($\bm{\lesssim10\%\,Z_\odot}$)}}\\ \hline
Leo P & ECG LP26 & G130M & \multicolumn{1}{c}{<\,0.10} & 0.84,0.06 & \multicolumn{1}{c}{<\,0.10} & \multicolumn{1}{c}{<\,0.08} & 0.17,0.05 & \multicolumn{1}{c}{<\,0.82} & \multicolumn{1}{c}{\nodata} & \multicolumn{1}{c}{\nodata} & \multicolumn{1}{c}{\nodata} & \multicolumn{1}{c}{\nodata} \\
 & & G160M & \multicolumn{1}{c}{\nodata} & \multicolumn{1}{c}{\nodata} & \multicolumn{1}{c}{\nodata} & \multicolumn{1}{c}{\nodata} & \multicolumn{1}{c}{\nodata} & \multicolumn{1}{c}{\nodata} & \multicolumn{1}{c}{<\,0.08} & \multicolumn{1}{c}{<\,0.20} & 0.65,0.10 & 0.35,0.09 \\
Leo A & GWS K1 & G130M & \multicolumn{1}{c}{<\,0.08} & 1.03,0.05 & 0.25,0.04 & 0.11,0.02 & \multicolumn{1}{c}{<\,0.10} & \multicolumn{1}{c}{<\,0.70} & \multicolumn{1}{c}{\nodata} & \multicolumn{1}{c}{\nodata} & \multicolumn{1}{c}{\nodata} & \multicolumn{1}{c}{\nodata} \\
 & & G160M & \multicolumn{1}{c}{\nodata} & \multicolumn{1}{c}{\nodata} & \multicolumn{1}{c}{\nodata} & \multicolumn{1}{c}{\nodata} & \multicolumn{1}{c}{\nodata} & \multicolumn{1}{c}{\nodata} & \multicolumn{1}{c}{<\,0.08} & \multicolumn{1}{c}{<\,0.20} & 0.62,0.08 & \multicolumn{1}{c}{\nodata} \\
Leo A & GWS K2 & G130M & \multicolumn{1}{c}{<\,0.08} & 1.10,0.04 & \multicolumn{1}{c}{<\,0.08} & 0.12,0.02 & \multicolumn{1}{c}{<\,0.10} & \multicolumn{1}{c}{<\,0.62} & \multicolumn{1}{c}{\nodata} & \multicolumn{1}{c}{\nodata} & \multicolumn{1}{c}{\nodata} & \multicolumn{1}{c}{\nodata} \\
 & & G160M & \multicolumn{1}{c}{\nodata} & \multicolumn{1}{c}{\nodata} & \multicolumn{1}{c}{\nodata} & \multicolumn{1}{c}{\nodata} & \multicolumn{1}{c}{\nodata} & \multicolumn{1}{c}{\nodata} & 0.10,0.04 & \multicolumn{1}{c}{<\,0.22} & 0.60,0.09 & \multicolumn{1}{c}{\nodata} \\
Sextans A & LGN s002 & G130M & 0.26,0.05 & 1.05,0.06 & 0.18,0.06 & 0.15,0.05 & 0.13,0.06 & \multicolumn{1}{c}{<\,0.84} & \multicolumn{1}{c}{\nodata} & \multicolumn{1}{c}{\nodata} & \multicolumn{1}{c}{\nodata} & \multicolumn{1}{c}{\nodata} \\
 & & G140L & \multicolumn{1}{c}{<\,0.16} & 0.88,0.07 & \multicolumn{1}{c}{<\,0.12} & \multicolumn{1}{c}{<\,0.06} & \multicolumn{1}{c}{<\,0.14} & \multicolumn{1}{c}{<\,0.72} & \multicolumn{1}{c}{<\,0.14} & \multicolumn{1}{c}{<\,0.40} & 0.48,0.17 & 0.37,0.14 \\
Sextans A & LGN s004 & G130M & 0.18,0.05 & 0.86,0.05 & 0.18,0.05 & 0.08,0.04 & 0.20,0.05 & \multicolumn{1}{c}{<\,0.78} & \multicolumn{1}{c}{\nodata} & \multicolumn{1}{c}{\nodata} & \multicolumn{1}{c}{\nodata} & \multicolumn{1}{c}{\nodata} \\
 & & G140L & \multicolumn{1}{c}{<\,0.14} & 0.84,0.09 & \multicolumn{1}{c}{<\,0.18} & 0.11,0.04 & 0.35,0.08 & \multicolumn{1}{c}{<\,0.96} & \multicolumn{1}{c}{<\,0.16} & \multicolumn{1}{c}{<\,0.46} & 0.65,0.20 & \multicolumn{1}{c}{<\,0.48} \\
Sextans A & LGN s007 & G130M & 0.14,0.05 & 0.97,0.06 & \multicolumn{1}{c}{<\,0.10} & \multicolumn{1}{c}{<\,0.22} & \multicolumn{1}{c}{<\,0.12} & \multicolumn{1}{c}{<\,0.78} & \multicolumn{1}{c}{\nodata} & \multicolumn{1}{c}{\nodata} & \multicolumn{1}{c}{\nodata} & \multicolumn{1}{c}{\nodata} \\
 & & G140L & \multicolumn{1}{c}{<\,0.12} & 1.17,0.09 & \multicolumn{1}{c}{<\,0.16} & \multicolumn{1}{c}{<\,0.06} & \multicolumn{1}{c}{<\,0.14} & \multicolumn{1}{c}{<\,0.80} & 0.16,0.06 & 0.39,0.19 & 0.89,0.15 & 0.37,0.17 \\
Sextans A & LGN s014 & G130M & 0.10,0.04 & 0.98,0.04 & 0.21,0.04 & 0.12,0.03 & 0.22,0.04 & \multicolumn{1}{c}{<\,0.58} & \multicolumn{1}{c}{\nodata} & \multicolumn{1}{c}{\nodata} & \multicolumn{1}{c}{\nodata} & \multicolumn{1}{c}{\nodata} \\
 & & G140L & \multicolumn{1}{c}{<\,0.10} & 0.94,0.05 & \multicolumn{1}{c}{<\,0.08} & 0.07,0.02 & \multicolumn{1}{c}{<\,0.12} & 0.74,0.26 & 0.17,0.05 & 0.30,0.14 & 0.49,0.11 & 0.48,0.11 \\
Sextans A & LGN s016 & G130M & 0.14,0.04 & 1.08,0.05 & 0.32,0.04 & 0.11,0.03 & \multicolumn{1}{c}{<\,0.10} & \multicolumn{1}{c}{<\,0.74} & \multicolumn{1}{c}{\nodata} & \multicolumn{1}{c}{\nodata} & \multicolumn{1}{c}{\nodata} & \multicolumn{1}{c}{\nodata} \\
 & & G140L & \multicolumn{1}{c}{<\,0.16} & 1.07,0.10 & 0.17,0.08 & \multicolumn{1}{c}{<\,0.08} & \multicolumn{1}{c}{<\,0.18} & \multicolumn{1}{c}{<\,0.98} & \multicolumn{1}{c}{<\,0.20} & 0.66,0.23 & 0.56,0.22 & 0.72,0.21 \\
Sextans A & LGN s021 & G130M & 0.13,0.06 & 0.84,0.07 & 0.17,0.06 & 0.16,0.05 & \multicolumn{1}{c}{<\,0.14} & \multicolumn{1}{c}{<\,0.96} & \multicolumn{1}{c}{\nodata} & \multicolumn{1}{c}{\nodata} & \multicolumn{1}{c}{\nodata} & \multicolumn{1}{c}{\nodata} \\
 & & G160M & \multicolumn{1}{c}{\nodata} & \multicolumn{1}{c}{\nodata} & \multicolumn{1}{c}{\nodata} & \multicolumn{1}{c}{\nodata} & \multicolumn{1}{c}{\nodata} & \multicolumn{1}{c}{\nodata} & 0.10,0.04 & \multicolumn{1}{c}{<\,0.20} & 0.57,0.11 & 0.29,0.10 \\
Sextans A & LGN s022 & G130M & 0.10,0.03 & 1.27,0.03 & 0.21,0.03 & 0.17,0.02 & 0.16,0.03 & \multicolumn{1}{c}{<\,0.46} & \multicolumn{1}{c}{\nodata} & \multicolumn{1}{c}{\nodata} & \multicolumn{1}{c}{\nodata} & \multicolumn{1}{c}{\nodata} \\
 & & G140L & \multicolumn{1}{c}{<\,0.10} & 1.18,0.06 & \multicolumn{1}{c}{<\,0.10} & \multicolumn{1}{c}{<\,0.08} & \multicolumn{1}{c}{<\,0.12} & \multicolumn{1}{c}{<\,0.62} & \multicolumn{1}{c}{<\,0.10} & \multicolumn{1}{c}{<\,0.34} & 0.51,0.13 & \multicolumn{1}{c}{<\,0.28} \\
Sextans A & LGN s029 & G130M & 0.17,0.07 & 1.18,0.08 & \multicolumn{1}{c}{<\,0.16} & 0.10,0.05 & \multicolumn{1}{c}{<\,0.16} & \multicolumn{1}{c}{<\,1.12} & \multicolumn{1}{c}{\nodata} & \multicolumn{1}{c}{\nodata} & \multicolumn{1}{c}{\nodata} & \multicolumn{1}{c}{\nodata} \\
 & & G160M & \multicolumn{1}{c}{\nodata} & \multicolumn{1}{c}{\nodata} & \multicolumn{1}{c}{\nodata} & \multicolumn{1}{c}{\nodata} & \multicolumn{1}{c}{\nodata} & \multicolumn{1}{c}{\nodata} & \multicolumn{1}{c}{<\,0.10} & \multicolumn{1}{c}{<\,0.26} & 0.51,0.12 & \multicolumn{1}{c}{<\,0.24} \\
Sextans A & LGN s037 & G130M & \multicolumn{1}{c}{<\,0.14} & 1.69,0.07 & 0.22,0.07 & 0.20,0.05 & \multicolumn{1}{c}{<\,0.16} & \multicolumn{1}{c}{<\,1.06} & \multicolumn{1}{c}{\nodata} & \multicolumn{1}{c}{\nodata} & \multicolumn{1}{c}{\nodata} & \multicolumn{1}{c}{\nodata} \\
 & & G160M & \multicolumn{1}{c}{\nodata} & \multicolumn{1}{c}{\nodata} & \multicolumn{1}{c}{\nodata} & \multicolumn{1}{c}{\nodata} & \multicolumn{1}{c}{\nodata} & \multicolumn{1}{c}{\nodata} & 0.21,0.04 & 0.28,0.12 & 0.64,0.11 & 0.42,0.10 \\
Sextans A & LGN s038 & G130M & \multicolumn{1}{c}{<\,0.06} & \multicolumn{1}{c}{\xmark{}} & 0.24,0.03 & 0.18,0.02 & 0.18,0.04 & 0.49,0.24 & \multicolumn{1}{c}{\nodata} & \multicolumn{1}{c}{\nodata} & \multicolumn{1}{c}{\nodata} & \multicolumn{1}{c}{\nodata} \\
 & & G140L & \multicolumn{1}{c}{<\,0.10} & \multicolumn{1}{c}{\xmark{}} & 0.11,0.04 & \multicolumn{1}{c}{<\,0.08} & 0.13,0.05 & \multicolumn{1}{c}{<\,0.54} & \multicolumn{1}{c}{<\,0.10} & \multicolumn{1}{c}{<\,0.28} & 0.58,0.12 & 0.27,0.13 \\
Sextans A & LGN s050 & G130M & 0.07,0.03 & 2.13,0.03 & 0.54,0.03 & 0.42,0.02 & 0.12,0.03 & \multicolumn{1}{c}{<\,0.48} & \multicolumn{1}{c}{\nodata} & \multicolumn{1}{c}{\nodata} & \multicolumn{1}{c}{\nodata} & \multicolumn{1}{c}{\nodata} \\
 & & G140L & \multicolumn{1}{c}{<\,0.12} & 2.05,0.05 & 0.29,0.04 & 0.35,0.03 & \multicolumn{1}{c}{<\,0.14} & \multicolumn{1}{c}{<\,0.62} & 0.17,0.06 & 0.52,0.15 & 0.67,0.14 & 0.63,0.14 \\
Sextans A & TEM 1 & G130M & 0.24,0.06 & 1.04,0.07 & \multicolumn{1}{c}{<\,0.14} & 0.14,0.06 & \multicolumn{1}{c}{<\,0.14} & \multicolumn{1}{c}{<\,0.96} & \multicolumn{1}{c}{\nodata} & \multicolumn{1}{c}{\nodata} & \multicolumn{1}{c}{\nodata} & \multicolumn{1}{c}{\nodata} \\
 & & G160M & \multicolumn{1}{c}{\nodata} & \multicolumn{1}{c}{\nodata} & \multicolumn{1}{c}{\nodata} & \multicolumn{1}{c}{\nodata} & \multicolumn{1}{c}{\nodata} & \multicolumn{1}{c}{\nodata} & \multicolumn{1}{c}{<\,0.08} & 0.27,0.10 & 0.69,0.10 & 0.31,0.10 \\\hline \multicolumn{13}{c}{\textbf{Low Metallicity ($\bm{\sim10-20\%\,Z_\odot}$)}}\\ \hline
WLM & BPU A11 & G130M & \multicolumn{1}{c}{<\,0.14} & \multicolumn{1}{c}{\xmark{}} & 0.52,0.04 & 0.40,0.02 & 0.21,0.04 & \multicolumn{1}{c}{<\,0.58} & \multicolumn{1}{c}{\nodata} & \multicolumn{1}{c}{\nodata} & \multicolumn{1}{c}{\nodata} & \multicolumn{1}{c}{\nodata} \\
 & & G160M & \multicolumn{1}{c}{\nodata} & \multicolumn{1}{c}{\nodata} & \multicolumn{1}{c}{\nodata} & \multicolumn{1}{c}{\nodata} & \multicolumn{1}{c}{\nodata} & \multicolumn{1}{c}{\nodata} & 0.21,0.03 & 0.70,0.09 & 0.82,0.06 & 0.64,0.06 \\
WLM & BPU A15 & G130M & \multicolumn{1}{c}{<\,0.10} & 1.00,0.06 & 0.22,0.05 & 0.18,0.03 & 0.22,0.06 & \multicolumn{1}{c}{<\,0.84} & \multicolumn{1}{c}{\nodata} & \multicolumn{1}{c}{\nodata} & \multicolumn{1}{c}{\nodata} & \multicolumn{1}{c}{\nodata} \\
 & & G160M & \multicolumn{1}{c}{\nodata} & \multicolumn{1}{c}{\nodata} & \multicolumn{1}{c}{\nodata} & \multicolumn{1}{c}{\nodata} & \multicolumn{1}{c}{\nodata} & \multicolumn{1}{c}{\nodata} & 0.21,0.04 & 0.60,0.10 & 0.89,0.10 & 0.61,0.09 \\
WLM & TEM 2 & G130M & \multicolumn{1}{c}{<\,0.08} & 1.04,0.05 & 0.19,0.04 & 0.16,0.02 & \multicolumn{1}{c}{<\,0.10} & \multicolumn{1}{c}{<\,0.66} & \multicolumn{1}{c}{\nodata} & \multicolumn{1}{c}{\nodata} & \multicolumn{1}{c}{\nodata} & \multicolumn{1}{c}{\nodata} \\
 & & G140L & \multicolumn{1}{c}{<\,0.12} & 0.78,0.08 & \multicolumn{1}{c}{<\,0.14} & 0.08,0.03 & \multicolumn{1}{c}{<\,0.14} & \multicolumn{1}{c}{<\,0.72} & 0.19,0.06 & \multicolumn{1}{c}{<\,0.38} & 0.82,0.15 & \multicolumn{1}{c}{<\,0.32} \\
NGC 3109 & EBU 20 & G130M & \multicolumn{1}{c}{<\,0.12} & \multicolumn{1}{c}{\xmark{}} & 0.23,0.06 & 0.38,0.04 & 0.24,0.07 & 1.33,0.43 & \multicolumn{1}{c}{\nodata} & \multicolumn{1}{c}{\nodata} & \multicolumn{1}{c}{\nodata} & \multicolumn{1}{c}{\nodata} \\
 & & G140L & \multicolumn{1}{c}{<\,0.16} & \multicolumn{1}{c}{\xmark{}} & \multicolumn{1}{c}{<\,0.16} & 0.12,0.04 & 0.41,0.07 & 1.51,0.41 & 0.31,0.06 & 0.74,0.19 & 0.62,0.18 & 0.55,0.15 \\
NGC 3109 & MTK 1 & G130M & 0.16,0.03 & 0.35,0.04 & \multicolumn{1}{c}{<\,0.08} & \multicolumn{1}{c}{<\,0.04} & 0.35,0.04 & 1.99,0.29 & \multicolumn{1}{c}{\nodata} & \multicolumn{1}{c}{\nodata} & \multicolumn{1}{c}{\nodata} & \multicolumn{1}{c}{\nodata} \\
 & & G140L & \multicolumn{1}{c}{<\,0.12} & 0.26,0.08 & \multicolumn{1}{c}{<\,0.08} & \multicolumn{1}{c}{<\,0.04} & 0.30,0.06 & 2.13,0.29 & 0.13,0.06 & \multicolumn{1}{c}{<\,0.32} & \multicolumn{1}{c}{\xmark{}} & \multicolumn{1}{c}{\xmark{}} \\
NGC 3109 & MTK 5 & G130M & 0.30,0.03 & 1.14,0.04 & 0.15,0.03 & 0.09,0.03 & 0.24,0.05 & 1.15,0.28 & \multicolumn{1}{c}{\nodata} & \multicolumn{1}{c}{\nodata} & \multicolumn{1}{c}{\nodata} & \multicolumn{1}{c}{\nodata} \\
 & & G140L & \multicolumn{1}{c}{<\,0.12} & 0.78,0.12 & \multicolumn{1}{c}{<\,0.12} & \multicolumn{1}{c}{<\,0.06} & 0.19,0.07 & 1.64,0.33 & \multicolumn{1}{c}{<\,0.12} & \multicolumn{1}{c}{<\,0.36} & \multicolumn{1}{c}{<\,0.34} & 0.45,0.16 \\
IC 1613 & BUG A13 & G130M & 0.16,0.04 & 0.43,0.04 & 0.19,0.03 & 0.06,0.02 & 0.57,0.03 & 2.65,0.23 & \multicolumn{1}{c}{\nodata} & \multicolumn{1}{c}{\nodata} & \multicolumn{1}{c}{\nodata} & \multicolumn{1}{c}{\nodata} \\
 & & G140L & 0.21,0.03 & 0.56,0.04 & 0.10,0.04 & \multicolumn{1}{c}{<\,0.02} & 0.53,0.04 & 2.89,0.18 & 0.40,0.04 & \multicolumn{1}{c}{<\,0.22} & 0.34,0.09 & \multicolumn{1}{c}{\xmark{}} \\
 & & G160M & \multicolumn{1}{c}{\nodata} & \multicolumn{1}{c}{\nodata} & \multicolumn{1}{c}{\nodata} & \multicolumn{1}{c}{\nodata} & \multicolumn{1}{c}{\nodata} & \multicolumn{1}{c}{\nodata} & 0.38,0.03 & \multicolumn{1}{c}{<\,0.18} & 0.34,0.06 & \multicolumn{1}{c}{\xmark{}} \\
IC 1613 & BUG B11 & G130M & \multicolumn{1}{c}{<\,0.08} & \multicolumn{1}{c}{\xmark{}} & 0.63,0.03 & 0.42,0.01 & 0.20,0.03 & \multicolumn{1}{c}{<\,0.46} & \multicolumn{1}{c}{\nodata} & \multicolumn{1}{c}{\nodata} & \multicolumn{1}{c}{\nodata} & \multicolumn{1}{c}{\nodata} \\
 & & G160M & \multicolumn{1}{c}{\nodata} & \multicolumn{1}{c}{\nodata} & \multicolumn{1}{c}{\nodata} & \multicolumn{1}{c}{\nodata} & \multicolumn{1}{c}{\nodata} & \multicolumn{1}{c}{\nodata} & 0.30,0.03 & 0.88,0.08 & 1.02,0.06 & 0.91,0.06 \\
IC 1613 & BUG B2 & G130M & 0.18,0.03 & 1.19,0.03 & 0.32,0.03 & 0.18,0.02 & 0.37,0.03 & 1.88,0.21 & \multicolumn{1}{c}{\nodata} & \multicolumn{1}{c}{\nodata} & \multicolumn{1}{c}{\nodata} & \multicolumn{1}{c}{\nodata} \\
 & & G140L & 0.24,0.05 & 1.04,0.06 & 0.14,0.04 & 0.08,0.02 & 0.41,0.05 & 1.14,0.27 & 0.27,0.05 & \multicolumn{1}{c}{<\,0.30} & 0.71,0.12 & 0.92,0.11 \\
IC 1613 & BUG B7 & G130M & 0.10,0.03 & \multicolumn{1}{c}{\xmark{}} & 0.28,0.03 & 0.15,0.02 & 0.16,0.03 & \multicolumn{1}{c}{<\,0.50} & \multicolumn{1}{c}{\nodata} & \multicolumn{1}{c}{\nodata} & \multicolumn{1}{c}{\nodata} & \multicolumn{1}{c}{\nodata} \\
 & & G140L & \multicolumn{1}{c}{<\,0.12} & \multicolumn{1}{c}{\xmark{}} & 0.11,0.05 & \multicolumn{1}{c}{<\,0.04} & 0.16,0.05 & \multicolumn{1}{c}{<\,0.54} & \multicolumn{1}{c}{<\,0.10} & \multicolumn{1}{c}{<\,0.32} & 0.54,0.12 & 0.38,0.12 \\
 & & G160M & \multicolumn{1}{c}{\nodata} & \multicolumn{1}{c}{\nodata} & \multicolumn{1}{c}{\nodata} & \multicolumn{1}{c}{\nodata} & \multicolumn{1}{c}{\nodata} & \multicolumn{1}{c}{\nodata} & 0.10,0.03 & \multicolumn{1}{c}{\nodata} & 0.64,0.07 & 0.31,0.07 \\
IC 1613 & GHV 62024 & G130M & 0.12,0.03 & 0.84,0.03 & 0.40,0.03 & 0.08,0.02 & 0.26,0.03 & 1.37,0.24 & \multicolumn{1}{c}{\nodata} & \multicolumn{1}{c}{\nodata} & \multicolumn{1}{c}{\nodata} & \multicolumn{1}{c}{\nodata} \\
 & & G140L & \multicolumn{1}{c}{<\,0.10} & 0.67,0.05 & 0.15,0.04 & 0.06,0.02 & 0.33,0.04 & 1.47,0.23 & 0.18,0.04 & \multicolumn{1}{c}{<\,0.28} & 0.41,0.11 & 0.62,0.11 \\
IC 1613 & GHV 64066 & G130M & 0.22,0.03 & 1.02,0.04 & 0.30,0.03 & 0.10,0.02 & 0.51,0.03 & 2.52,0.25 & \multicolumn{1}{c}{\nodata} & \multicolumn{1}{c}{\nodata} & \multicolumn{1}{c}{\nodata} & \multicolumn{1}{c}{\nodata} \\
 & & G140L & \multicolumn{1}{c}{<\,0.14} & 0.68,0.10 & \multicolumn{1}{c}{<\,0.16} & \multicolumn{1}{c}{<\,0.06} & 0.56,0.08 & 3.24,0.42 & \multicolumn{1}{c}{<\,0.18} & \multicolumn{1}{c}{<\,0.56} & \multicolumn{1}{c}{<\,0.46} & \multicolumn{1}{c}{<\,0.46} \\
IC 1613 & GHV 67559 & G130M & 0.07,0.03 & 1.44,0.04 & 0.44,0.03 & 0.24,0.02 & 0.30,0.04 & 1.13,0.24 & \multicolumn{1}{c}{\nodata} & \multicolumn{1}{c}{\nodata} & \multicolumn{1}{c}{\nodata} & \multicolumn{1}{c}{\nodata} \\
 & & G140L & \multicolumn{1}{c}{<\,0.10} & 1.34,0.05 & 0.26,0.04 & 0.16,0.02 & 0.27,0.04 & \multicolumn{1}{c}{<\,0.52} & 0.17,0.04 & 0.75,0.13 & 0.60,0.13 & 0.69,0.11 \\
IC 1613 & GHV 67684 & G130M & 0.16,0.03 & 1.22,0.03 & 0.32,0.03 & 0.16,0.02 & 0.63,0.03 & 1.79,0.21 & \multicolumn{1}{c}{\nodata} & \multicolumn{1}{c}{\nodata} & \multicolumn{1}{c}{\nodata} & \multicolumn{1}{c}{\nodata} \\
 & & G140L & \multicolumn{1}{c}{<\,0.10} & 0.84,0.09 & 0.21,0.05 & \multicolumn{1}{c}{<\,0.06} & 0.54,0.06 & 1.64,0.34 & 0.14,0.06 & \multicolumn{1}{c}{<\,0.38} & \multicolumn{1}{c}{\xmark{}} & 0.54,0.15 \\\hline
\end{tabular}}

\tablecomments{Equivalent widths (EWs) of ten photospheric absorption features in the HST/COS spectra (see Section~\ref{sec:photlines}). We require measurements greater than two times the uncertainty to be considered detected; otherwise the 2$\sigma$ upper limit is reported. EWs are not reported for lines that are clearly impacted by the stellar wind, and such cases are marked with an \xmark{} symbol.}
\end{table*}

\bibliography{tempos}
\bibliographystyle{aasjournalv7}

\end{document}